\documentclass[11pt]{article}
\usepackage[colorlinks=true,linkcolor=blue,citecolor=blue,urlcolor=blue]{hyperref}

\usepackage[english]{babel}
\usepackage{nomencl}
\makenomenclature
\usepackage{verbatim,amsmath,amssymb}
\usepackage{epsfig,float,color}
\usepackage{enumitem}
\usepackage[font=small,labelfont=bf]{caption}
\usepackage{subcaption}
\usepackage{geometry}
\usepackage{amsmath}
\usepackage{booktabs}  
\usepackage{titlesec}
\usepackage{natbib}
\usepackage[utf8]{inputenc}
\usepackage[T1]{fontenc}
\usepackage{refcount}  
\usepackage{graphicx}
\usepackage [english]{babel}
\usepackage [autostyle, english = american]{csquotes}
\usepackage{hyperref,url}
\usepackage{setspace}         
\newcommand {\bgap}{\textbf{{\textit{g}}}}

\DeclareUnicodeCharacter{0301}{\'{e}}
\usepackage[labelfont=bf]{caption}
\usepackage{array}
\usepackage{makecell}
\usepackage{xcolor}

\definecolor{darkblue}{rgb}{0,0,1}
\definecolor{col1}{rgb}{1,0,1}
\definecolor{col2}{rgb}{0,0.5,0}
\definecolor{col3}{rgb}{0.5,0,1}
\definecolor{col4}{rgb}{0.1,.75,0}

\hypersetup{pdftex=true, colorlinks=true, breaklinks=true, linkcolor=darkblue, menucolor=darkblue, pagecolor=darkblue, citecolor=darkblue, urlcolor=darkblue}

\newcommand{\bitm}{\begin{itemize}}
\newcommand{\eitm}{\end{itemize}}
\newcommand{\bnumr}{\begin{enumerate}}
\newcommand{\enumr}{\end{enumerate}}

\newcommand {\eqb}[1]{\begin{equation}\begin{array}{#1}}
\newcommand {\eqe}{\end{array}\end{equation}}

\newcommand {\esb}[1]{\begin{equation*}\begin{array}{#1}}
\newcommand {\ese}{\end{array}\end{equation*}}
\newcommand {\ds}{\displaystyle}

\newcommand {\pa}[2]{\frac{\partial{#1}}{\partial{#2}}}

\newcommand {\II}{{I\kern-.3em I}}
\newcommand {\III}{{I\kern-.3em I\kern-.3em I}}

\newcommand {\mra}{\mathrm{a}}
\newcommand {\mrb}{\mathrm{b}}
\newcommand {\mrc}{\mathrm{c}}
\newcommand {\mrd}{\mathrm{d}}
\newcommand {\mre}{\mathrm{e}}

\newcommand {\mrg}{\mathrm{g}}

\newcommand {\mrt}{\mathrm{t}}

\newcommand {\mrz}{\mathrm{z}}

\newcommand {\bx}{\boldsymbol{x}}

\newcommand {\IR}{{\rm\kern.24em
   \vrule width.02em height1.53ex depth-.05ex
   \kern-.3em R}}
\newcommand {\ic}{{\rm\kern.20em
   \vrule width.02em height1.0ex depth-.05ex
   \kern-.22em c}}
\newcommand {\ia}{{\rm\kern.20em
   \vrule width.02em height1.05ex depth-.0ex
   \kern-.25em a}}
\newcommand {\IC}{{\rm\kern.24em
   \vrule width.02em height1.4ex depth-.05ex
   \kern-.26em C}}
\newcommand {\ID}{{\rm\kern.34em
   \vrule width.02em height1.5ex depth-.05ex
   \kern-.36em D}}
\newcommand {\IS}{{\rm\kern.24em
   \vrule width.02em height1.6ex depth.05ex
   \kern-.26em S}}
\newcommand {\IT}{{\rm\kern.50em
   \vrule width.02em height1.55ex depth-.05ex
   \kern-.52em T}}

\newcommand {\IE}{{\rm\kern.24em
   \vrule width.02em height1.55ex depth-.05ex
   \kern-.33em E}}
\newcommand {\IEa}{{\rm\kern.24em
   \vrule width.02em height1.55ex depth-.05ex
   \kern-.33em E}^{1}_{ijkl}}
\newcommand {\IEb}{{\rm\kern.24em
   \vrule width.02em height1.55ex depth-.05ex
   \kern-.33em E}^{2}_{ijkl}}

\newcommand {\Ass}[2]{\kern 0.9ex \vrule width0.45em height0.2ex depth0ex \kern -2.1ex \bigwedge_{#1}^{#2}}
\newcommand {\ASS}[2]{\kern 1.45ex \vrule width0.5em height0.2ex depth0ex \kern -2.65ex \bigwedge_{#1}^{#2}}

\graphicspath{ {images/} }

\begin{document}

\begin{center}
\Large{\bf{Investigating the sliding behavior of graphene nanoribbons }}\\

\end{center}

\renewcommand{\thefootnote}{\fnsymbol{footnote}}

\begin{center}
\large{Gourav Yadav$^{\mra}$, Aningi Mokhalingam$^{\mrb}$,  Roger A.~Sauer$^{\mrc,\mrd,\mre}$\footnote[1]{Corresponding author, email: roger.sauer@rub.de} and Shakti S.~Gupta$^{\mra}$\footnote[7]{Corresponding author, email: ssgupta@iitk.ac.in} 
}\\
\vspace{4mm}

\small{\textit{
$^\mra$Department of Mechanical Engineering, Indian Institute of Technology Kanpur, UP 208016, India \\[1.1mm]
$^\mrb$Department of Mechanical Engineering, Maulana Azad National Institute of Technology Bhopal, MP 462003, India \\[1.1mm]
$^\mrc$Institute for Structural Mechanics, Ruhr University Bochum, 44801 Bochum, Germany \\[1.1mm]
$^\mrd$Department of Structural Mechanics, Gda\'{n}sk University of Technology, 80-233 Gda\'{n}sk, Poland \\[1.1mm]
$^\mre$Department of Mechanical Engineering, Indian Institute of Technology Guwahati, Assam 781039, India
\\
}}
\end{center}

\vspace{1mm}

\renewcommand{\thefootnote}{\arabic{footnote}}

\vspace{-5mm}

\vspace{2mm}

\vspace{-1mm}


\rule{\linewidth}{.15mm}
{\bf Abstract} \\
This work presents a Euler-Bernoulli beam finite element (FE) model to study the interlayer interaction mechanics of graphene nanoribbon (GNR) over a graphene substrate. The FE model is calibrated using molecular dynamics (MD) simulations employing the potential of \citet{Kolmogorov2005}. This study focuses mainly on the effect of boundary conditions on the sliding behavior, and on the strain transfer between layers when the substrate is subjected to uniform biaxial deformations. The interlayer sliding behavior is found to depend on the presence of critical parameters, namely, the length of the GNR and the applied strain to the substrate. The FE results indicate that the applied strain transferred from the substrate to the GNR varies linearly up to a critical value $\epsilon_{\text{c}}$ beyond which it decreases suddenly. Further, $\epsilon_{\text{c}}$ is found to appear beyond a critical GNR length, $L_{\text{e}}$ $\approx$14~nm. Furthermore, a length parameter $L_{\text{d}}$ $\approx$ 10~nm is computed, beyond which the sliding of GNR is dissipative. Through FE simulations, it is also found that for a GNR length $\geq$ 17~nm, the edge pulling force saturates. Our results also highlight the importance of the inertia of GNR on its sliding for different boundary conditions. It is also concluded that the maximum strain that can be transferred to GNR lies between 0.59$\%$ and 1.15$\%$. The results of the FE approach align with MD simulations within an error of approximately 10$\%$ that can be attributed to the choice of material parameters and the simulation setup. \\


\vspace{-5mm}
\rule{\linewidth}{.15mm}

\section{Introduction} \label{sec_1}
Stacked layers of two-dimensional (2D) materials, such as bilayer graphene sheets and graphene-hBN heterojunctions, have drawn extensive attention due to their exceptional mechanical \citep{G.Wang2017, Dai2020} and tribological properties \citep{Grattan1967, Bowden2001, Liechti2015, Verhoeven2004, Zheng2008, Berman2014}. These materials exhibit strong intralayer covalent bonding and relatively weak interlayer van der Waals (vdW) interactions, facilitating easy sliding between neighboring atomic layers. Their high in-plane stiffness enables incommensurable contact with most solids, and their weak interaction with other materials results in ultra-low friction during interfacial sliding. Moreover, interfacial sliding properties of bilayer graphene can be further tuned through strain engineering and variations in stacking arrangements \citep{Dienwiebel2004, Feng2013, Mandelli2017, Cao2018, Morovati2022}.

In the last decade, strain engineering has emerged as a powerful strategy for tuning not only tribological properties but also optoelectronic properties of 2D materials \citep{Liu2014, Roldan2015, Yang2016, Lin2019, Peng2020, Xu2022, Mallik2022, Basu2023, Rodriguez2024, Michail2024, Cenker2025}. Various methods have been developed to apply strain, including tensile strain induced by substrate bending \citep{Carrascoso2021, Rodriguez2024, Michail2024}, biaxial strain through thermal expansion of the substrate \citep{Plechinger2015}, and localized strain generated by external features such as nanobubbles, wrinkles, or nanopillars \citep{Sun2019, Hou2024}. These methods often face significant stability challenges (i.e., interlayer slippage, rippling, and kink formation) \citep{Chen2015, Liu2015, Deng2016}, resulting in inefficient strain transfer from the substrate to the 2D material, limiting the effective utilization of strain engineering in such systems.

The development of high-precision manufacturing has enabled the integration of sub-micrometer graphene structures into large-scale tribological systems \citep{Zhang2005, Heersche2007, Cai2010}. However, during the manufacturing and transfer processes of bilayer graphene (BLG), particularly at large length scales, defects are often unavoidable. Among these, incommensurate domain walls, are of particular significance \citep{Yin2016, Alden2013}. These defects have been shown to substantially influence the local electronic \citep{Shallcross2017} and optical \citep{Wang2010} properties of BLG, notably due to Dirac point transport phenomena. Recent advancements have also enabled the manipulation of individual dislocations in BLG via localized mechanical forces \citep{Jiang2018}. Beyond its utilization in friction reduction mechanisms, graphene also exhibits significant potential for diverse applications, including nanoelectromechanical systems (NEMS) \citep{Wei2009}, nanofillers \citep{Rafiee2010}, transistors \citep{Li2008}, and various electronic, spintronic devices \citep{Terrones2010}, surface coating, and composites \citep{KAMBOJ2023}. Given the broad scope of applications and the complex phenomena associated with such materials, a thorough understanding of interlayer interaction mechanics is essential for the continued development of 2D materials.

To advance the contact understanding, experimental studies have investigated the sliding mechanics of graphene layers using friction force microscopy (FFM) \citep{Dienwiebel2004, Dienwiebel2005, Zheng2008, Feng2013}. \citet{Verhoeven2004} studied the friction between an FFM tip with an attached graphene flake and a graphite substrate using the modified Prandtl-Tomlinson model \citep{Tomlinson1929} and found that the frictional forces depend on the flake size and relative rotation between the graphene flake and substrate, with an angular periodicity of 60$^{\circ}$. Similar angular direction sliding dependency was observed by \citet{Dienwiebel2005}. They measured ultra-low friction or `superlubricity' due to incommensurability and formation of Moiré superlattices arising from lattice mismatch. \citet{Guo2007} examined the effect of the interlayer gap on static friction between a graphene flake and a graphite substrate in both commensurate and incommensurate stacking. Their findings indicate that incommensurate contact still results in ultralow friction at reduced gaps (higher contact loads) of approximately 0.24 nm. 

Atomistic simulation techniques, including density functional theory (DFT), which incorporates exchange-correlation functionals into the Hamiltonian of the Schrödinger's equation \citep{kohn1996}, and molecular dynamics (MD), which employs Newton’s second law, offer robust alternatives to experimental methods by enabling detailed, atomic-scale investigation of interlayer mechanics. Among these approaches, MD simulations are especially advantageous, striking an effective balance between computational efficiency and predictive accuracy. Several studies have used MD to investigate the role of interlayer interactions and the resulting structural deformations on frictional sliding behavior and strain transfer between graphene layers \citep{Guo2007, Xu2011, Popov2011,  ZHANG2015, Popov2017, G.Wang2017, wang2017size, Mandelli2017, Ouyang2018, Li2020, Yang2020}. 
For instance, in the context of graphene nanoribbons (GNRs) -- which are essentially narrow strips of graphene with width much smaller than their longitudinal dimension, \citet{Kawai2016} conducted edge-driven sliding experiments over a gold substrate and demonstrated that the friction force per unit length decreases with increasing GNR length. Subsequently, \citet{wang2017size} and \citet{Ouyang2018} employed MD simulations to investigate length-dependent friction behavior in edge-driven bilayer graphene systems. Their results show that the maximum shear force initially increases with length but saturates beyond approximately 20 nm.

Owing to its atomically smooth surface, graphene lends itself well to precise modeling using theoretical and finite element (FE) methods. FE models based on shell, plate, beam, rod, or truss elements provide useful approximations, but replacing a discrete atomic lattice with a homogeneous medium remains a challenging task. An alternative is the atomistic FE approach, in which carbon–carbon bonds are represented by energetically equivalent beam elements forming a space-frame network \citep{li2003structuralb}. This equivalence is established by matching interatomic potential energy with the strain energy of beam elements. Unlike classical shell and plate element models, this formulation is thickness-free, directly parameterized by bond-level force constants \citep{Scarpa_Adhikari_Phani_2009, scarpa2010bending,  Chandra_Chowdhury_Adhikari_Scarpa_2011_PhysicaE}, and preserves lattice topology, thereby capturing orientation-dependent and crystalline properties with higher fidelity \citep{cao2018review}. This approach has been extensively applied to investigate the mechanical properties of graphene and related nanostructures. \citet{li2003structuralb} and \citet{Gajbhiye2016} investigated the elastic properties of graphene sheets, while \citet{Tsiamaki2014} applied the method to monolayer graphene resonators for mass detection. \citet{Tserpes2015} extended the approach to study both pristine and defect graphene. For multilayer systems, \citet{Chandra2020} analyzed the vibrational response of bilayer graphene, and \citet{namnabat2021nonlinear} examined their buckling characteristics. Beyond graphene, \citet{Lu2012} investigated the mechanical behavior of carbon nanotubes, and more recently, \citet{Nickabadi2024} applied the approach to vibrational analysis of double-walled silicon carbide nanocones, demonstrating its adaptability to non-carbon two-dimensional materials.


Extending the classical Frenkel-Kontorova model \citep{Kontorova1938}, \citet{Lebedeva2019} theoretically demonstrated the formation of an equilateral triangular network of incommensurate domain walls separating commensurate regions, triggered by a critical homogeneous strain ($\approx$0.3$\%$) applied to the bottom layer. However, this study did not consider the influence of system size on strain transmission.
For sliding behavior, \citet{Xue2022} employed finite element model to investigate edge-driven sliding of GNRs, highlighting the size-dependent nature of the response, the formation of strain solitons, and their gliding mechanisms, which collectively lead to stick-slip motion. However, their analysis did not account time dependency, which can significantly influence the overall sliding response. Furthermore, to validate their results for the unconstrained sliding of armchair GNR, they compared their results to those reported by \citet{Mandelli2017}. However, this comparison is flawed, as the peak magnitudes they sought to match correspond to different wavelengths of the applied pulling forces, implying differing boundary conditions.

In this study, we develop a 1D Euler-Bernoulli beam-based FE framework to study the influence of GNRs' size under different boundary conditions and time dependency on the sliding behavior. As highlighted in previous studies \citep{Ouyang2018, Xue2022}, the width of graphene primarily affects interlayer sliding behavior quantitatively. Consequently, in this work, a GNR with a width of approximately $\approx 0.7$ nm is adopted as the representative system instead of the full graphene sheet. Our FE model, calibrated from MD simulations, is employed to investigate critical length scales that characterize energy dissipation mechanisms during sliding and strain transmission. Certain length scales are determined by identifying the associated energy conversion pathways that distinguish dissipative and non-dissipative sliding regimes, while other length scales are obtained by studying strain transfer to the GNR by uniformly straining the bottom graphene sheet (substrate). Using the calibrated FE model, we examine edge-driven sliding behavior across different GNR lengths, loading directions, boundary conditions, and both static and dynamic conditions. Additionally, the FE results are validated against MD simulations, confirming the robustness and reliability of the proposed modeling approach. In summary, the key novelties of the current work are:
\begin{enumerate}[label=(\roman*), noitemsep, topsep=0pt]
    \item Comprehensive investigation of the static and dynamic interfacial sliding behavior under various boundary conditions.
    \item Identification and provision of critical length scales and bounds governing interfacial sliding behavior.
    \item Validation and verification of the FE results from MD data.
\end{enumerate}

The remainder of this paper is organized as follows: Section~\ref{Section 2} outlines the modeling approach, the applied boundary conditions, and the details of the FE formulation. Section \ref{Section 3} presents the FE simulation results for both edge-pulled GNR sliding and strain transmission. Finally, conclusions are provided in Section \ref{Section 4}.

\section{Methodology} \label{Section 2}

This section first derives a continuum expression for the vdW interaction forces between the substrate and an atom. Three boundary conditions, along with the representative sliding paths of the GNR atoms, are discussed. Finally, the FE method formulation is presented. 
\subsection{Interactions and material modeling}\label{subsection 2.1}
The potential energy of an atom interacting with a graphene sheet is found to be a function of the gap vector $\textbf{{\textit{g}}}$ and is commonly written in the form \citep{Steele1973, Lebedeva2010, Lebedeva2011, Mokhalingam2024}
\eqb{l}

\begin{figure}[H]
\begin{center} \unitlength1cm
\begin{picture}(0,5.3)
\put(-8,0.5){\includegraphics[height=4cm]{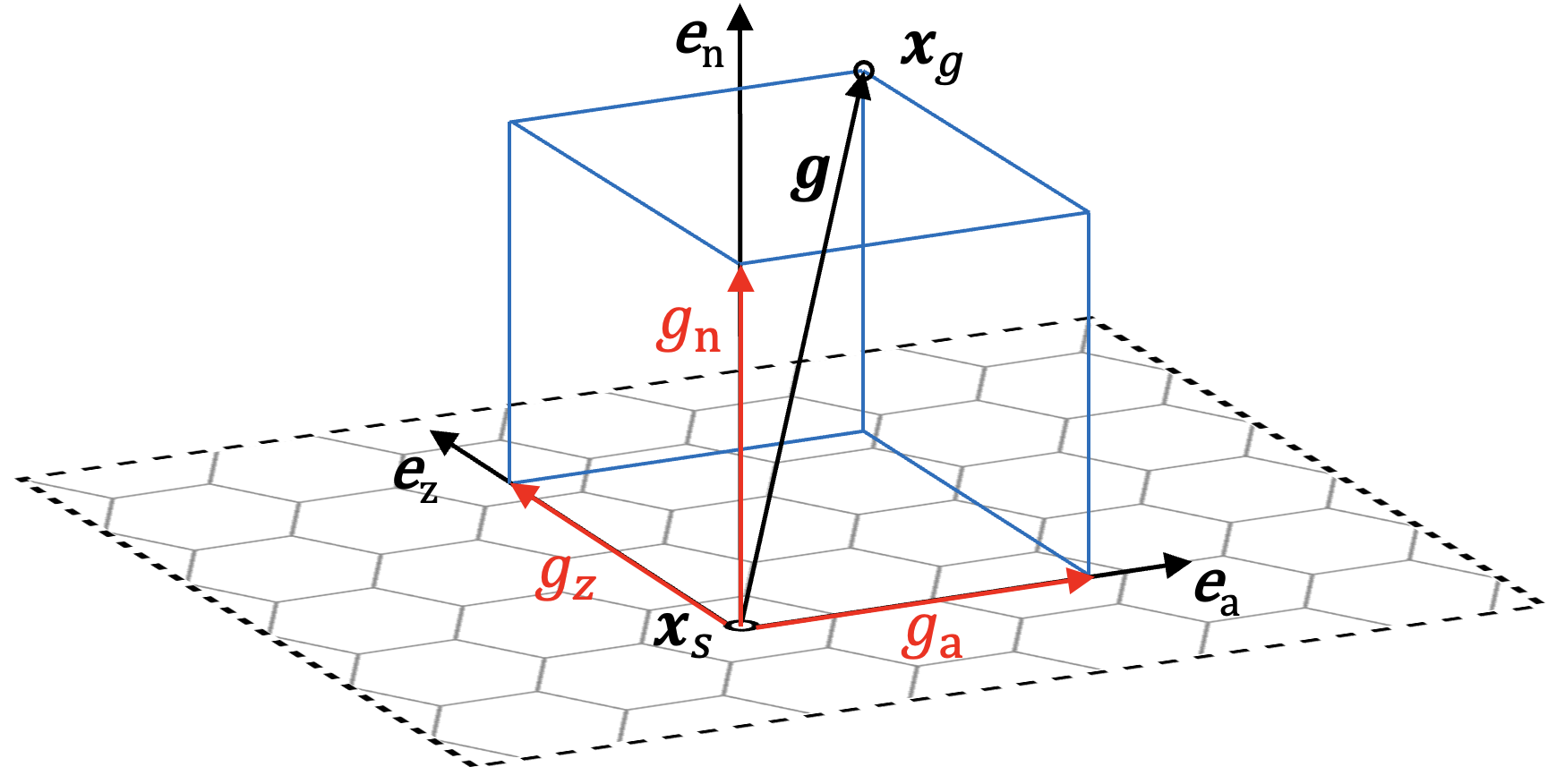}}
\put(0,0){\includegraphics[height=5.5cm]{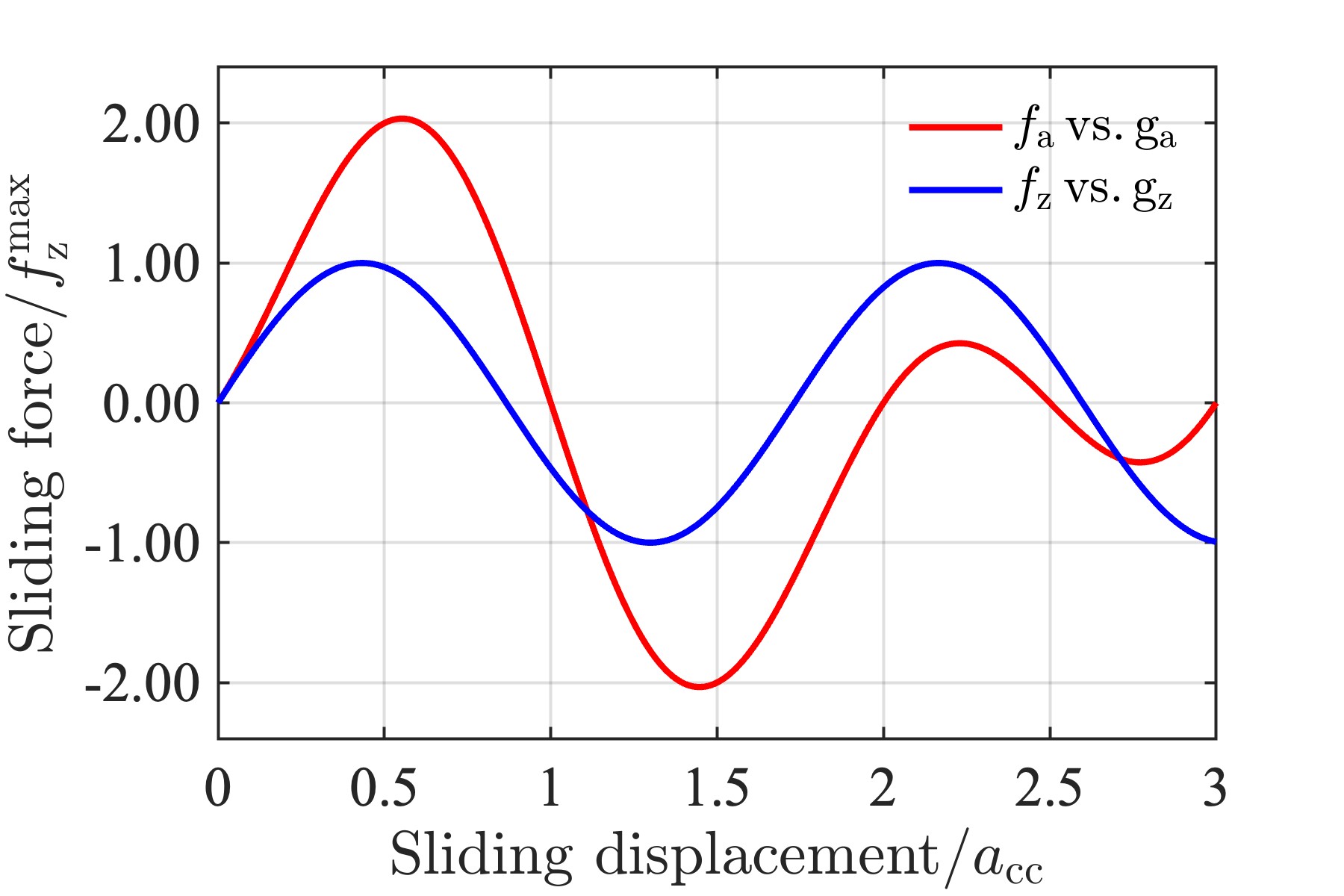}}
\put(-7.5,0){(a)}
\put(0,0){(b)}
\end{picture}
\caption{(a) Illustration of the gap vector $\bgap$ defined between the position $\mathcal{\bx}_{g}$ of a GNR atom and the center $\mathcal{\bx}_{s}$ of a unit cell in the underlying graphene substrate.
(b) Shear force on a carbon atom as a function of the sliding displacement. (Here $f_z^{\text{max}}$ = $8\pi\Psi_2/l_z$)} 
\label{gap_vector_schematic}
\end{center}
\end{figure}

\ds \Psi(\bgap) = \ds \Psi_1(\mathit{g}_\mathrm{n}) +  \Psi_2(\mathit{g}_\mathrm{n})\,\bar\Psi_\mrt(\mathit{g}_\mathrm{a},\mathit{g}_\mathrm{z})~. \,
\label{energy expression}\eqe
Here, $\bgap$ consists of components $\mathit{g}_\mathrm{a}$, $\mathit{g}_\mathrm{z}$, and $\mathit{g}_\mathrm{n}$, which represent the atomic displacements in the armchair  ${ \boldsymbol{e}}_\mathrm{a}$, and zigzag ${ \boldsymbol{e}}_\mathrm{z}$, and normal  ${ \boldsymbol{e}}_\mathrm{n}$,  directions, respectively, of the graphene substrate as shown in Fig.~\ref{gap_vector_schematic}a.
$\Psi_1(\mathit{g}_\mathrm{n})$ describes the normal contact potential, and is commonly expressed in the form \citep{Hamaker1937}

\begin{equation}
\ds \Psi_1(\mathit{g}_\mathrm{n}) = \Psi_{01}\Bigg(\frac{1}{10}\bigg(\frac{\mathit{g}_{01}}{\mathit{g}_\mathrm{n}}\bigg)^{\!\!10} \!\!- \frac{1}{4}\bigg(\frac{\mathit{g}_{01}}{\mathit{g}_\mathrm{n}}\bigg)^{\!\!4}\Bigg)\,,
    \label{Psi_1 from}
\end{equation}

where $\Psi_{01}$ and $\mathit{g}_{01}$ denote the homogenized energy constant and the homogenized gap constant, respectively. 

The tangential contact behavior is described by $\Psi_2\,\bar\Psi_{\mrt}$. Here $\Psi_2(\mathit{g}_\mathrm{n})$ represents the gap-dependent amplitude of energy modulations and is obtained using MD simulations following the procedure mentioned in \citet{Mokhalingam2024}. The MD-calibrated expression is

\begin{equation}
    \ds \Psi_2(\mathit{g}_\mathrm{n}) := 
\Psi_{02}\exp\bigg(\!\!-\ds\frac{\mathit{g}_\mathrm{n}}{\mathit{g}_{02}}\bigg)~,
\label{Psi2}
\end{equation}
where $\Psi_{02}$ and $\mathit{g}_{02}$ denote the modulating energy constant and the modulating gap constant, respectively.
The function $\bar\Psi_\mrt(\mathit{g}_\mathrm{a},\mathit{g}_\mathrm{z})$, referred to as the energy modulation function, captures the variation of the interaction energy in the tangential (in-plane) directions. It accounts for the two-dimensional periodicity of the substrate and is typically expressed in terms of the first Fourier components in reciprocal space \citep{Steele1973, Carlos1979, Verhoeven2004, Lebedeva2010, Lebedeva2011},
\eqb{l}
\bar\Psi_\mrt(\mathit{g}_\mathrm{a},\mathit{g}_\mathrm{z}) = \ds2\cos\frac{4\pi\mathit{g}_\mathrm{a}}{\ell_\mra} + 4\cos\frac{2\pi\mathit{g}_\mathrm{a}}{\ell_\mra} \cos\frac{2\pi\mathit{g}_\mathrm{z}}{\ell_\mrz}~.
\label{modulating function}\eqe

The corresponding force on each carbon atom due to the interaction with the substrate then follows as

\begin{subequations}
\begin{align} \label{force acting on individual atom}
  \ds  f_\mra & :=  -\ds\pa{\Psi}{\mathit{g}_\mathrm{a}}  = \ds\frac{8\pi\Psi_2}{\ell_\mra}\bigg(\!\sin\frac{4\pi\mathit{g}_\mathrm{a}}{\ell_\mra} + \sin\frac{2\pi\mathit{g}_\mathrm{a}}{\ell_\mra} \cos\frac{2\pi\mathit{g}_\mathrm{z}}{\ell_\mrz}\bigg)\,, \\
\ds   f_\mrz  & := -\ds\pa{\Psi}{\mathit{g}_\mathrm{z}}  = \ds\frac{8\pi\Psi_2}{\ell_\mrz} \cos\frac{2\pi\mathit{g}_\mathrm{a}}{\ell_\mra} \sin\frac{2\pi\mathit{g}_\mathrm{z}}{\ell_\mrz}~.
  \end{align}
  \label{force acting on individual atom}
\end{subequations}

Here, $f_\mra$ and $f_\mrz$ refer to the force components acting in the armchair and zigzag direction, respectively, and $\ell_\mra$ and $\ell_\mrz$ are the graphene lattice constants (see Table~\ref{table1}). The corresponding force variation is shown in Fig.~\ref{gap_vector_schematic}b. As observed from the figure, the maximum value of $f_\mra$ is approximately twice that of $f_\mrz$, and the slopes are equal for small displacements.  In this work, sliding between the layers is considered at the constant normal gap $\mathit{g}_\mathrm{n}$ = 0.34 nm. The constants in $\Psi_1$ and $\Psi_2$ listed in Table~\ref{table1} are determined by fitting the single atom interaction energy from MD simulation for two in-plane locations: ($\mathit{g}_\mathrm{a}$ = 0, $\mathit{g}_\mathrm{z}$ = 0) and ($\mathit{g}_\mathrm{a}$ = 0, $\mathit{g}_\mathrm{z}$ = $a_{\text{cc}}/\sqrt3$). The procedure of the MD simulation is discussed subsequently. 

The GNR is modeled such that each atom is represented as a nodal point mass and each covalent bond is a massless 1D Euler-Bernoulli beam element. The equivalent beam parameters are obtained through the equivalence between structural mechanics and atomistic mechanics \citep{Chandra2020}, and the corresponding mass, lattice, and structural parameters are summarized in Table~\ref{table1}. It is assumed that the elemental strains are small, allowing the beam parameters to be considered constant.

\setlength{\tabcolsep}{4pt}         

\begin{table}[H]
    \centering
    {\fontsize{10}{12}\selectfont 
   
    \renewcommand{\arraystretch}{1} 
    \begin{tabular}{@{}|>{\rule{0pt}{12pt}}l|>{\rule{0pt}{12pt}}l|>{\rule{0pt}{12pt}}l|@{}}
        \hline
        Model parameters & Symbol& Value  \\
        \hline
        Carbon-Carbon bond length & $a_{\text{cc}}$ & 0.1397 \, \text{nm}  \\
Unit cell area & $A_0 ${\vphantom{$^{2}$}} & $3\sqrt{3}a_{\text{cc}}^2/2 \, \text{nm}^2 $  \\
Armchair lattice parameter &  $\bar{l}_{\mathrm{a}}$  & 3   \\
Zigzag lattice parameter &  $\bar{l}_{\mathrm{z}}$ & $\sqrt{3}$  \\
Armchair lattice constant &  $l_{\mathrm{a}}$  & $\bar{l}_{\mathrm{a}}\,a_{\text{cc}}$\\
Zigzag lattice constant &  $l_{\mathrm{z}}$ & $\bar{l}_{\mathrm{z}}\,a_{\text{cc}}$  \\
 Beam Young's modulus &  $E$  & 34.2533 $\times$ $10^3$ $\text{eV}\,\text{nm}^{-3}$\\
   Beam second moment of area &  $I$  & 0.22682 $\times$ $10^{-4}$ $\text{nm}^4$ \\
        Beam cross-sectional area &  $A$ &  1.687 $\times$ $10^{-2}$ $\text{nm}^2$\\
   Mass of carbon atom &  $m_\text{c}$ & 0.1237 $\text{eV}$ $\text{ps}^2$\,$\text{nm}^{-2}$ (1.99 $\times$ $10^{-26}$ $\text{kg}$)\\
        Integration step size &  $\Delta t$ & 1 $\text{ps}$\,\,\footnotemark[1] \\
        Damping &  $c$ &0.1237 $\text{meV}$ \,$\text{ps}$ $\text{nm}^{-2}$ (1.99 $\times$ $10^{-17}$ $\text{kg/s}$)\,\,\footnotemark[1]\\
         Homogenized energy constant &  $\Psi_{01}$  & 264 meV (per atom)\\
        Homogenized gap constant  &  $\mathit{g}_{01}$ &  0.34 nm \\
         Modulating energy constant &  $\Psi_{02}$  & 93.23 eV (per atom)\\
         Modulating gap constant &  $\mathit{g}_{02}$ &  3.14 $\times$ $10^{-2}$ nm \\
        \hline
    \end{tabular}
    }
    \caption{Material and potential parameters used in the graphene model. ($E$, $I$, and $A$ are taken from the works of \citet{Baykasoglu2012}, \citet{Resketi2018} and \citet{Chandra2020}.}
    \label{table1}
\end{table}
\footnotetext[1]{Variations of these parameters were examined to verify the convergence of the FE results (see Appendix \ref{Appendix A}). The selected values of $\Delta t$ and $c$ were found to be adequate to guarantee convergence.}

\subsection{Boundary conditions and sliding paths} \label{subsection 2.2}

\begin{figure}[H]
\begin{center} \unitlength1cm
\begin{picture}(0,4.1)
\put(-7,0.5){\includegraphics[height=3.5cm]{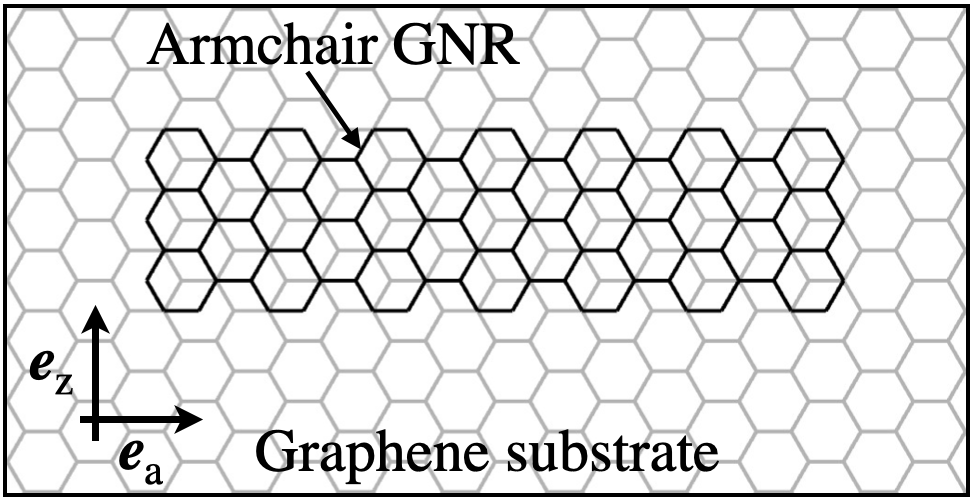}}
\put(1,0.5){\includegraphics[height=3.5cm]{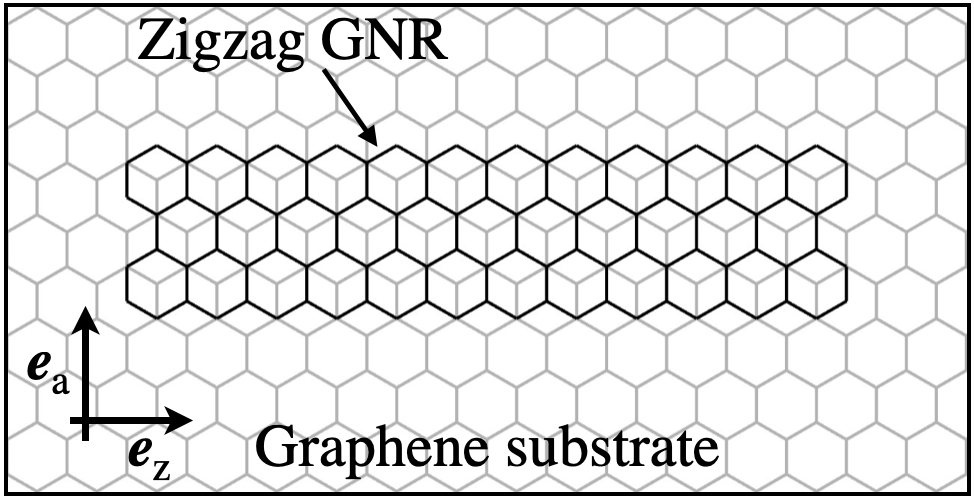}}
\put(-7.5,0){(a)}
\put(0.5,0){(b)}
\end{picture}
\caption{(a) Armchair GNR. (b) Zigzag GNR, each aligned with the corresponding direction of the graphene substrate.} 
\label{two bilayer graphene configuration}
\end{center}
\end{figure}

To investigate the interfacial behavior between a GNR and a graphene substrate, two commensurate configurations are considered. The first configuration is an armchair GNR aligned with the armchair direction ${\boldsymbol{e}}_\mathrm{a}$ (see Fig.~\ref{two bilayer graphene configuration}a), and the second is a zigzag GNR aligned with the zigzag direction ${\boldsymbol{e}}_\mathrm{z}$ (see Fig.~\ref{two bilayer graphene configuration}b). The nomenclature of these directions follows from the orientation of the GNR edges. All other orientations of a GNR on the graphene substrate lead to incommensurate configurations, characterized by an irrational ratio of the layers’ lattice constants. Therefore, only these two commensurate configurations are considered in this study.

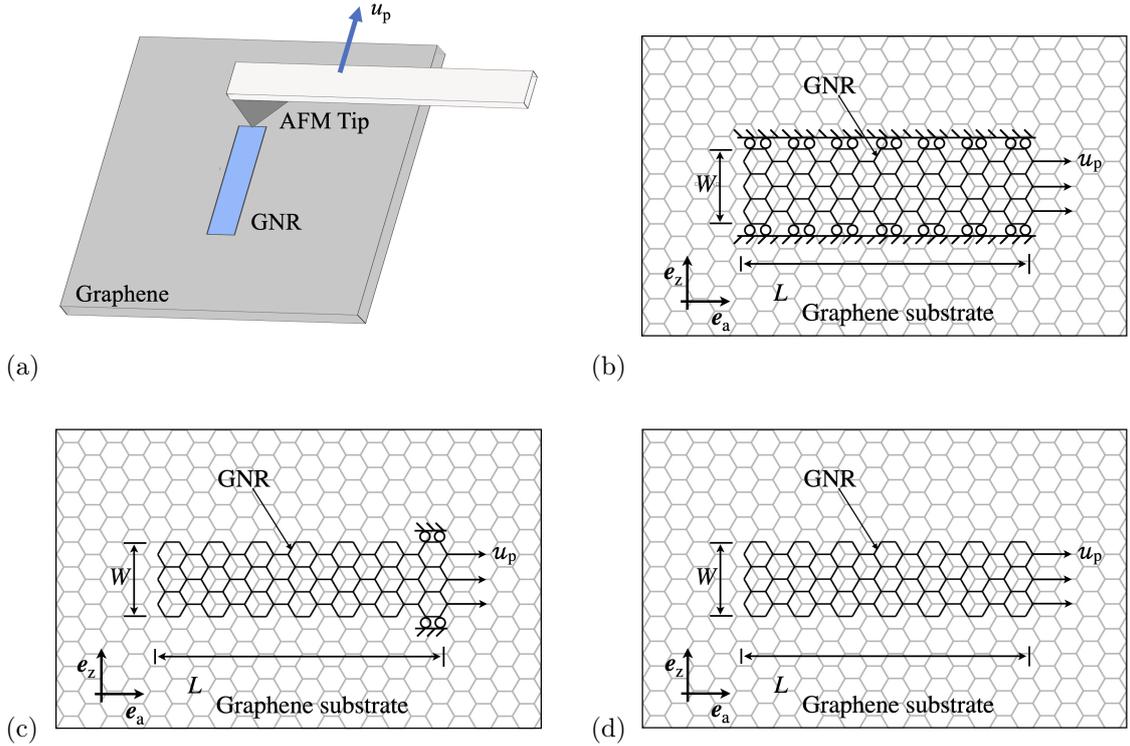
\begin{figure}[H]
\begin{center} \unitlength1cm
\begin{picture}(0,11)
\put(-6.8,5.5){\includegraphics[height=45mm]{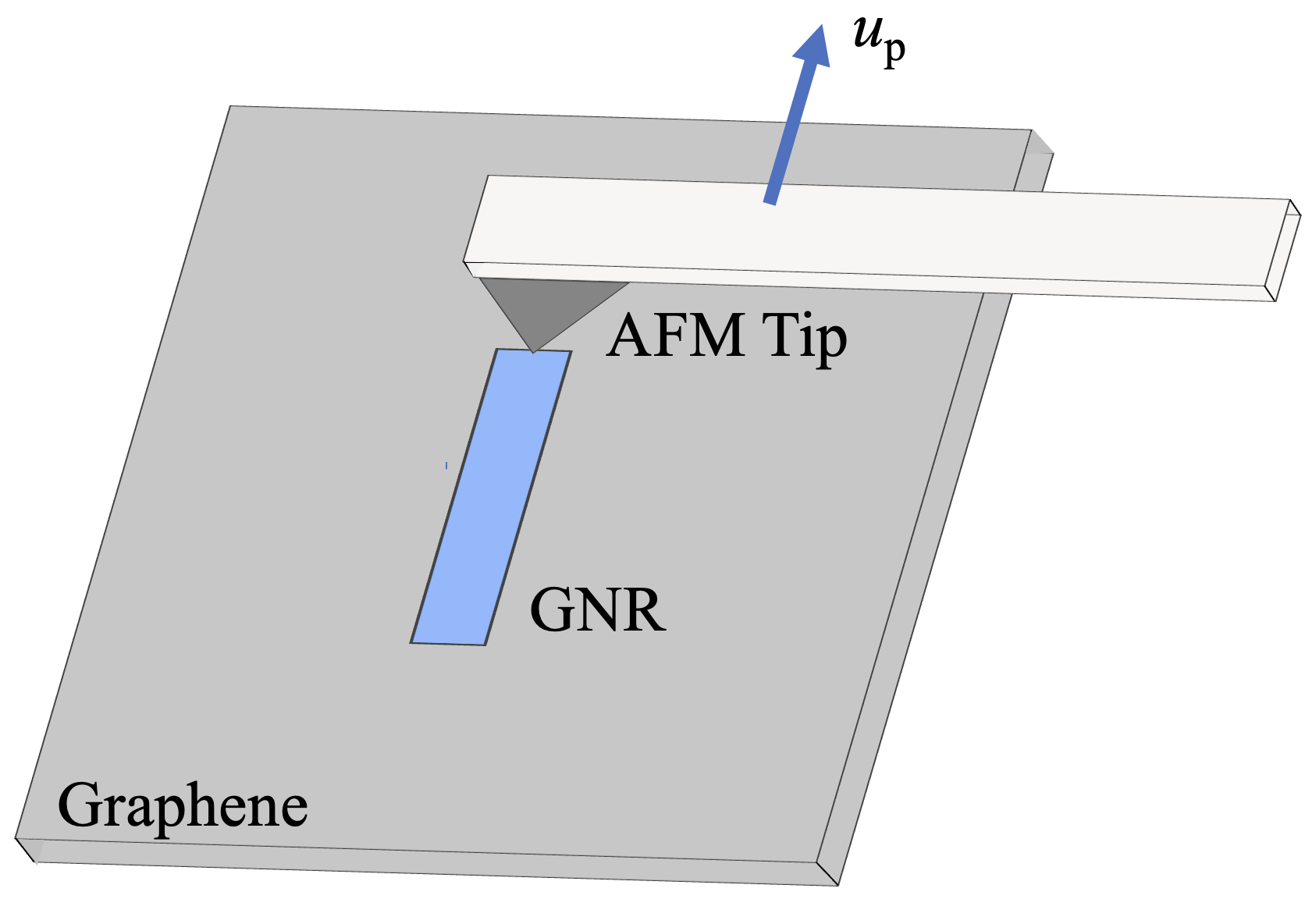}}
\put(0.9,5.5){\includegraphics[height=40mm]{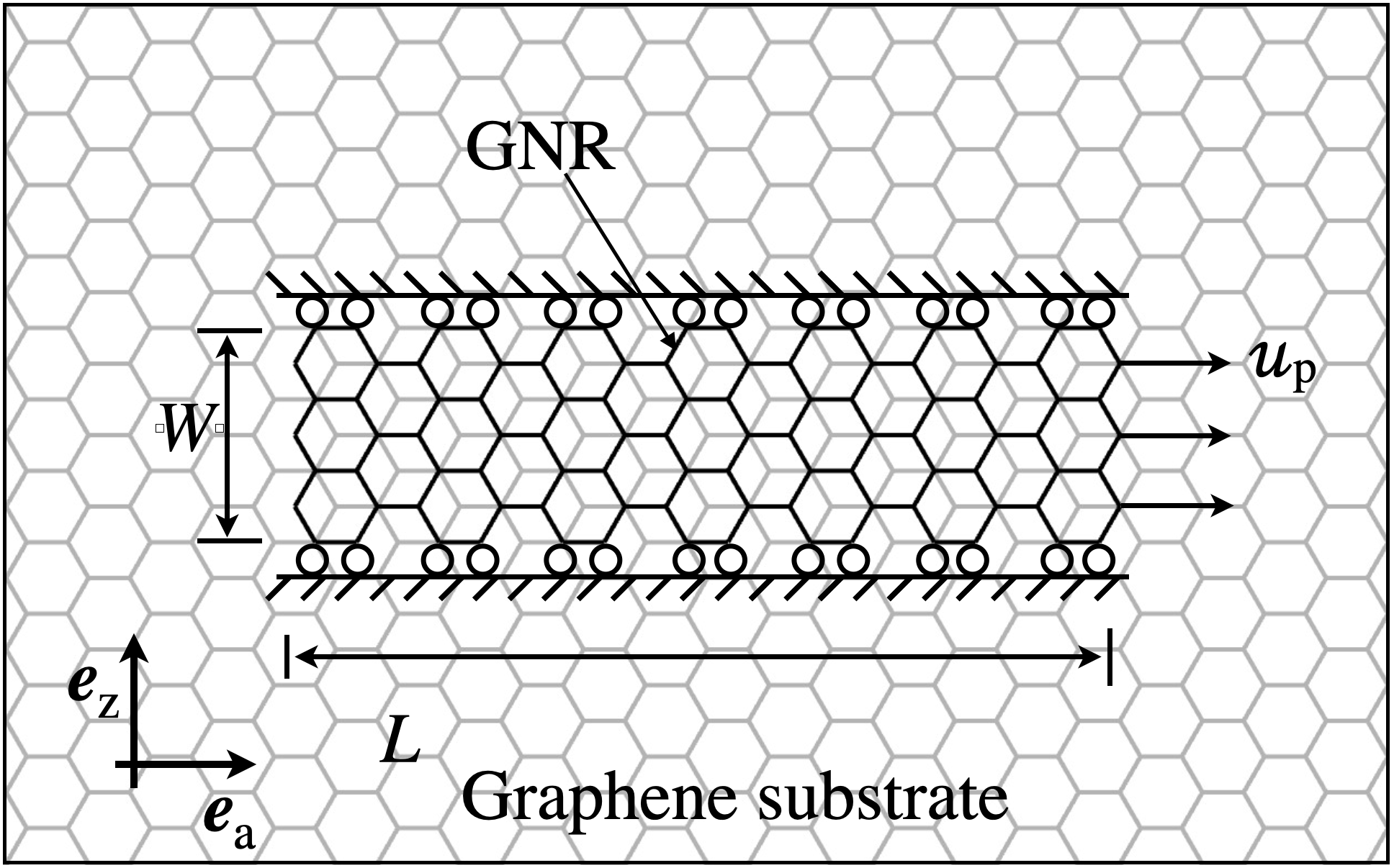}}

\put(-6.8,0.3){\includegraphics[height=40mm]{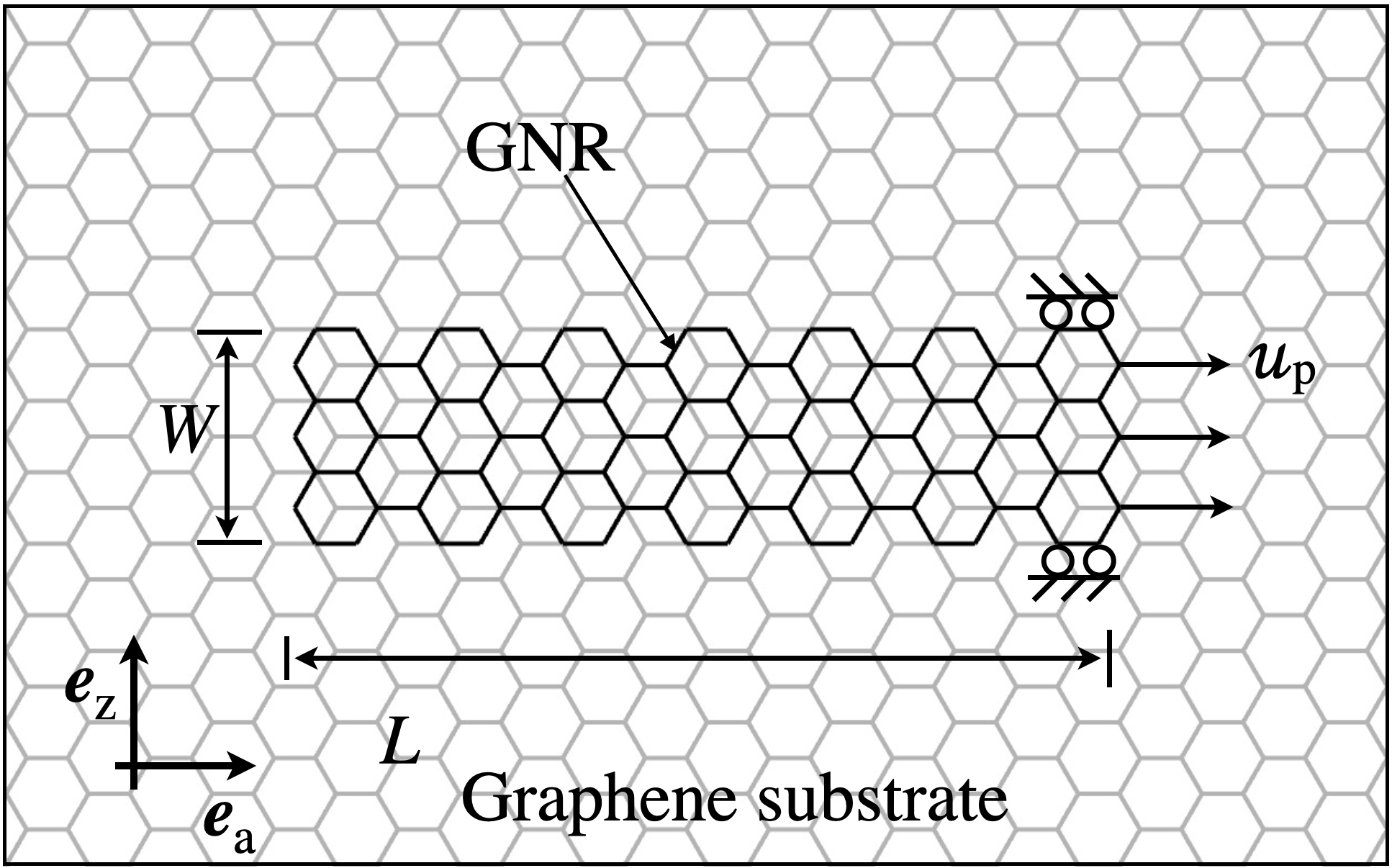}}
\put(0.9,0.3){\includegraphics[height=40mm]{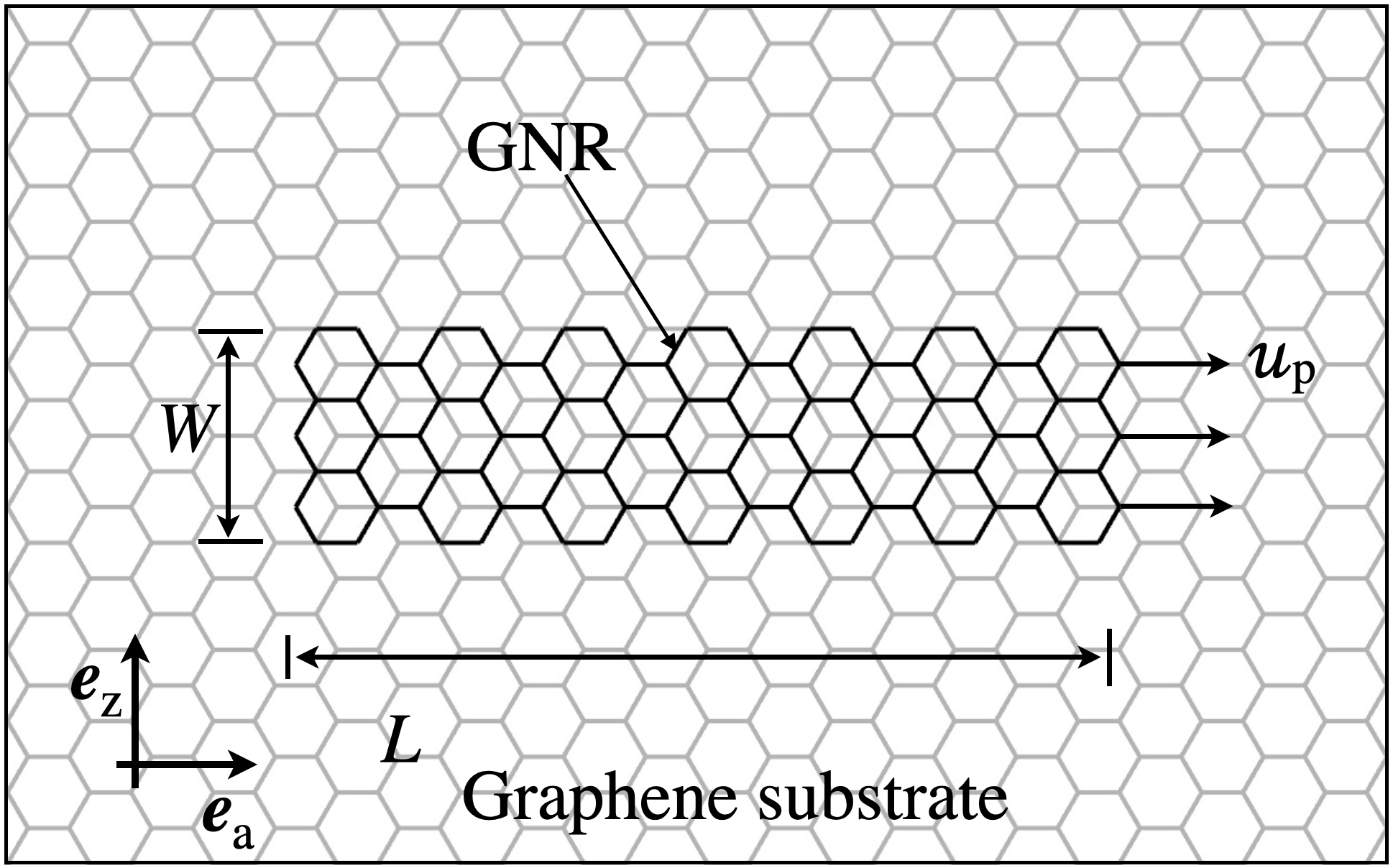}}

\put(-7.45,5){\fontsize{10pt}{20pt}\selectfont (a)}
\put(0.25,5){\fontsize{10pt}{20pt}\selectfont (b)}
\put(-7.45,0.2){\fontsize{10pt}{20pt}\selectfont (c)}
\put(0.25,0.2){\fontsize{10pt}{20pt}\selectfont (d)}

\end{picture}
\caption{ (a) Sliding of a GNR on a graphene substrate by pulling with an AFM tip. (b-c) top view of an edge-driven armchair GNR of width 
$W$ and length $L$, with prescribed displacement $u_\mathrm{p}$ at the pulling end along the armchair axis of the graphene substrate. Three boundary conditions are illustrated: (b)
BC 1, (c) BC 2, and (d) BC 3. }
\label{schematic setup}
\end{center}
\end{figure}

\newpage
In friction experiments of 2D materials, the motion of the material is highly influenced by its elastic properties, which have been well documented in the literature, particularly in the review article by \citet{Wang2024}. For graphene flakes, the probing tip is typically attached at the center of mass, whereas for a GNR, its one end is anchored to the probing tip of the AFM (see schematic shown in Fig.~\ref{schematic setup}a) and dragged back and forth in a controlled manner while the friction force is recorded \citep{Verhoeven2004, Dienwiebel2004, Dienwiebel2005, Lee2010, Filleter2009}. The resulting dynamics are governed by the interplay of material elasticity, interfacial interaction energy, and boundary conditions.

Here, the following three boundary conditions for edge pulling of GNR initially in commensurate `AB' stacking with the substrate are considered as shown in Fig.~\ref{schematic setup}b-\ref{schematic setup}d:
\begin{enumerate}[label=(\arabic*)]
    \item BC 1: Lateral edge constraint 
    \item BC 2: Lateral constraint of GNR head and body free
    \item BC 3: Lateral free
\end{enumerate}

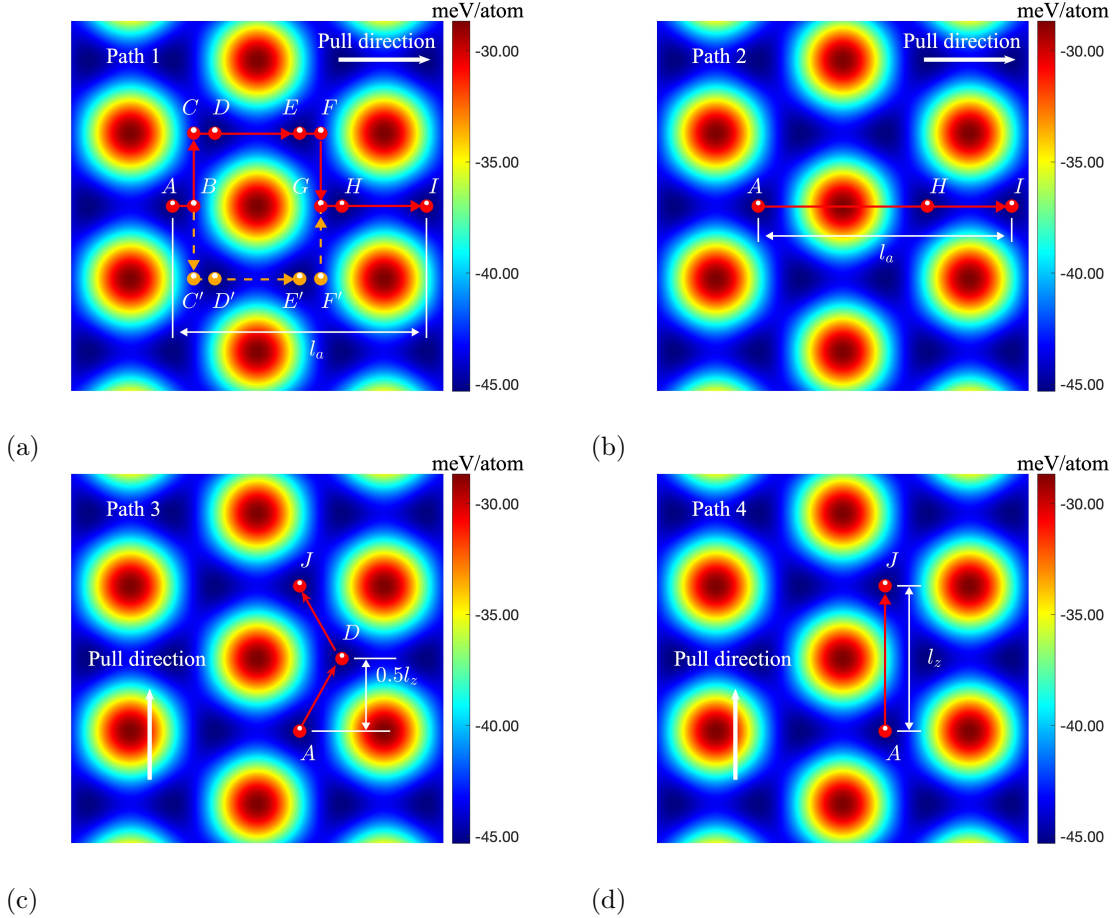
\begin{figure}[h]
\begin{center} \unitlength1cm
\begin{picture}(0,13)
\put(-7.7,6.5){\includegraphics[height=60mm]{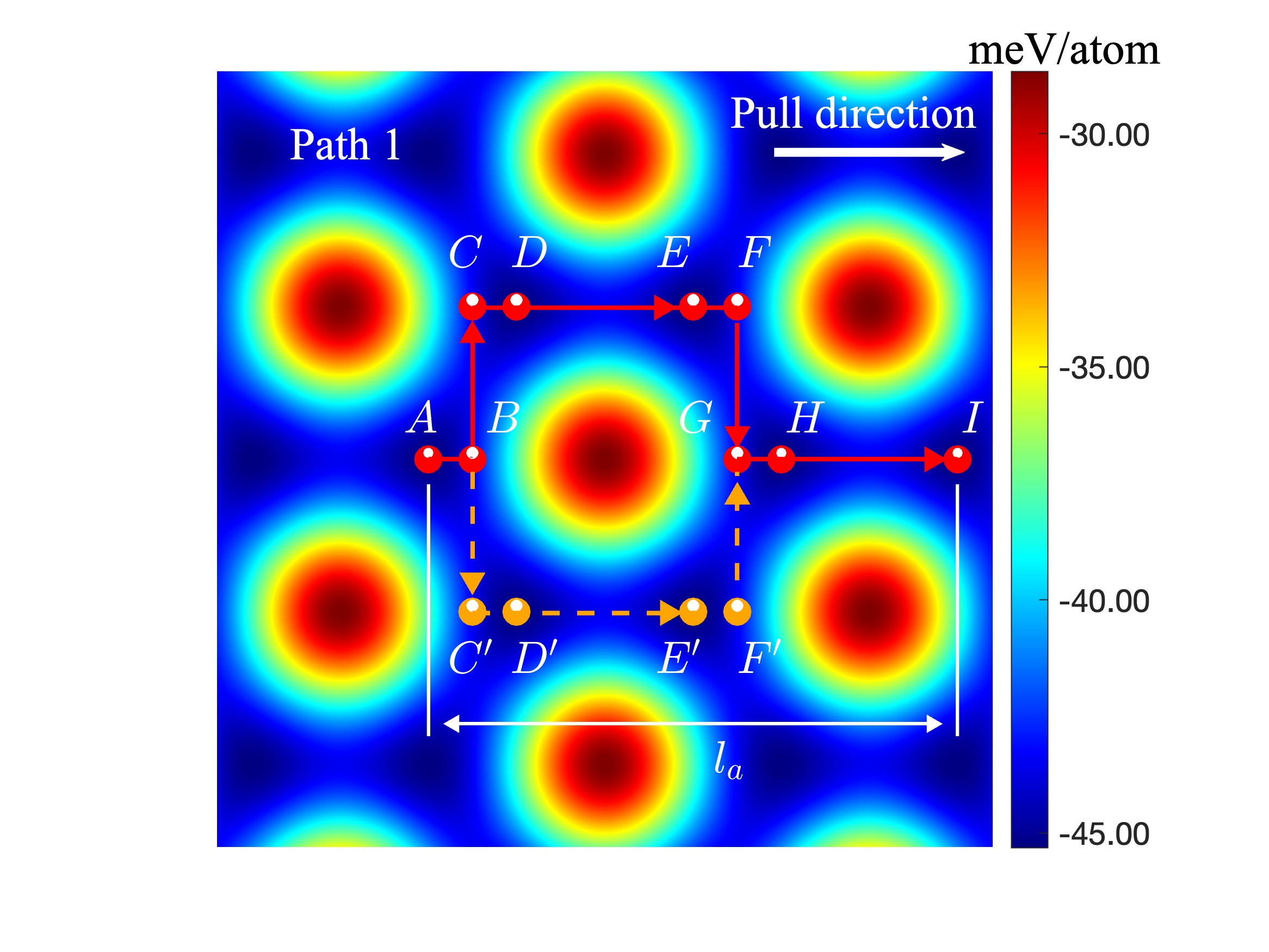}}
\put(0,6.5){\includegraphics[height=60mm]{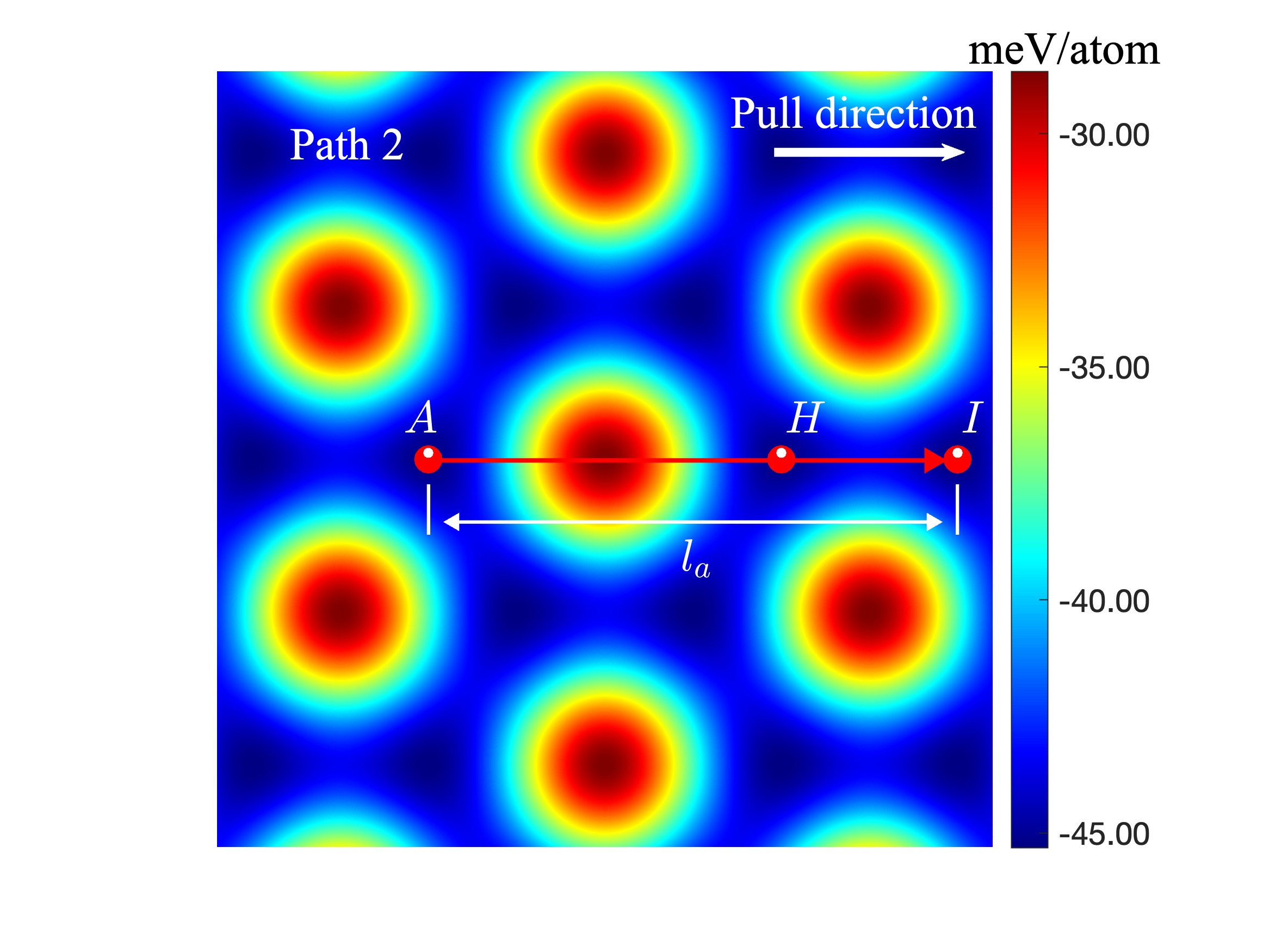}}
\put(-7.7,0.5){\includegraphics[height=60mm]{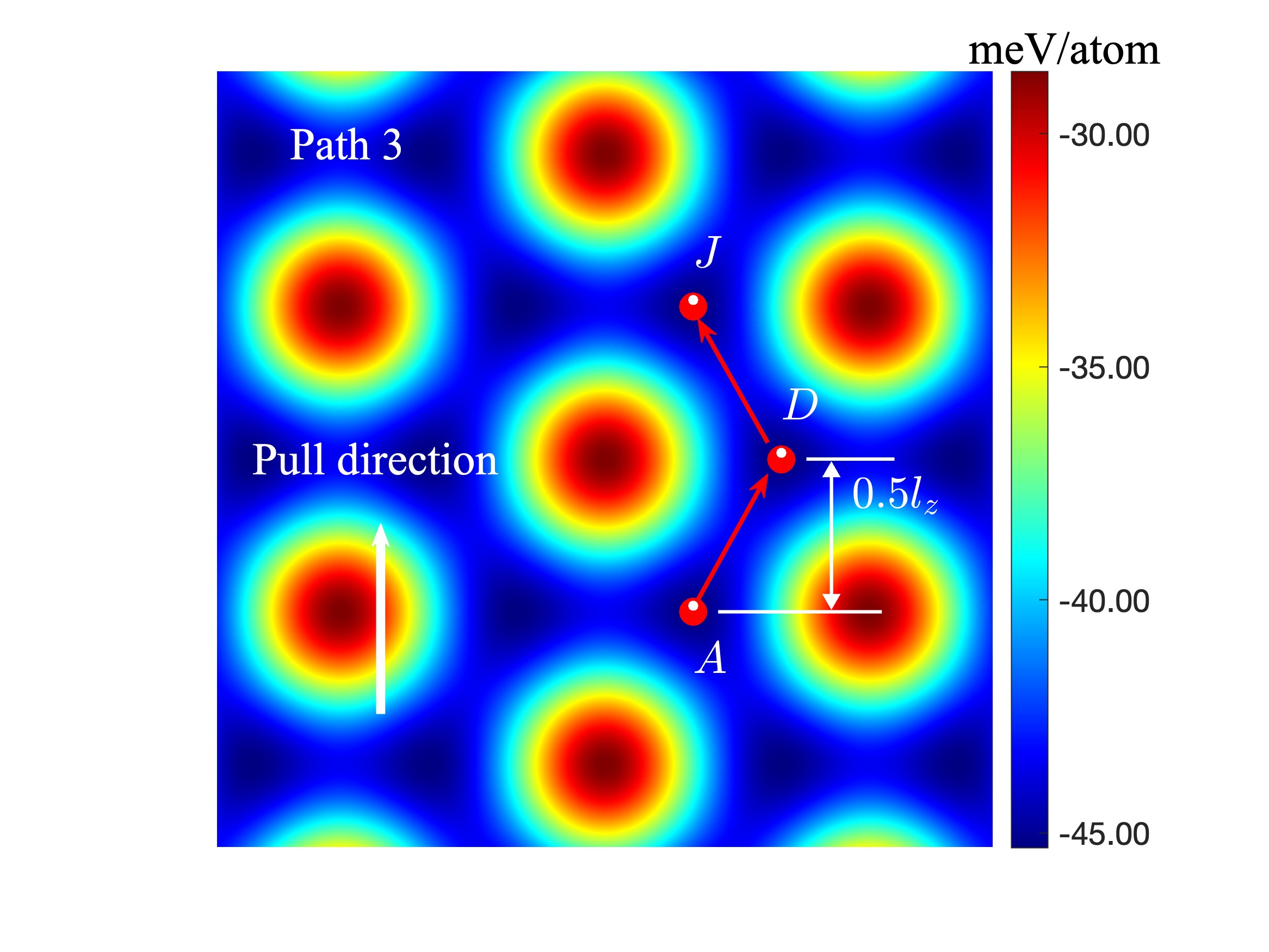}}
\put(0,0.5){\includegraphics[height=60mm]{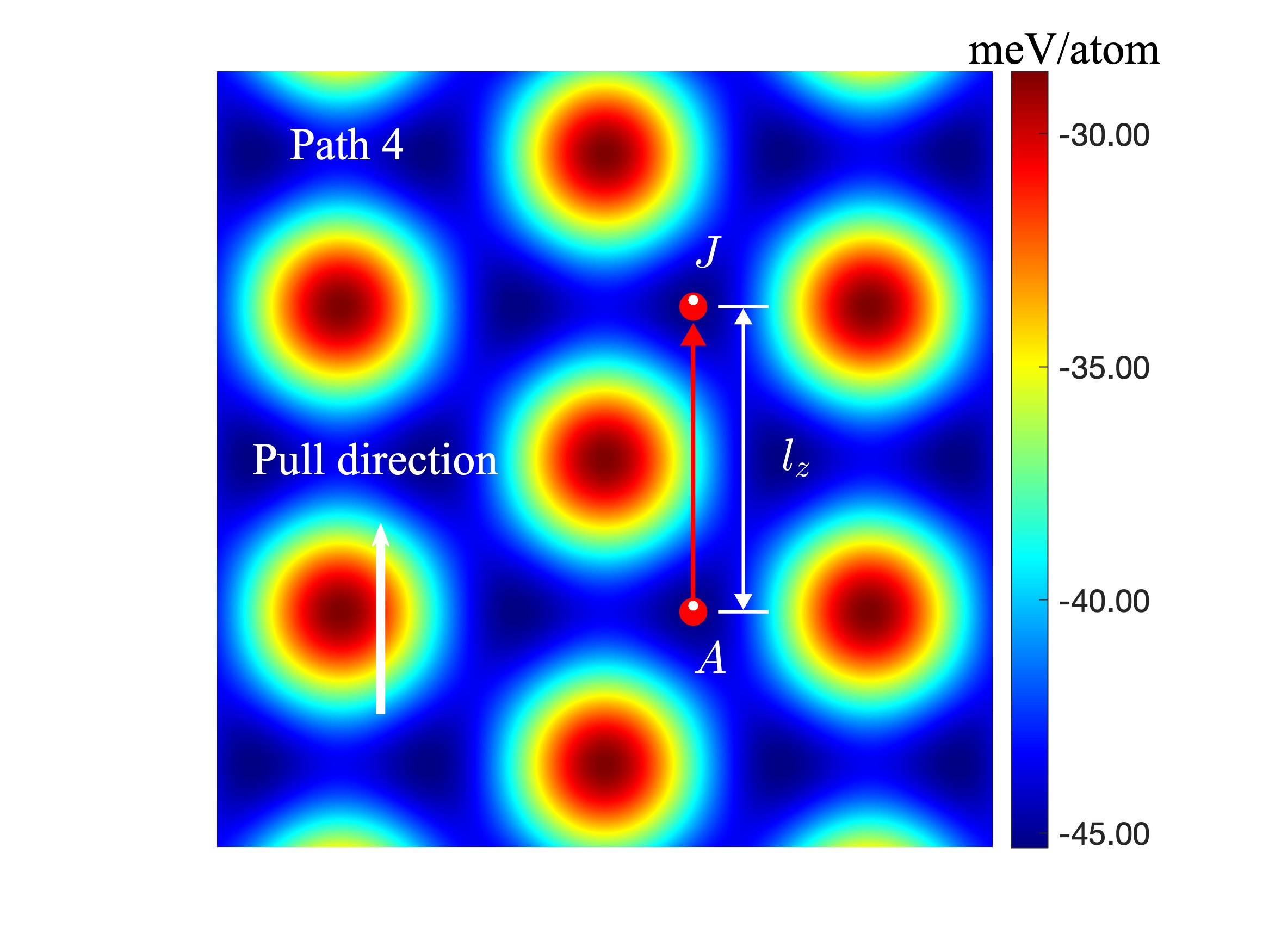}}
\put(-7.2,6.3){\fontsize{10pt}{20pt}\selectfont (a)  }
\put(0.5,6.3){\fontsize{10pt}{20pt}\selectfont (b) }
\put(-7.2,0.3){\fontsize{10pt}{20pt}\selectfont (c)  }
\put(0.5,0.3){\fontsize{10pt}{20pt}\selectfont (d) }
\end{picture}
\caption{ Energy landscape for carbon atoms moving across a graphene substrate along representative paths for armchair and zigzag pulling direction. For armchair direction: (a) Path 1 with low energy barrier, (b) Path 2 with high energy barrier.  For zigzag direction: (c) Path 3 with low energy barrier, (d) Path 4 with high energy barrier.}
\label{Sliding paths}
\end{center}
\end{figure}

During edge-driven sliding, different regions of the GNR exhibit various deformations depending on the study type (static or dynamic) and boundary conditions. Interfacial forces acting on the GNR depend on its lattice registry with the substrate (Eq.~\eqref{force acting on individual atom}), which evolves due to elastic deformations during sliding. Lattice registry relates the relative position of the GNR atoms with respect to the `AB' stacking arrangement. To study the effect of lattice registry, some representative sliding paths of a GNR atom\footnotemark[2] are identified following the approach of \citet{wang2017size}, for sliding along the armchair and zigzag directions as shown in Fig.~\ref{Sliding paths}. These paths are selected based on the associated energy barrier, which follows from Eq.~\eqref{energy expression}.

 \footnotetext[2]{Although this discussion focuses on a single atom, it can be generalized to a GNR unit cell, as all GNR atoms experience the same energy barrier. This statement holds exactly for a rigid GNR and approximately for an elastic one, where minor axial and lateral deformations occur. Hence, the proposed sliding path theory may be regarded as an idealized representation.}

\begin{figure}[H]
\begin{center} \unitlength1cm
\begin{picture}(0,7)
\put(-4,0){\includegraphics[height=70mm]{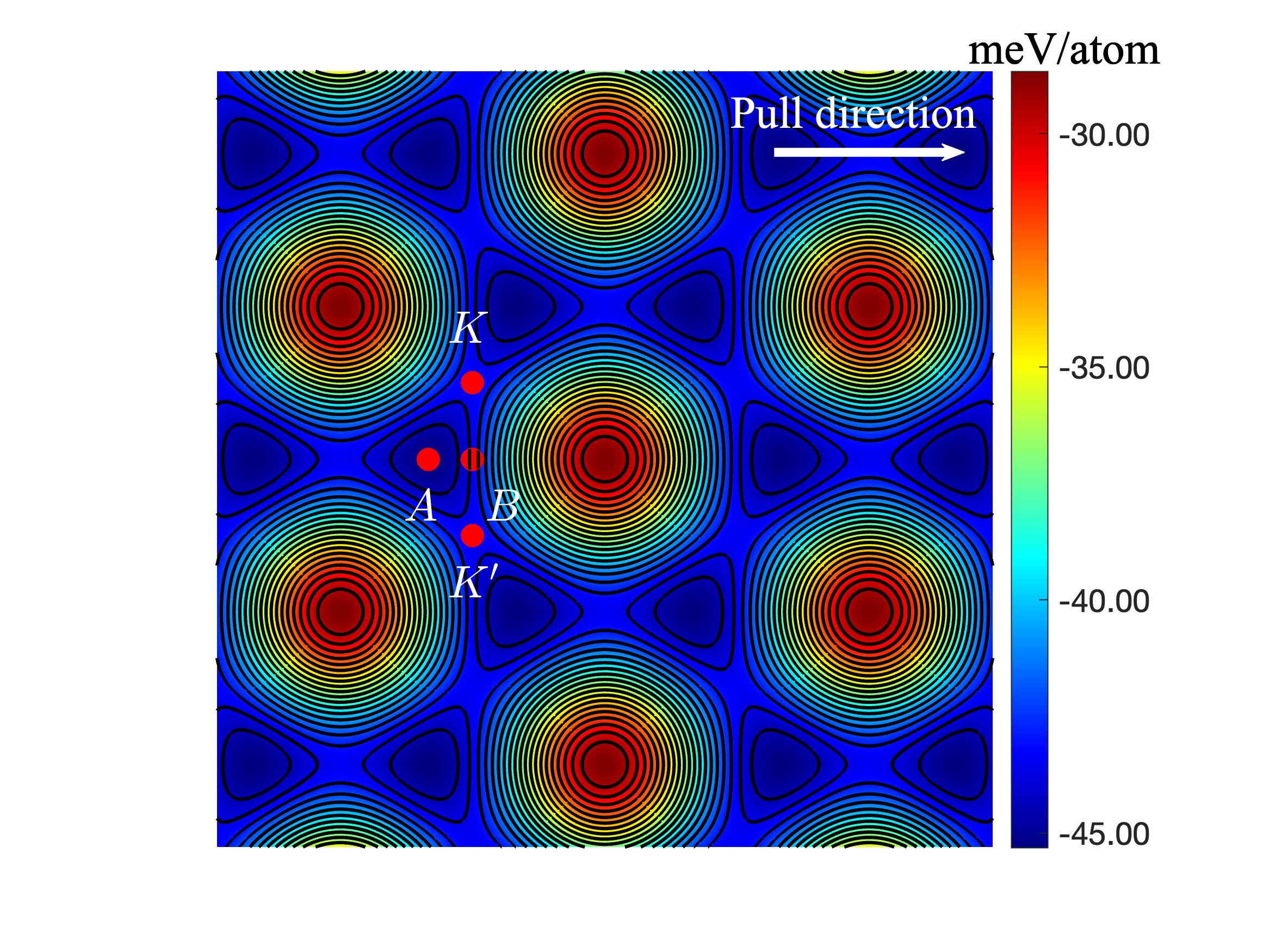}}
\end{picture}
\caption{Energy contour plot with atomic Positions $A$, $B$, $K$, and $K'$. States $B$, $K$, and $K'$ share the same energy level.}
\label{Same energy points}
\end{center}
\end{figure}

When the GNR atom in its lowest-energy configuration, state $\textit{A}$, and is pulled along the armchair direction ${\boldsymbol{e}}_\mathrm{a}$, the symmetry along ${\boldsymbol{e}}_\mathrm{a}$ (see Figs.~\ref{Sliding paths}a and \ref{Sliding paths}b) allows the atoms to transition to either state $\textit{B}$ or $\textit{K}$ or $K'$ (see Fig.~\ref{Same energy points}), since any displacement along $A$–$K$, $A$–$K'$, or $A$–$B$ is energetically equivalent.
For pulling along ${\boldsymbol{e}}_\mathrm{a}$, an atom ideally follows the $A$–$B$ path. The following points summarize the resulting behavior:
\begin{enumerate}[label=(\roman*), noitemsep, topsep=0pt]
 \item In the presence of imperfections and the absence of lateral constraints -- which breaks the symmetry of the energy barrier -- the atom will transition from state $B$ to either $C$ or $C'$. This occurs because states $B$, $C$, and $C'$ have equal energy (see Fig.~\ref{Sliding paths}a), but the energy gradient at $C$ or $C'$ is more favorable than at $B$.
 \item However, these jumps from $B$ to $C$, and subsequently from $F$ to $H$, may not be as sharp as in the idealized case of Fig.~\ref{Sliding paths}a. Due to inertia, which increases with GNR size, the actual trajectory will follow a curved path between Path 1 and Path 2.
 \item Path 2 represents an unstable trajectory that can be followed only under perfectly symmetric conditions or when the system inherently possesses a large lateral stiffness (greater than the critical stiffness obtained from the second derivative of $\Psi$ with respect to $g_z$) or when such stiffness is applied externally, as in the case of BC 1.
 \item Path 1 corresponds to the low-energy barrier, while Path 2 corresponds to the high-energy barrier.
 
\end{enumerate}

While for pulling along the zigzag direction, atoms can move along two extreme paths, Path 3 and Path 4, as shown in Fig.~\ref{Sliding paths}c and Fig.~\ref{Sliding paths}d, respectively. Atoms, free along the lateral direction, prefer to move along Path 3 with states $\textit{A-D-J}$, while atoms constrained in the lateral direction move along Path 4 with states $\textit{A-J}$. Along this direction, the paths of the GNR atoms remain unaffected by the nature of sliding due to the absence of symmetry in the energy barrier surface. These two paths can provide an upper and lower estimate of the shear force between the GNR and the substrate, for relative motion along the zigzag direction. 

In a dynamically deforming system, some atoms may follow intermediate paths between these extreme paths, as discussed in detail by \citet{wang2017size}. Nevertheless, these extreme paths provide a useful basis for understanding the general sliding behavior. Further elaboration is made in Section \ref{Section 3} on the specific paths chosen by the GNR.

\subsection{Finite element formulation} \label{subsection 2.3}
As evident from the interlayer energy expression in Eq.~\eqref{energy expression} and supported by previous studies \citep{wang2017size, Ouyang2018, Xue2022}, the sliding behavior of GNRs is highly nonlinear. Due to the large number of degrees of freedom involved, solving such systems analytically becomes intractable, thereby necessitating the use of numerical methods such as the finite element method (FEM). While MD simulations provide high accuracy in capturing atomic-scale behavior, they are computationally expensive, particularly for large-scale systems. On the other hand, continuum approaches based on 3D shell and plate element models offer a more computationally efficient alternative, as they require significantly fewer degrees of freedom. However, this efficiency comes with a trade-off in accuracy, since such models cannot capture the discrete nature of individual atomic bonds. This issue can be somewhat resolved by using planar 1D beam elements \citep{Hollerer2014}. This approach aims at retaining the essential mechanical characteristics of the system while reducing the computational cost over MD. The cost reduction is two-fold: (i) elimination of non-neighboring atomic intrasolid interactions to determine the internal forces and (ii) elimination of pairwise summations of atomic intersolid interaction energies
by employing Eq.~\eqref{energy expression}.

\begin{figure}[H]
\begin{center} \unitlength1cm
\begin{picture}(0,4)
\put(-5,0){\includegraphics[height=30mm]{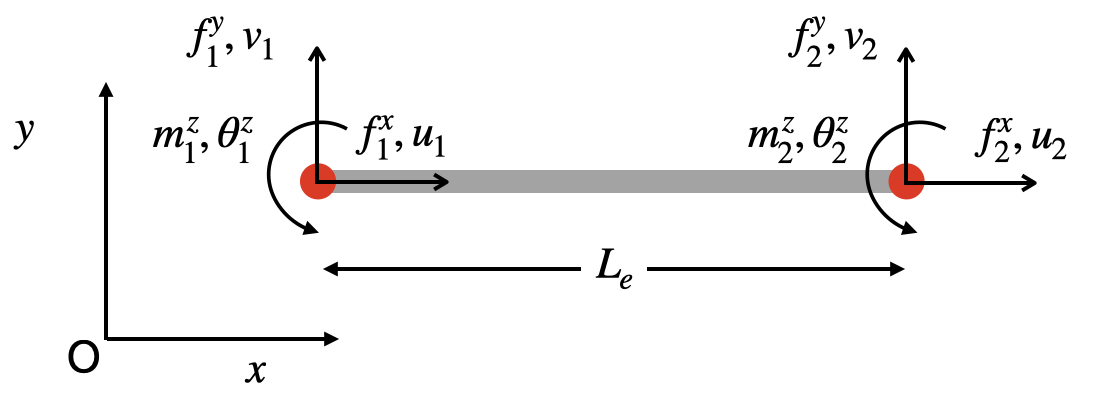}}
\end{picture}
\caption{Nodal displacements and forces of a two-noded Euler Bernoulli beam element in physical coordinate system. }
\label{beam_element}
\end{center}
\end{figure}

Since the entire study is carried out at a fixed normal gap, the out-of-plane degrees of freedom are inherently constrained. Therefore, a 2-noded Euler-Bernoulli planar beam element, as shown in Fig.~\ref{beam_element}, is used to model the bond of the GNR. Each node of the beam element has three degrees of freedom
$\mathbf{u}_A$ = $\{u_A,v_A,\theta_A\}^T$, corresponding to the horizontal and vertical displacements and the rotation of the beam element axis, respectively.

The current and reference configurations are discretized using standard finite element interpolation 
\begin{equation}
    \ds \textbf{\textit{x}} \approx \sum_{A=1}^{n_e} N_A \mathbf{x}_{A} = \mathbf{N}_e \mathbf{x}_e, \hspace{1cm} \textbf{\textit{X}} \approx \sum_{A=1}^{n_e} N_A \mathbf{X}_{A} = \mathbf{N}_e \mathbf{X}_e,
\end{equation}

where $\mathbf{N}_e $ is the element wise shape function matrix defined as [$\mathbf{1} $ ${N}_1 $, $\mathbf{1} $${N}_2 $, . . . , $\mathbf{1} $${N}_{n_e} $], with $\mathbf{1}$ denoting the 3 × 3 identity
matrix, contains the $n_e$ nodal shape functions $N_A$. The FE approximation for the displacement within element $e$ is then

\begin{equation}
    \ds \textbf{\textit{u}} := \textbf{\textit{x}} - \textbf{\textit{X}} \approx\sum_{A=1}^{n_e} N_A \mathbf{u}_{A} = \mathbf{N}_e \mathbf{u}_e, 
\end{equation}
Here,
\begin{equation}
    \mathbf{x}_e:=\left[ \begin{array}{c}
\textbf{\textit{x}}_1 \\
\textbf{\textit{x}}_2 \\
\vdots \\
\textbf{\textit{x}}_{n_e}
\end{array} \right]~, \hspace{1cm}\mathbf{X}_e:=\left[ \begin{array}{c}
\textbf{\textit{X}}_1 \\
\textbf{\textit{X}}_2 \\
\vdots \\
\textbf{\textit{X}}_{n_e}
\end{array} \right]~, \hspace{1cm}\mathbf{u}_e:=\left[ \begin{array}{c}
\textbf{\textit{u}}_1 \\
\textbf{\textit{u}}_2 \\
\vdots \\
\textbf{\textit{u}}_{n_e}
\end{array} \right]~,
\end{equation}

\begin{equation}
    \ds \textbf{\textit{u}}_A = \textbf{\textit{x}}_A - \textbf{\textit{X}}_A ~.
\end{equation}
represent the arrays of the $n_e$ nodal positions and displacements
associated with element $e$.

The elemental nodal contact force vector $\mathbf{f}_\text{c}^e$ is determined based on the nodal forces described in Eq.~\eqref{force acting on individual atom}. The components of this vector, $-$$f_A^x$ and $-$$f_A^y$, represent the force contributions in the armchair  ${ \boldsymbol{e}}_\mathrm{a}$, and zigzag ${ \boldsymbol{e}}_\mathrm{z}$-directions at node $A$. Since no contact moment acts at the nodes, the third component (moment) of this vector is zero. Since only Dirichlet boundary conditions are prescribed, the external force vector $\mathbf{f}_{\text{ext}}^e$ for the element is equal to zero.

\nomenclature{$v_A$}{Nodal displacement in the y-direction}
\nomenclature{$u_A$}{Nodal displacement in the x-direction}
\nomenclature{$E$}{Young's modulus}
\nomenclature{$\nu$}{Poisson's ratio}

The elemental force vector $\mathbf{f}^e$ is then given as \citep{Sauer2007}
\begin{equation}
     \ds \mathbf{f}^e(\textbf{u})= \mathbf{f}^e_{\text{int}}(\mathbf{u}_e)+ \mathbf{f}^e_{\text{c}}(\mathbf{u}_e)~,
\end{equation}
and the corresponding elemental tangent stiffness matrix is given by
\begin{equation}
    \ds \mathbf{k}_e = \frac{\partial  \mathbf{f}^e}{\partial \mathbf{u}_e}~.
\end{equation}

The elemental contributions are assembled into global residual force vector $\mathbf{F}$ and global stiffness matrix $\mathbf{K}$ defined as
\begin{equation}
    \ds \mathbf{F}= \sum_{e=1}^{n_{\text{el}}} \mathbf{f}^e , \hspace{1cm} \mathbf{K}= \sum_{e=1}^{n_{\text{el}}} \mathbf{k}^e ,
\end{equation}
These quantities are used in the discretized weak form and its linearization, which are essential for solving the nonlinear system via the Newton-Raphson (NR) solution procedure. The linearized form of the static equilibrium equation at iteration $k$+1 is 
\begin{equation}
    \mathbf{F}( \mathbf{u}^{k}) +  \mathbf{K}(\mathbf{u}^{k})\cdot \Delta \mathbf{u}^{k+1}+... =0 
    \label{linearized form}
\end{equation}

where $\Delta \mathbf{u}^{k+1}$ = $ \mathbf{u}^{k+1}$ $-$  $ \mathbf{u}^{k}$ is the displacement increment at iteration 
$k+1$. Since the observed behavior involves limit points associated with snap-through and snap-back instabilities, the standard Newton–Raphson (NR) solver fails to capture them. At these points, the tangent stiffness matrix becomes singular, leading to either no displacement update or unrealistic updates. As a result, the solver fails to find an equilibrium, causing non-convergence. To accurately trace an equilibrium path with limit points, it is necessary to control the magnitude as well as the direction of loading. This dual control can be effectively achieved in numerical simulations by introducing a third parameter -- the $\textit{arc\,length}$ -- allowing independent regulation of both force and displacement \citep{Riks1979}. The equilibrium curves obtained using this solver provide complete insight into the solution.

Alternatively, this issue of non-convergence can be successfully addressed by incorporating relaxation into the system through inertia and linear damping. Thus, Eq.~\eqref{linearized form} is modified as
\begin{equation}
  \ds  \mathbf{M} \ddot{\mathbf{u}}_{n+1}^{k+1} + \mathbf{C}\dot{\mathbf{u}}_{n+1}^{k+1} +   \mathbf{F}( \mathbf{u}_{n+1}^{k}) +  \left(\mathbf{K}(\mathbf{u}^{k})\cdot \Delta \mathbf{u}^{k+1}\right)_{n+1} =\mathbf{0}~.
    \label{dynamic equation}
\end{equation}
Here, $\mathbf{M}$, $\mathbf{C}$, $k$, and $n$ are the mass matrix, damping matrix, load step, and time step, respectively. The obtained equation is solved using the standard Newmark scheme, e.g. see \citep{Chopra2007}. This scheme provides an unconditionally stable result for the non-zero damping case. The damping matrix $\mathbf{C}$, is taken as $\alpha\mathbf{M}$, where $\alpha$ is the Rayleigh damping coeffcient.

For the subsequent analysis involving edge-driven sliding behavior and shear characterization of GNRs, either the arc-length continuation approach or the dynamic relaxation method (DRM), as described in Eq.~\eqref{dynamic equation}, is employed. 

\section{ Results and discussions }\label{Section 3}

First, the FE model is validated against MD simulations by comparing the edge-pulled sliding behavior. Subsequently, the edge-pulled sliding of both armchair and zigzag GNRs is examined under various boundary conditions and time dependency (static or dynamic). To characterize the sliding response, the energy evolution of the GNR during sliding is analyzed. Additionally, strain transmission between the GNR and the graphene substrate is investigated under uniform biaxial stretching of the graphene substrate. The initial geometry parameters of these GNRs, such
as length, width, and number of unit cells are listed in Appendix \ref{Appendix B}, Table \ref{table2}.

\subsection{FE model verification} \label{subsection 3.1}

Here, we verify the MD-calibrated FE model by comparing its predictions to MD simulation results for interlayer sliding. To perform this verification, boundary condition BC 2 is chosen, as it represents an intermediate case and permits in-plane bending. Details of the MD simulation setup and sliding results are provided in Appendix \ref{Appendix C} and \ref{Appendix D}, respectively.

\begin{figure}[H]
\begin{center} \unitlength1cm
\begin{picture}(0,11.6)
\put(-8.2,6){\includegraphics[height=54mm]{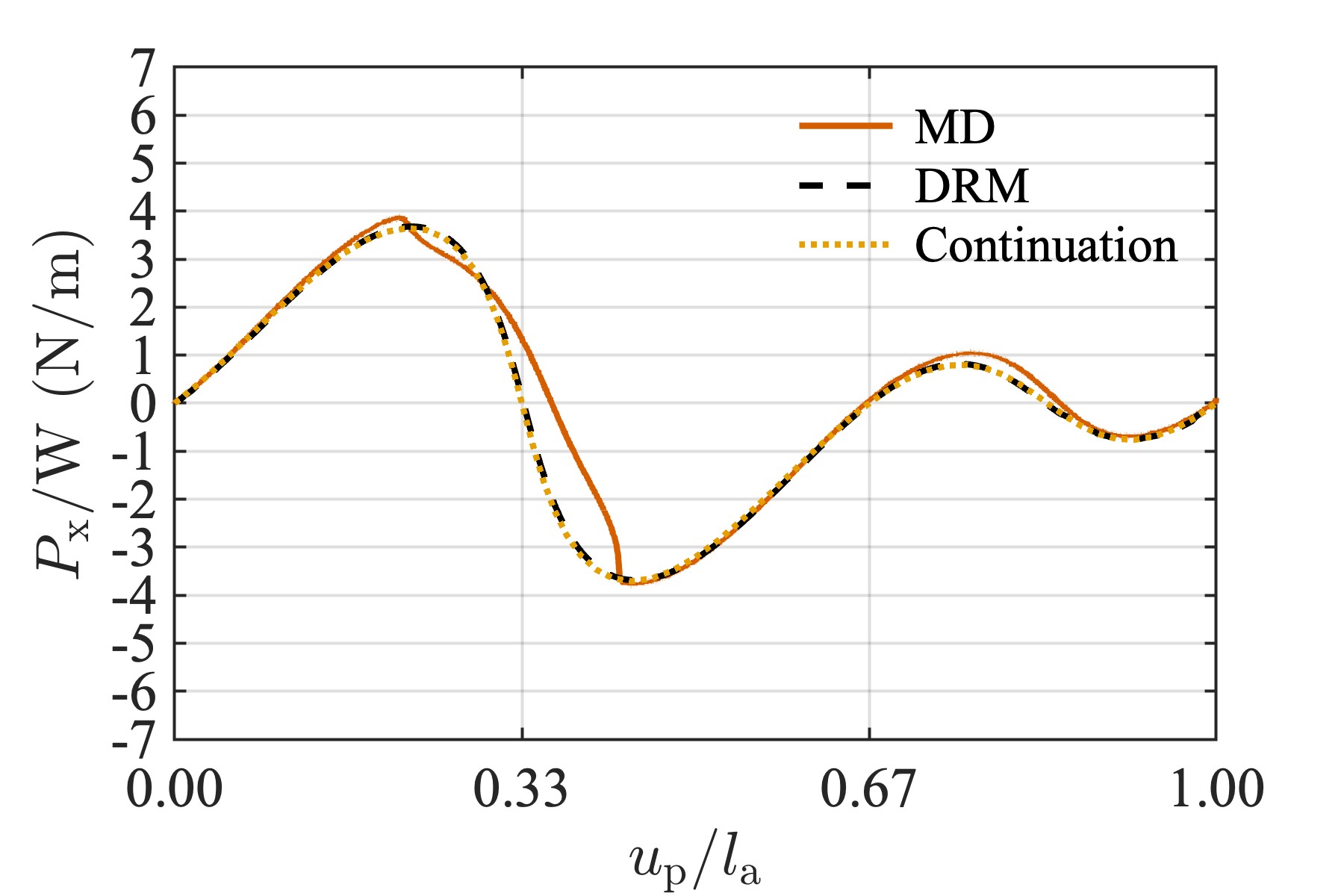}}
\put(-0.5,6){\includegraphics[height=54mm]{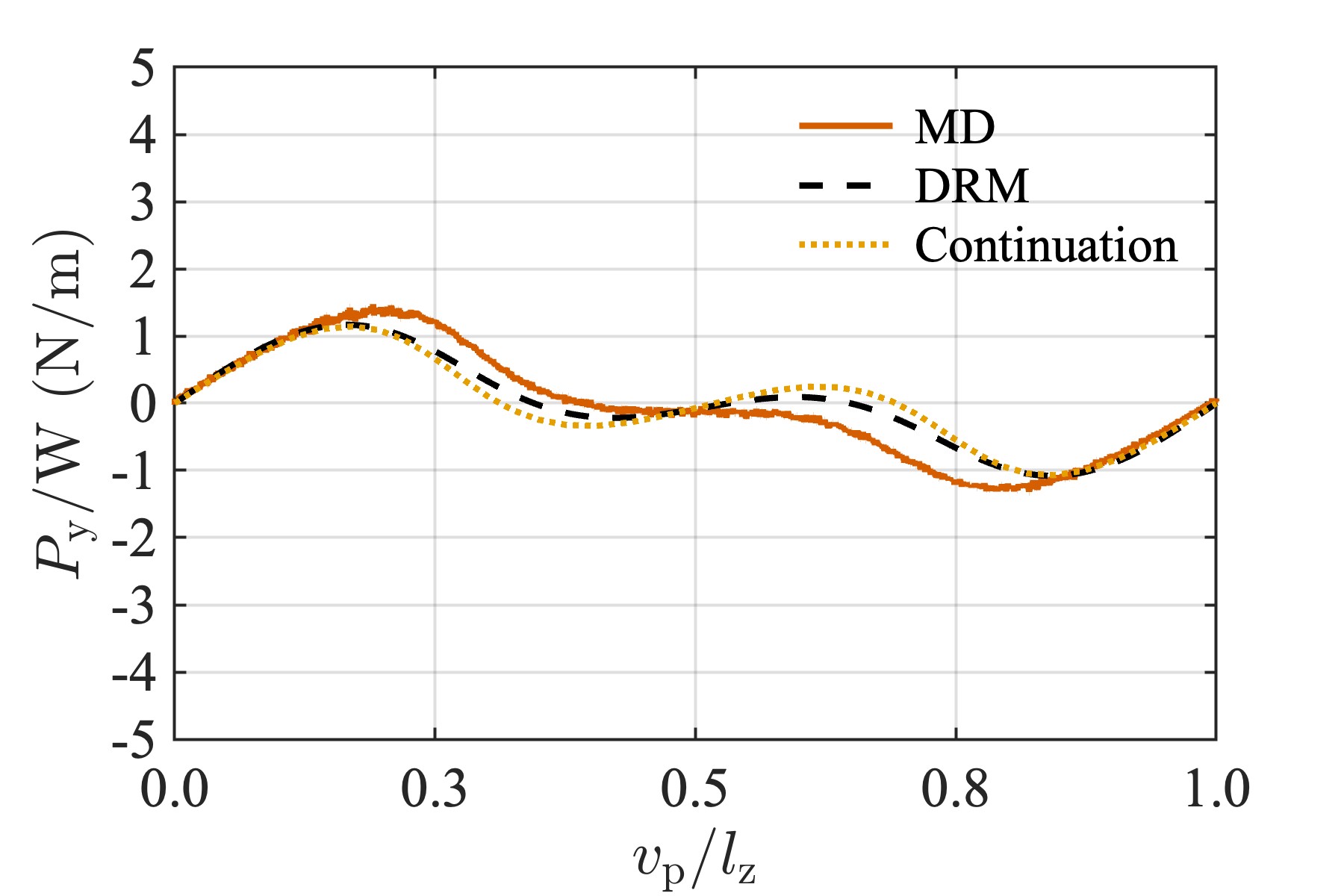}}

\put(-8.2,0){\includegraphics[height=54mm]{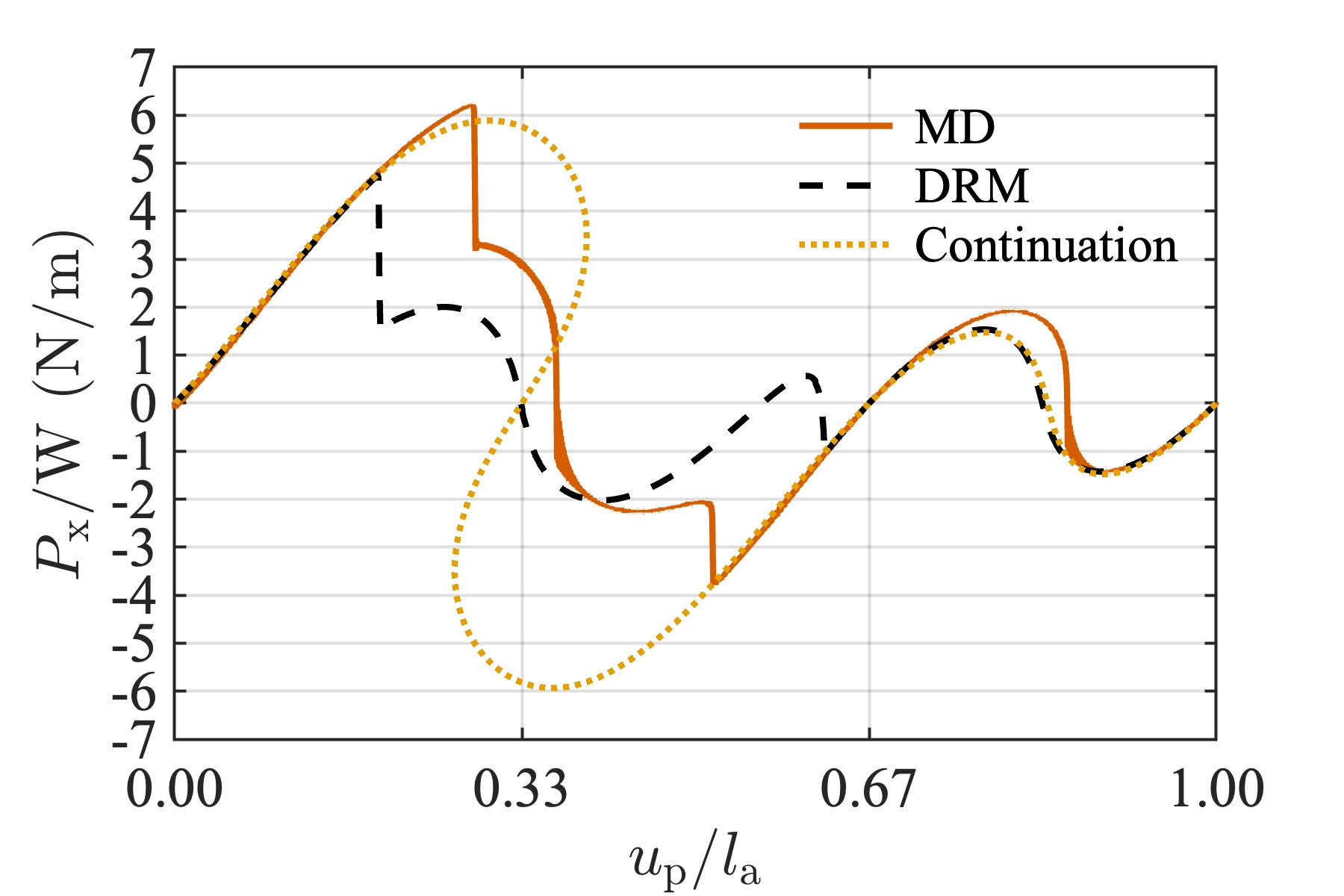}}
\put(-0.5,0){\includegraphics[height=54mm]{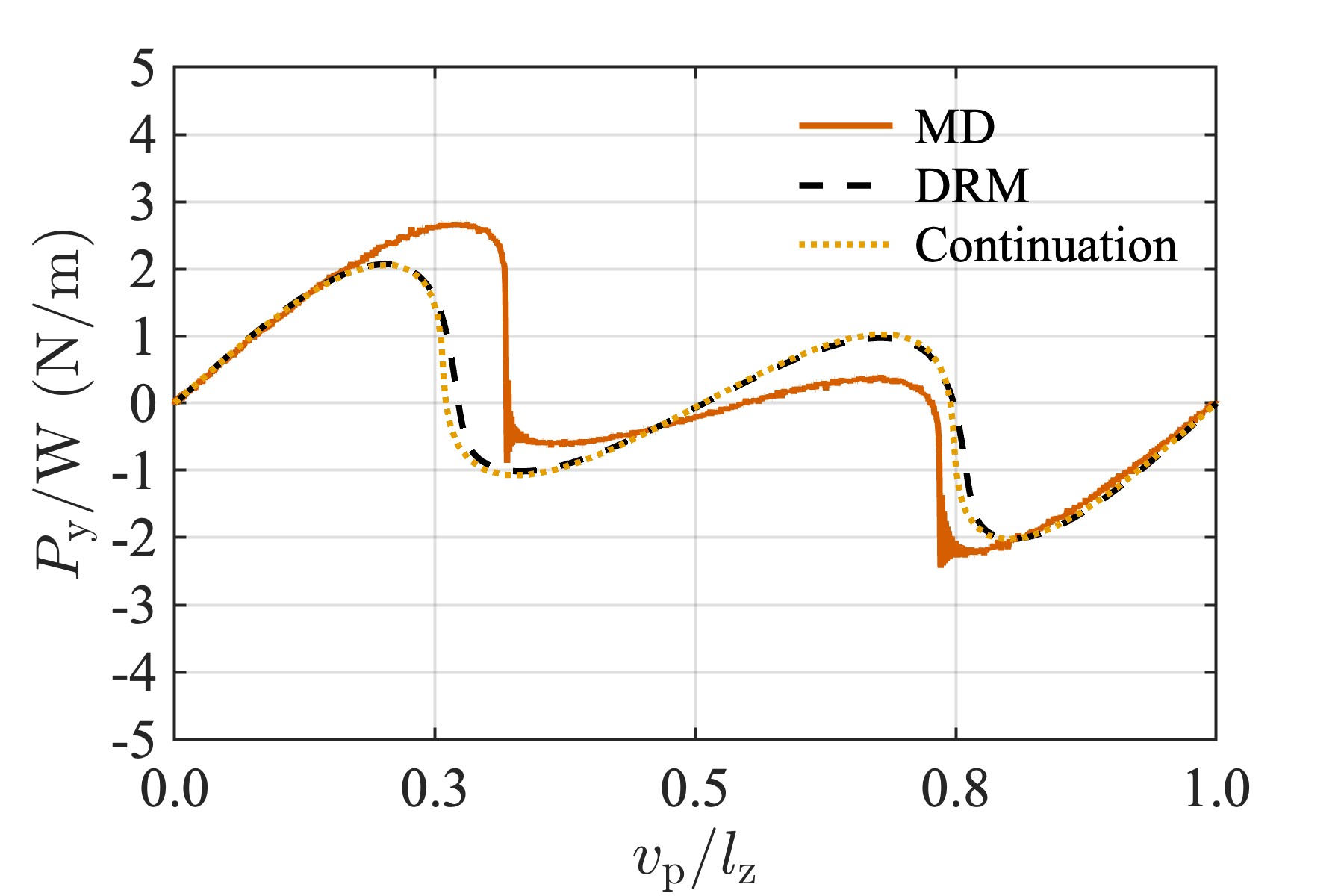}}

\put(-7.4,6){\fontsize{10pt}{20pt}\selectfont (a)}
\put(-0.1,6){\fontsize{10pt}{20pt}\selectfont (b)}
\put(-7.4,-0.10){\fontsize{10pt}{20pt}\selectfont (c)}
\put(-0.1,-0.10){\fontsize{10pt}{20pt}\selectfont (d)}

\end{picture}
\caption{ Comparison of FE results for sliding of GNRs under BC 2: (a,b) 5 nm and (c,d) 10 nm lengths; along (a,c) the armchair and (b,d) the zigzag directions. }
\label{comparison of different methods}
\end{center}
\end{figure}
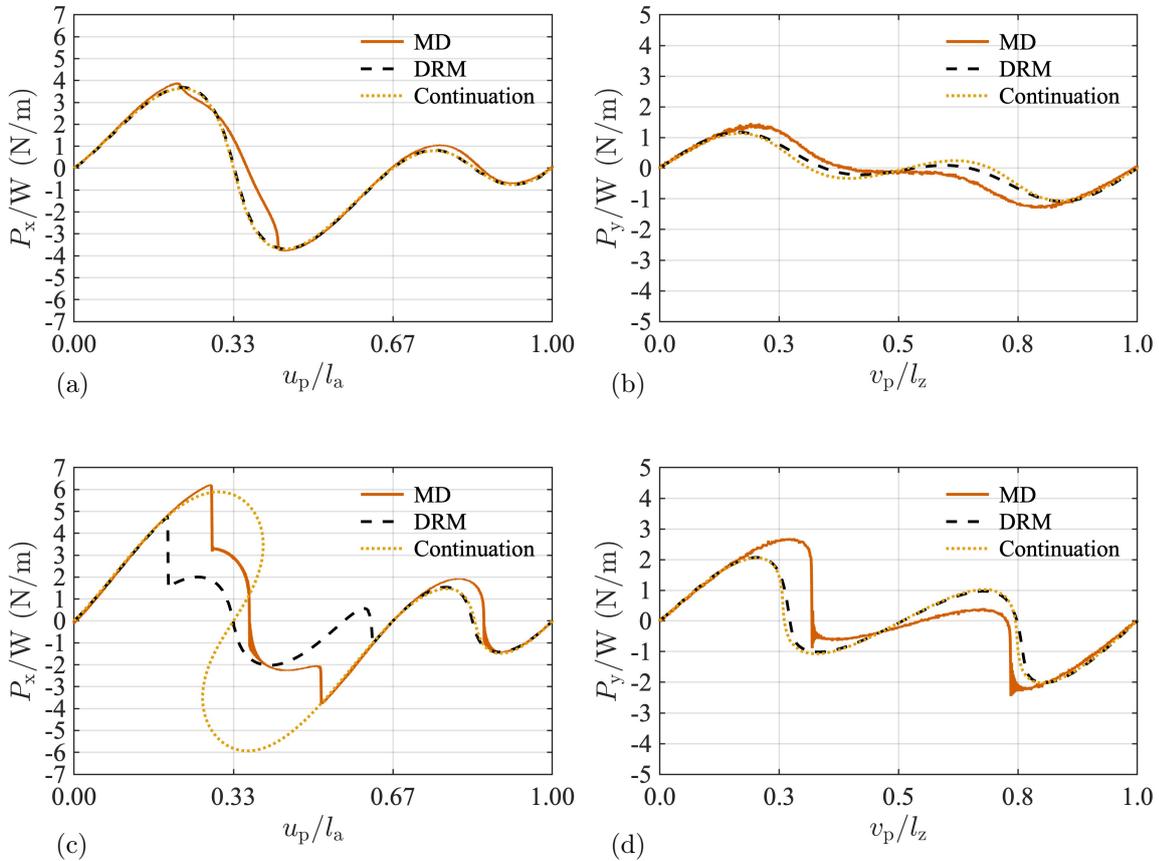

Fig.~\ref{comparison of different methods} presents a comparison of the sliding response obtained from FE simulations -- using both the arc-length continuation and the DRM solver -- with the corresponding MD data. The two methods show excellent agreement for most of the curve, particularly in the small deformation regime, where the sliding stiffness agrees. This strong correlation verifies the accuracy of the FE results. 

\begin{figure}[H]
\begin{center} \unitlength1cm
\begin{picture}(0,6.2)
\put(-4,0){\includegraphics[height=52mm]{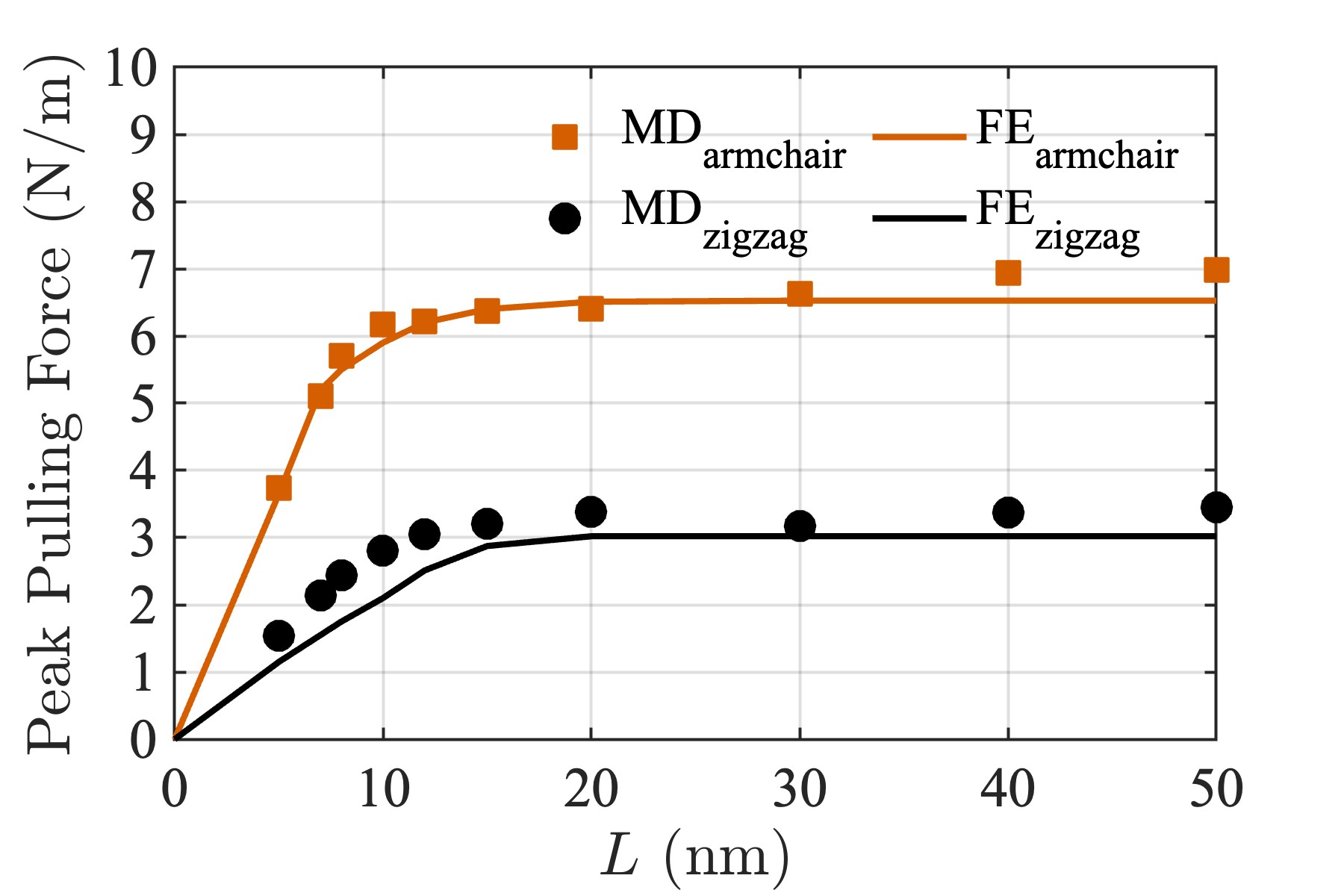}}
\end{picture}
\caption{Comparison of the peak pulling force, $P_{\mathrm{y}}^{\mathrm{max}}$, obtained from FE and MD simulation. The values of $P_{\mathrm{y}}^{\mathrm{max}}$ are plotted against the GNR length for sliding under boundary condition BC 2.}
\label{Comparison of force saturation for zigzag BC 2}
\end{center}
\end{figure}

The comparison of the peak pulling force, $P_{\mathrm{y}}^{\mathrm{max}}$, shown in Fig.~\ref{Comparison of force saturation for zigzag BC 2}, shows a quantitative difference of up to 10 $\%$. This difference is partly attributed to material properties and system parameters. Additionally, localized pinning effects at the leading and trailing edges -- well documented in the literature \citep{Zhang2021, Liu2023} -- may also contribute to this mismatch. Also, it is observed from both the MD and FE simulations that the region near the pulling end undergoes distortion, with a localized strain of approximately 1–2 $\%$. This distortion causes the structure to reconfigure in axial and lateral directions. Consequently, this may give rise to an emergent longer-wavelength modulation -- similar to a Moiré pattern \citep{Yan2024}. However, the effect of this local strain is not accounted for in the current form of the interaction energy expression (Eq.~\ref{energy expression}), a limitation also noted by \citet{Mokhalingam2024}.

Despite these differences, we believe that the overall behavior captured by the FE model reflects the essential physics of interlayer sliding. Rather than undermining the model, the observed discrepancies underscore the potential for further refinement, such as parameter calibration informed by DFT data. Importantly, the current model provides a computationally efficient and versatile framework for exploring GNR sliding. Building on this foundation, we proceed to study the full sliding response, frictional characterization, and interlayer strain transmission using the proposed FE model.

\subsection{Edge-pulled GNR sliding} \label{subsection 3.2}

\begin{figure}[H]
\begin{center} \unitlength1cm
\begin{picture}(0,5.5)
\put(-4,0){\includegraphics[height=54mm]{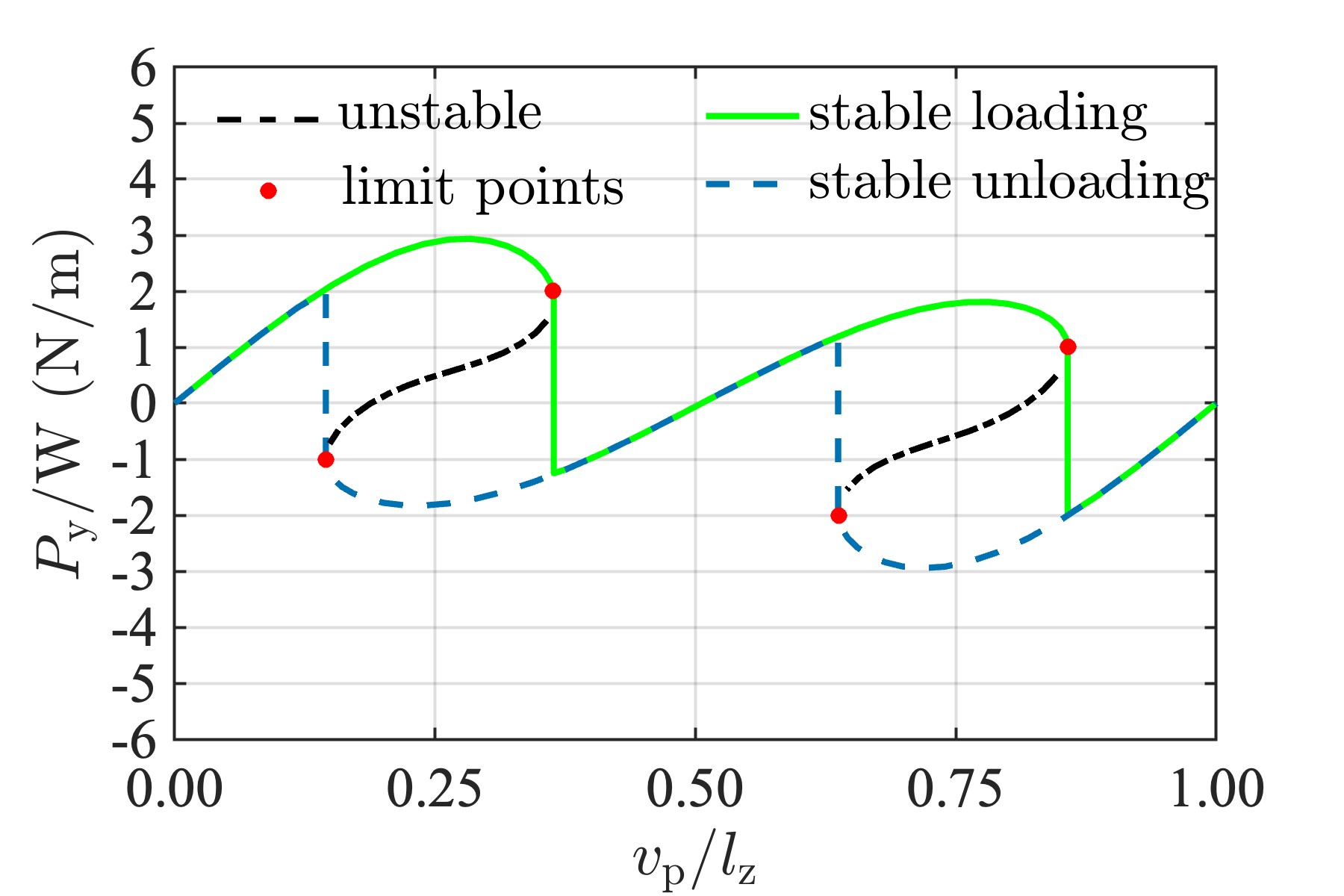}}
\end{picture}
\caption{Exemplary load-displacement curve for a 20 nm long zigzag GNR pulled across a graphene substrate along the zigzag direction using BC 2. The stable loading path (increasing $v_\mathrm{p}$) is shown in green; the stable unloading path (decreasing $v_\mathrm{p}$) is shown in dashed blue. The remaining parts of the curve are unstable and only accessible by the arc-length (continuation) method.}
\label{stability}
\end{center}
\end{figure}

The sliding behavior of the GNR exhibits strong nonlinearity, with unstable segments evident in the force–displacement response, as shown in Fig.~\ref{stability} and \ref{All cases}. In monotonic, displacement-controlled simulations (such as those using the DRM method), the appearance of limit points as seen in Fig.~\ref{stability} causes the structure to snap-through, while maintaining the prescribed displacement at the actuation point(s). At the limit points, the FE tangent matrix becomes singular, and the simulation solution jumps to another stable configuration, rendering portions of the force-displacement curve inaccessible for displacement control. However, the continuation method captures the full snap-back curve beyond the limit point. The limit point effectively partitions the equilibrium path into stable and unstable segments, as shown in Fig.~\ref{stability}. The stability of the force-displacement curve segments obtained from the continuation method is determined by confirming the positive definiteness of the tangent stiffness matrix. Due to the presence of instabilities during loading and unloading, a hysteresis is observed. 

A thorough investigation is conducted to examine boundary effects under both static and dynamic sliding. First, static sliding is discussed for the three boundary condition, and a continuation method was employed to overcome potential singularities that may arise during sliding. Subsequently, the results are compared with those from dynamic sliding, for which the DRM solver is used.

\subsubsection{BC 1}\label{subsubsection 3.2.1}
The variation of the pulling force is presented in Fig.~\ref{All cases}a and \ref{All cases}b for the sliding of armchair GNR along the armchair direction of the substrate and zigzag GNR along the zigzag direction of the substrate, respectively. Under BC 1, the unit cells of the GNR are constrained laterally, meaning that an armchair GNR is constrained to go along Path 2 (see Fig.~\ref{Sliding paths}b) and a zigzag GNR is constrained to go along Path 4 (see Fig.~\ref{Sliding paths}d). Due to this restricted sliding, the structure is not able to relax and follow a lower energy barrier path, resulting in the sliding force wavelength being the same as the lattice spacing. For long GNRs, it is found that the maximum interlayer shear force when pulling along the armchair direction is $\approx$ 30-50$\%$ higher than that when pulling along the zigzag direction, compare Fig.~\ref{All cases}a and \ref{All cases}b. 

\subsubsection{BC 2}\label{subsubsection 3.2.2}

\begin{figure}[H]
\begin{center} \unitlength1cm
\begin{picture}(0,16)
\put(-8,12){\includegraphics[height=5.5cm]{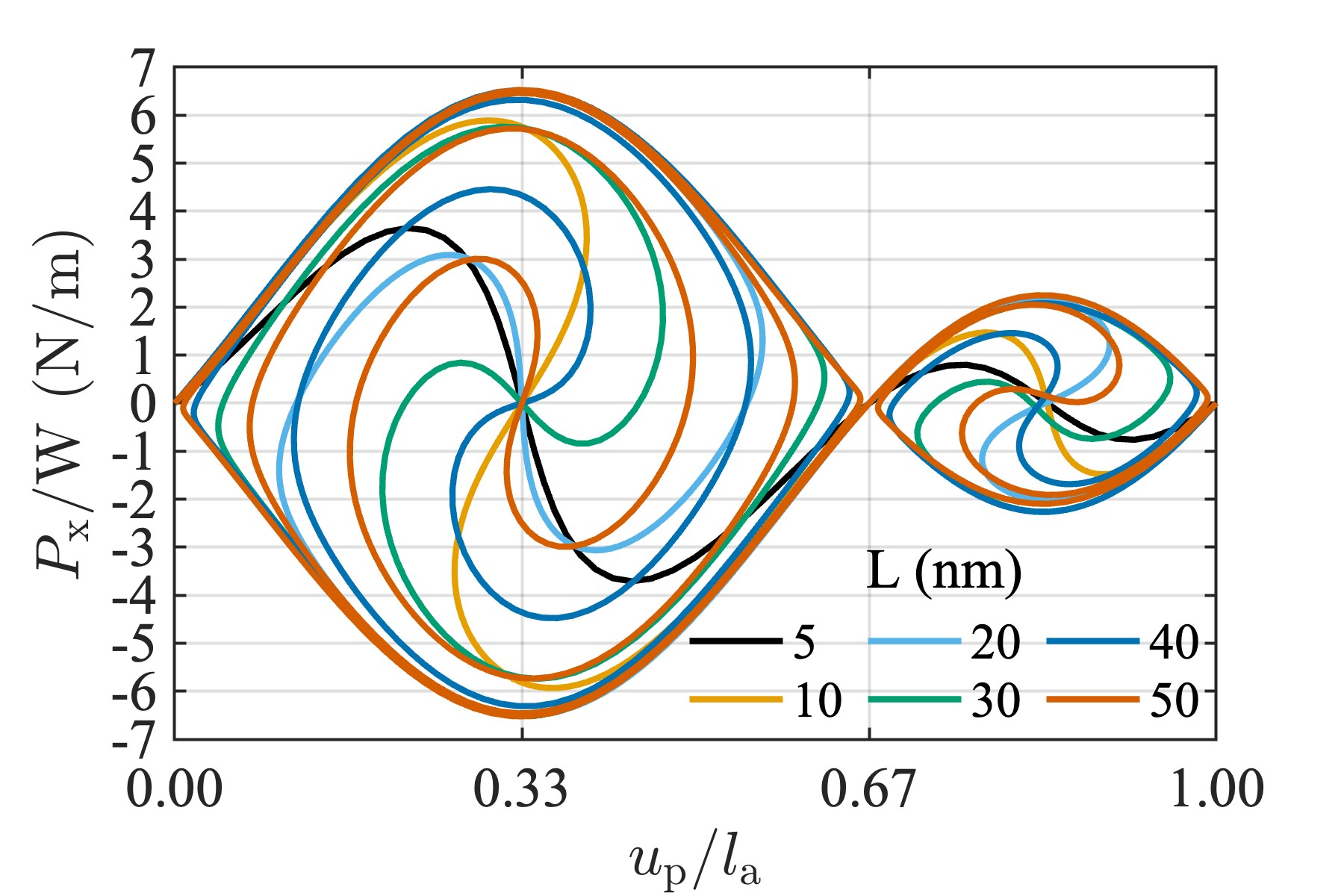}}
\put(0,12){\includegraphics[height=5.5cm]{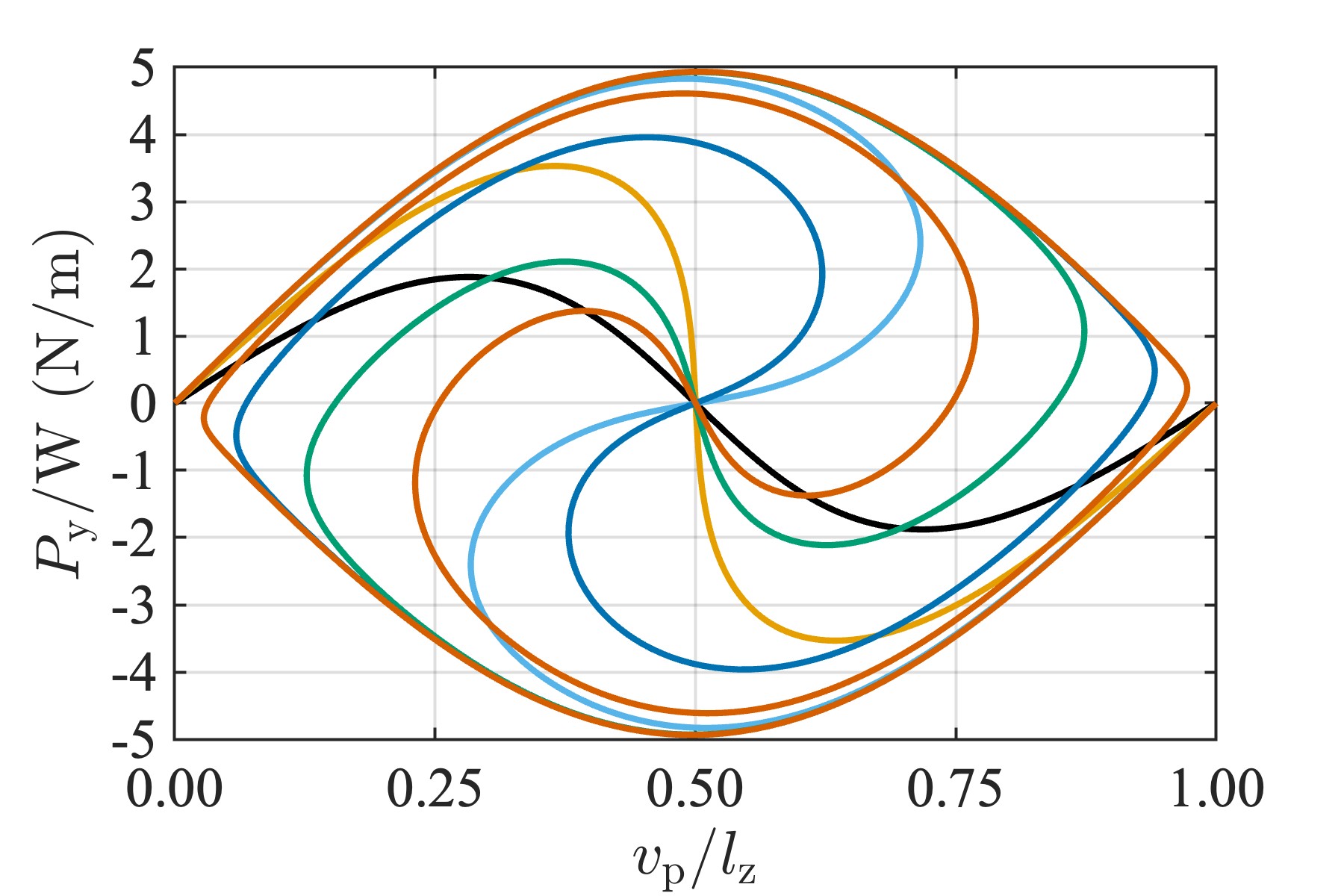}}
\put(-8,6){\includegraphics[height=5.5cm]{Armchair_AC.jpeg}}
\put(0,6){\includegraphics[height=5.5cm]{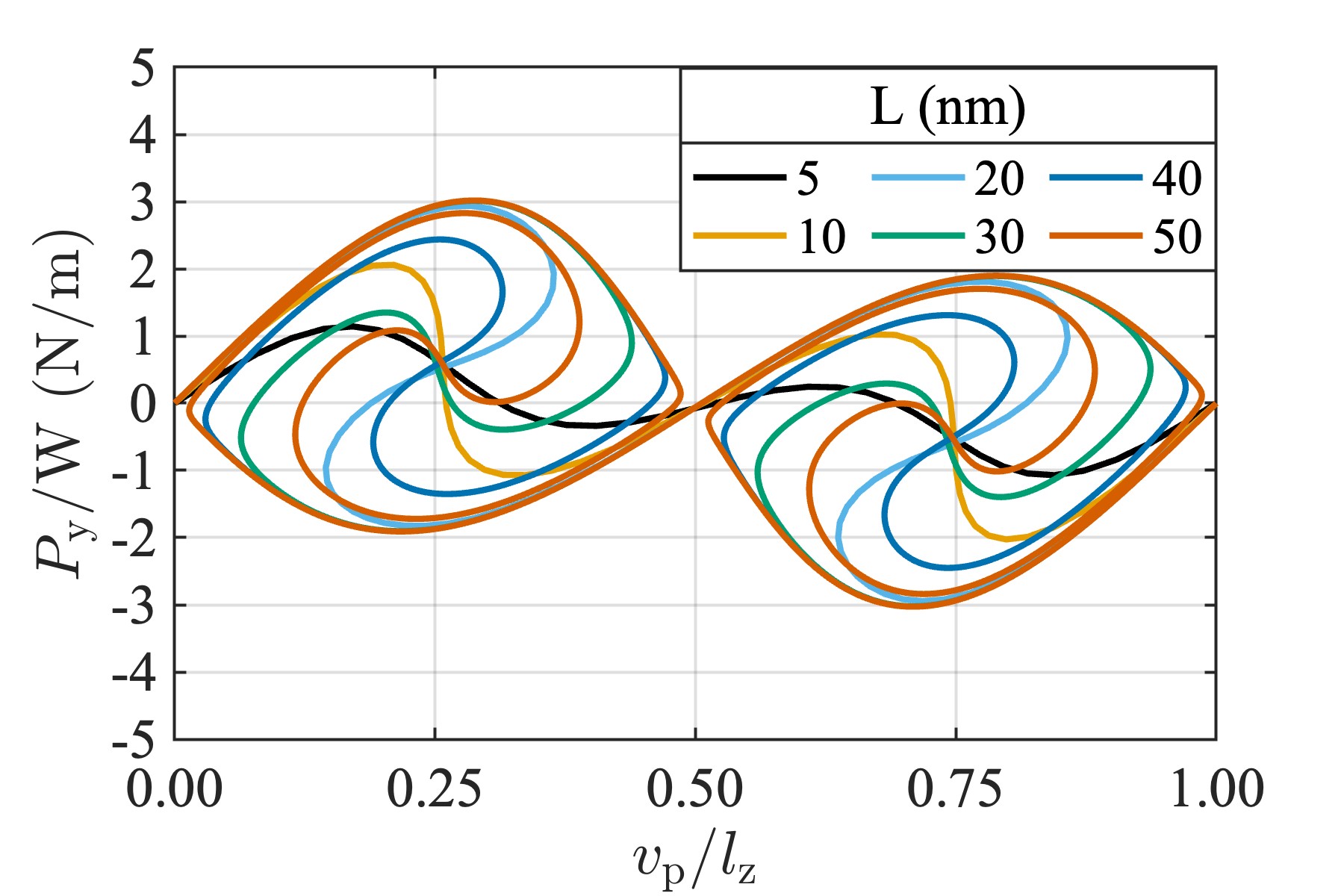}}
\put(-8,0){\includegraphics[height=5.5cm]{Armchair_AC.jpeg}}
\put(0,0){\includegraphics[height=5.5cm]{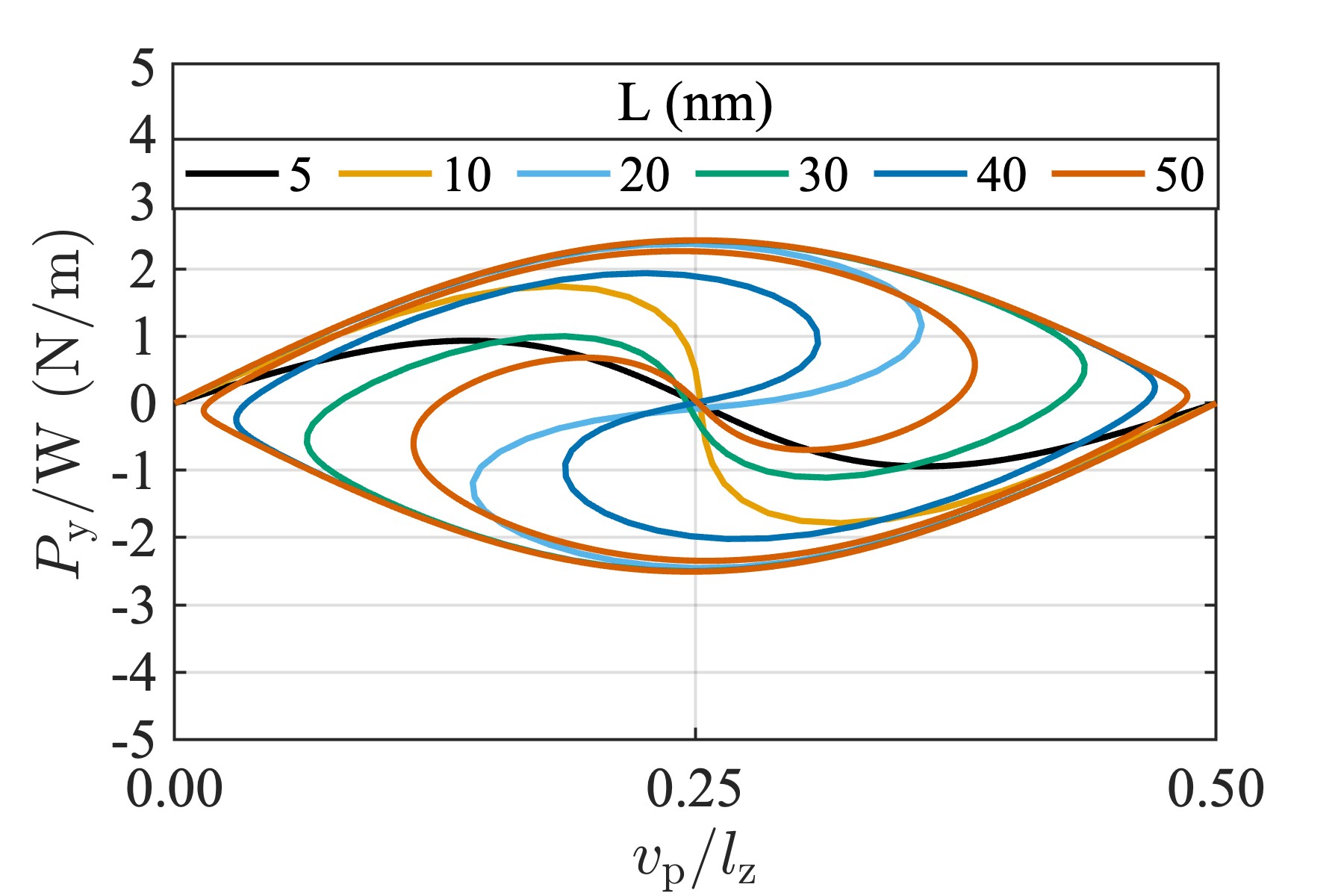}}
\put(-7.5,12){(a)}
\put(0,12){(b)}
\put(-7.5,6){(c)}
\put(0,6){(d)}
\put(-7.5,0){(e)}
\put(0,0){(f)}
\end{picture}
\caption{Pulling force-displacement diagram of sliding GNR obtained from FE simulations for various ribbon lengths (in nm). Results are obtained using the arc-length continuation solver to obtain the entire static solution paths. Results are shown for the armchair sliding direction in (a) for BC 1, (c) for BC 2, and (e) for BC 3, while the corresponding results for the zigzag sliding direction are shown in (b) for BC 1, (d) for BC 2, and (f) for BC 3. (Here, $u_\mathrm{p}$ and $v_\mathrm{p}$ are the prescribed displacements in the armchair and zigzag directions. The stable and unstable (i.e, snap back part of curve) are not shown, but they are exactly the same as in Fig.~\ref{stability}.)} 
\label{All cases}
\end{center}
\end{figure}

Fig.~\ref{All cases}c shows that, the sliding behavior of the armchair GNR along the armchair direction of the substrate is found to be similar to the case when BC 1 is imposed (Fig.~\ref{All cases}a). This is because under perfectly symmetric static sliding conditions all atoms will follow Path 2 as already discussed in Section \ref{subsection 2.2}. This behavior is analogous to the classic case of a perfectly straight column under compression, which remains unbuckled until perturbed. To further verify this, laterally perturbed static sliding studies for armchair GNR are presented in Appendix \ref{Appendix E}.
Fig.~\ref{All cases}d shows that when sliding along the zigzag direction, most of the GNR follows Path 3 due to the energy-favorable `BA' stacking state, causing the body of the GNR to buckle in-plane (see \href{https://doi.org/10.6084/m9.figshare.30061102.v1}{Supplementary Movie 1}). This, in turn, reduces the net pulling force compared to BC 1. However, because of the lateral constraint on the head of the GNR, the wavelength of the sliding force remains the same as in BC 1 (see Figs.~\ref{All cases}d and~\ref{All cases}b).

\subsubsection{BC 3}\label{subsubsection 3.2.3}

Fig.~\ref{All cases}e shows that during the pulling of armchair GNRs along the armchair direction of the substrate, the behavior remains the same as in BC 1 and BC 2, i.e., sliding occurs along Path 2. However, when pulling zigzag GNRs along the substrate’s zigzag axis, the GNR head follows a zigzag motion, as Path 3 represents the energetically preferred sliding path. The rest of the GNR adapts to this motion, resulting in a snake-like motion. This results in the reduction of the friction force by $\approx \, $50$\%$ as the path passes through another local stacking energy minimum (the `SP' stacking configuration), and the $P_y$ wavelength becomes exactly half, as seen in Fig.~\ref{All cases}f, compared to the zigzag sliding cases with BC 1 and BC 2, shown in Figs.~\ref{All cases}b and \ref{All cases}d, respectively.

\begin{figure}[H]
\begin{center} \unitlength1cm
\begin{picture}(0,5)
\put(-7.8,0){\includegraphics[height=52mm]{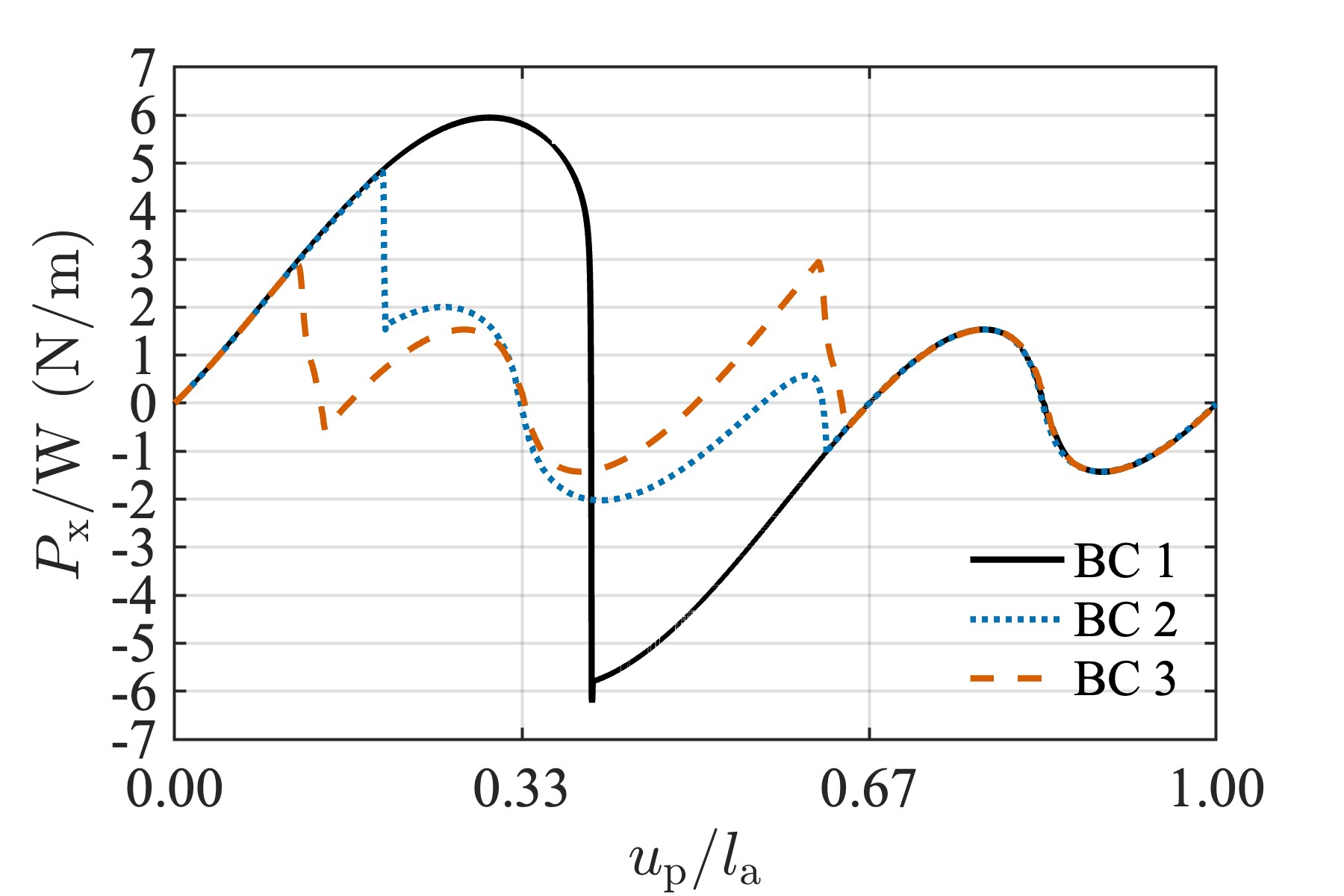}}
\put(-0.1,0){\includegraphics[height=52mm]{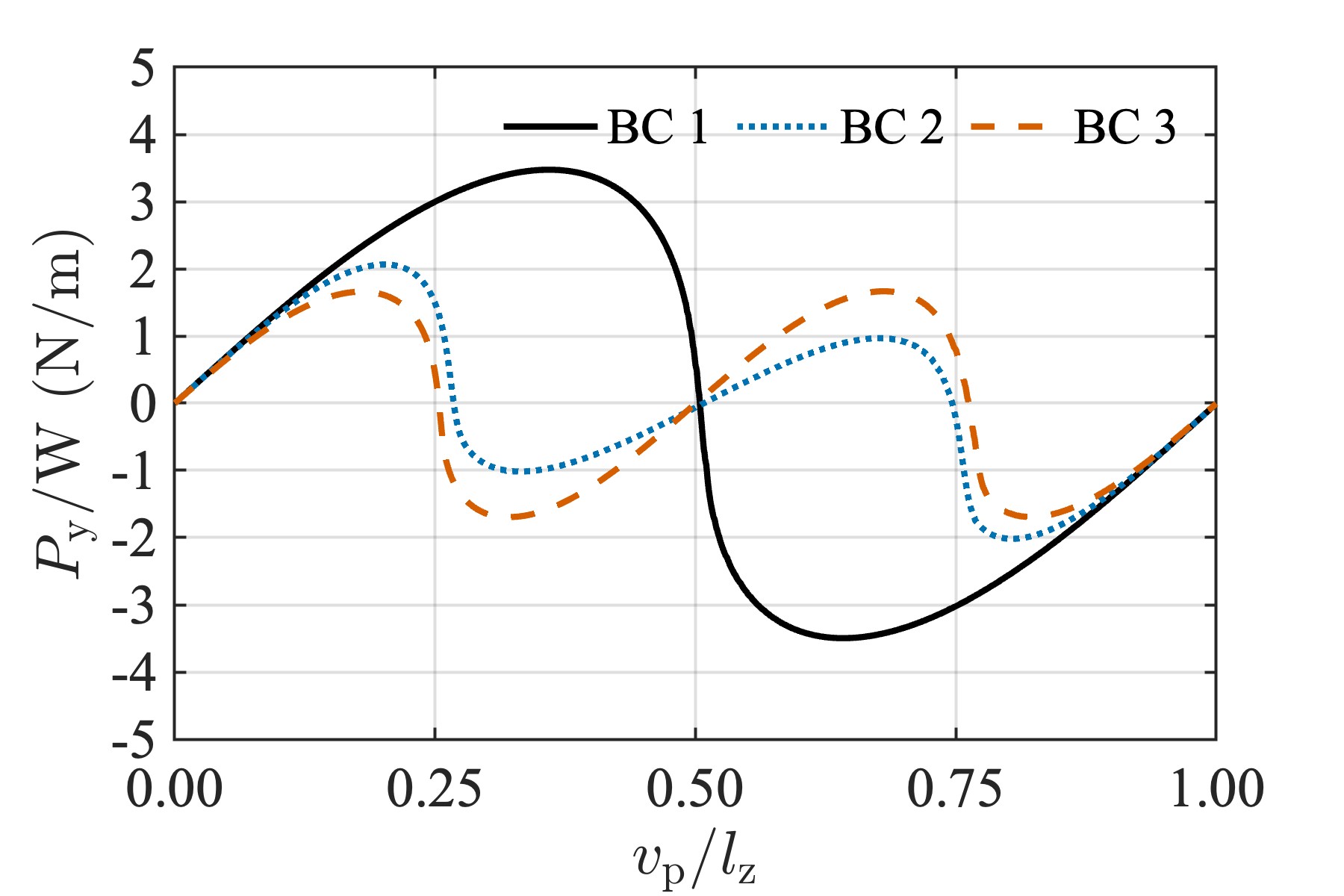}}
\put(-7,0){\fontsize{10pt}{20pt}\selectfont (a)  }
\put(0.3,0){\fontsize{10pt}{20pt}\selectfont (b) }
\end{picture}
\caption{ Pulling force-displacement curve of a sliding GNR for a 10 nm long ribbon subjected to the three different boundary conditions. FE results for sliding along (a) armchair and (b) zigzag directions, obtained with the DRM solver.}
\label{DRM 10 nm all bcs}
\end{center}
\end{figure}

On the other hand, during dynamic sliding, atoms follow Path 1 for BC 2 and BC 3 when armchair GNRs slide along the ${\boldsymbol e}_\mathrm{a}$-direction. This behavior arises because dynamic forces and oscillations perturb the perfect symmetry present along the ${\boldsymbol e}_\mathrm{a}$-direction. Now, the sliding behavior of both GNR types is boundary condition affected, as shown in Fig.~\ref{DRM 10 nm all bcs}a and \ref{DRM 10 nm all bcs}b for sliding along the ${ \boldsymbol{e}}_\mathrm{a}$- and  ${ \boldsymbol{e}}_\mathrm{z}$-direction, respectively. For BC 2 and BC 3, armchair GNRs also buckle, as indicated by the earlier drop in pulling force in Fig.~\ref{DRM 10 nm all bcs}a, compared to BC 1. Similar to zigzag GNRs, they follow a more energy-favorable sliding path, such as Path 1. For BC 3, buckling leads to a lowering in the peak forces by a factor of $\approx$ 2 compared to the case with BC 1 clearly seen in Fig.~\ref{DRM 10 nm all bcs}. Also, the wavelength of the pulling forces reduces to half for the BC 3 case, as stacking states $E$, $D$, and $A$ for armchair sliding correspond to the same energy state, i.e., representing the same structural configuration.

\begin{figure}[H]
\begin{center} \unitlength1cm
\begin{picture}(0,4.5)
\put(-7.8,0){\includegraphics[height=52mm]{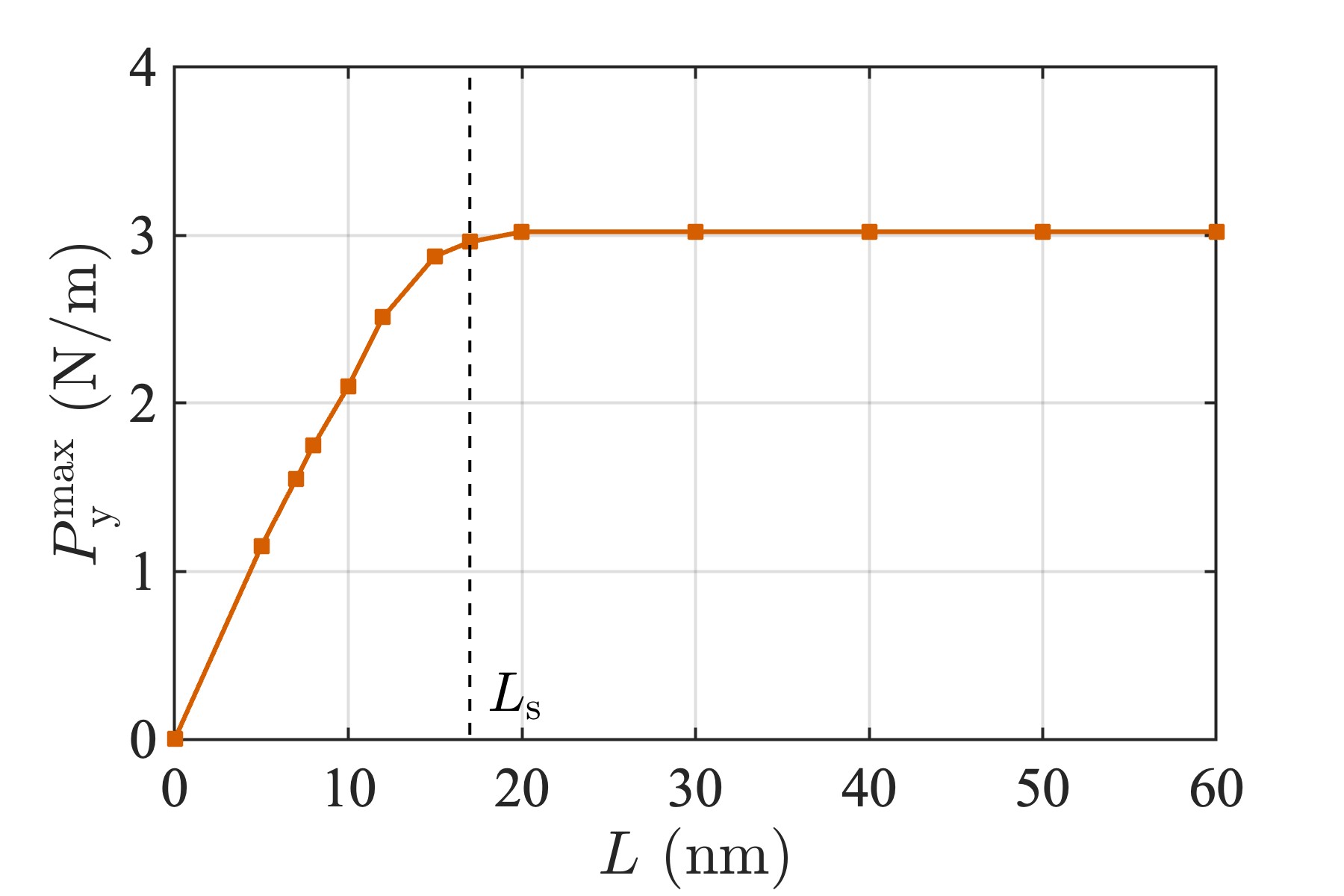}}
\put(-0.1,0){\includegraphics[height=52mm]{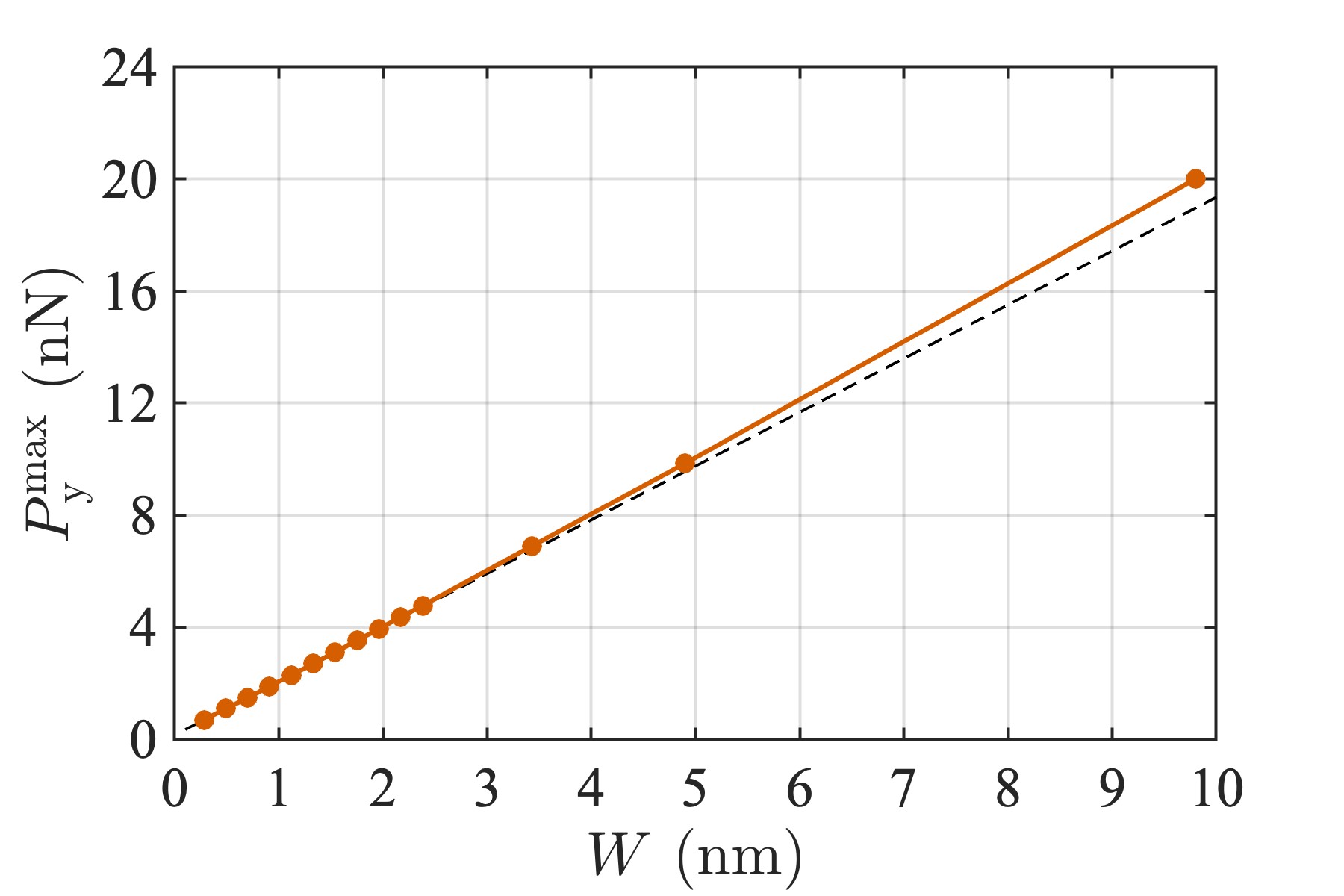}}
\put(-7,0){\fontsize{10pt}{20pt}\selectfont (a)  }
\put(0.3,0){\fontsize{10pt}{20pt}\selectfont (b) }
\end{picture}
\caption{Influence of GNR size on the peak pulling force, $P_{\mathrm{y}}^{\mathrm{max}}$, during sliding along the zigzag direction under BC 2. (a) Effect of GNR length for a 0.7 nm wide GNR; (b) Effect of GNR width for a 10 nm long GNR, (the linear black dashed line is for reference). }
\label{force saturation for zigzag BC -B}
\end{center}
\end{figure}

These observations provide an explanation for the variation of the friction force with ribbon length. For both armchair GNR and zigzag GNR sliding along armchair and zigzag direction of the graphene substrate, respectively, it is observed that the sliding behavior depends on the length of the ribbon $L$ $\leq $50 nm, as seen in Figs.~\ref{All cases} and ~\ref{MD simualtion sliding results}. For shorter GNRs, i.e., $L$ $<$ $L_{\text{c}}$, a linear increase of friction forces $P_x$ and $P_y$ is observed. Here, $L_\text{c}$ is the characteristic length scale for the stress decay in the GNR, see Appendix \ref{Appendix F}. For $L$ $>$ $L_{\text{c}}$, atoms close to and at the pulling edge experience the majority of the stretching, and the rest of the GNR remains almost in its original configuration, because of the shear-lag, until slip occurs. Fig.~\ref{force saturation for zigzag BC -B}a illustrates the critical ribbon length, $L_\text{s}$, beyond which the peak pulling force, $P_y^{\mathrm{max}}$, saturates. This saturation occurs because the elastic energy stored during sliding reaches its maximum, beyond which no further deformation can be accommodated by the GNR. This limiting behavior is discussed in more detail in Section \ref{subsection 3.3}. The value of $L_\text{s}$ from the FE model is $\approx$ 17 nm, which is consistent with previous theoretical calculations that predicted $L_\text{s}$ = 3.3 $L_\text{c}$ \citep{Liu2012, Xia2016, Feng2021}. The sliding behavior predicted by the FE simulations also shows strong consistency with the fully atomistic results reported in the literature \citep{wang2017size, Ouyang2018}. In particular, the peak force values are in excellent agreement with the MD simulations of \citet{wang2017size} for sliding along the ${\boldsymbol{e}}_\mathrm{z}$-direction with BC 3, with both approaches predicting a peak force of $\approx$ 2.4 N/m. Similarly, for the ${\boldsymbol{e}}_\mathrm{a}$-direction with BC 3, the FE model predicts a peak force of 3.54 N/m, which matches the value 3.65 N/m reported by \citet{Ouyang2018}. Additionally, the effect of GNR width on the sliding response is examined, as shown in Fig.~\ref{force saturation for zigzag BC -B}b. This result indicates that the width primarily affects the response quantitatively, through a linear increase. However, at larger widths, an increase in the in-plane bending stiffness is observed, which reduces the tendency for buckling. This trend is consistent with the observations reported by \citet{Korhonen2016, Xue2022}.

\subsubsection{Energy transitions and reversibilty}\label{subsubsection 3.2.4}

\begin{figure}[H]
\begin{center} \unitlength1cm
\begin{picture}(0,5.6)
\put(-4,0){\includegraphics[height=54mm]{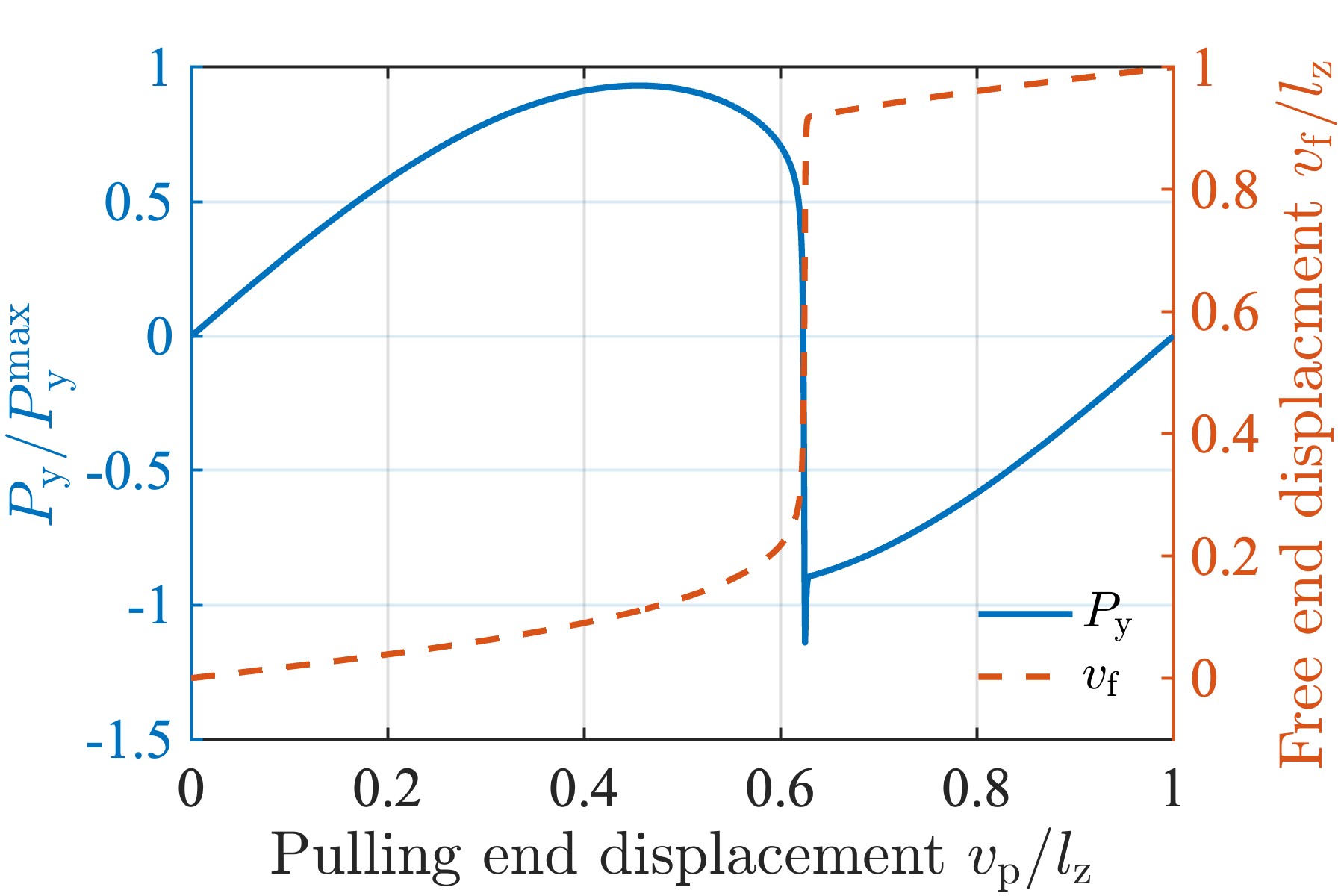}}
\end{picture}
\caption{ The pulling force and the free end displacement of a 16 nm GNR under laterally constrained boundary condition (BC 1). (It should be noted that the dip observed in the pulling force–displacement curve at the end of slip does not represent the (static) equilibrium points. Rather, its occurrence and magnitude depend on the chosen parameters of the DRM).}
\label{force_slip_10nm}
\end{center}
\end{figure}

In Fig.~\ref{force_slip_10nm}, the displacement of the free end is plotted against the pulling displacement at the front end. A sudden increase in the relative displacement (i.e., slip) of the free end is observed, which is accompanied by a sharp drop in the pulling force. This behavior is akin to stick-slip sliding and is found to be dependent on the length of the GNR. In this section, we investigate how the length of the GNR influences the extent of slip and how the system's energetics evolve during stick-slip. Here, we aim to develop a theoretical understanding of dissipation and irreversibility within the FE framework. The quantification of dissipation and irreversibility is carried out in terms of the amount of slip. Among the considered boundary conditions, slip is most clearly observed under BC 1, whereas other boundary conditions introduce in-plane bending effects that complicate the interpretation. For these reasons, a GNR subjected to BC 1, with its 
${ \boldsymbol{e}}_\mathrm{z}$-axis aligned along the loading direction, is considered in this study.

During the pull force drop, energy transitions occur: The conversion of potential energy to kinetic energy, followed by its dissipation. The total energy of the system, $\mathrm{E}_{\mathrm{tot}}$ can be divided into four parts: The surface potential $\Pi_{\mathrm{surf}}$, the elastic energy $\Pi_{\mathrm{el}}$, the kinetic energy  $ T$, and the dissipated energy $\mathcal{D}$ such that $\mathrm{E}_{\mathrm{tot}}$ = $\Pi_{\mathrm{surf}} + \Pi_{\mathrm{el}} + T + \mathcal{D}$. The first three are given by

\begin{equation}
  \ds  \Pi_{\mathrm{surf}} = W\int_0^L \Psi \, \text{d}x \, ,
  \label{interaction energy}
\end{equation}

\begin{equation}
 \ds   \Pi_{\mathrm{el}} = \frac{1}{2} \textbf{u}   \cdot \textbf{K}\textbf{u} \, ,
 \label{strain energy}
\end{equation}

\begin{equation}
  \ds  T = \frac{1}{2} \dot{\textbf{u}}  \cdot  \textbf{M} \dot{\textbf{u}}\, .
  \label{Kinetic energy}
\end{equation}

\begin{figure}[h]
\begin{center} \unitlength1cm
\begin{picture}(0,6)
\put(-7.8,0){\includegraphics[height=54mm]{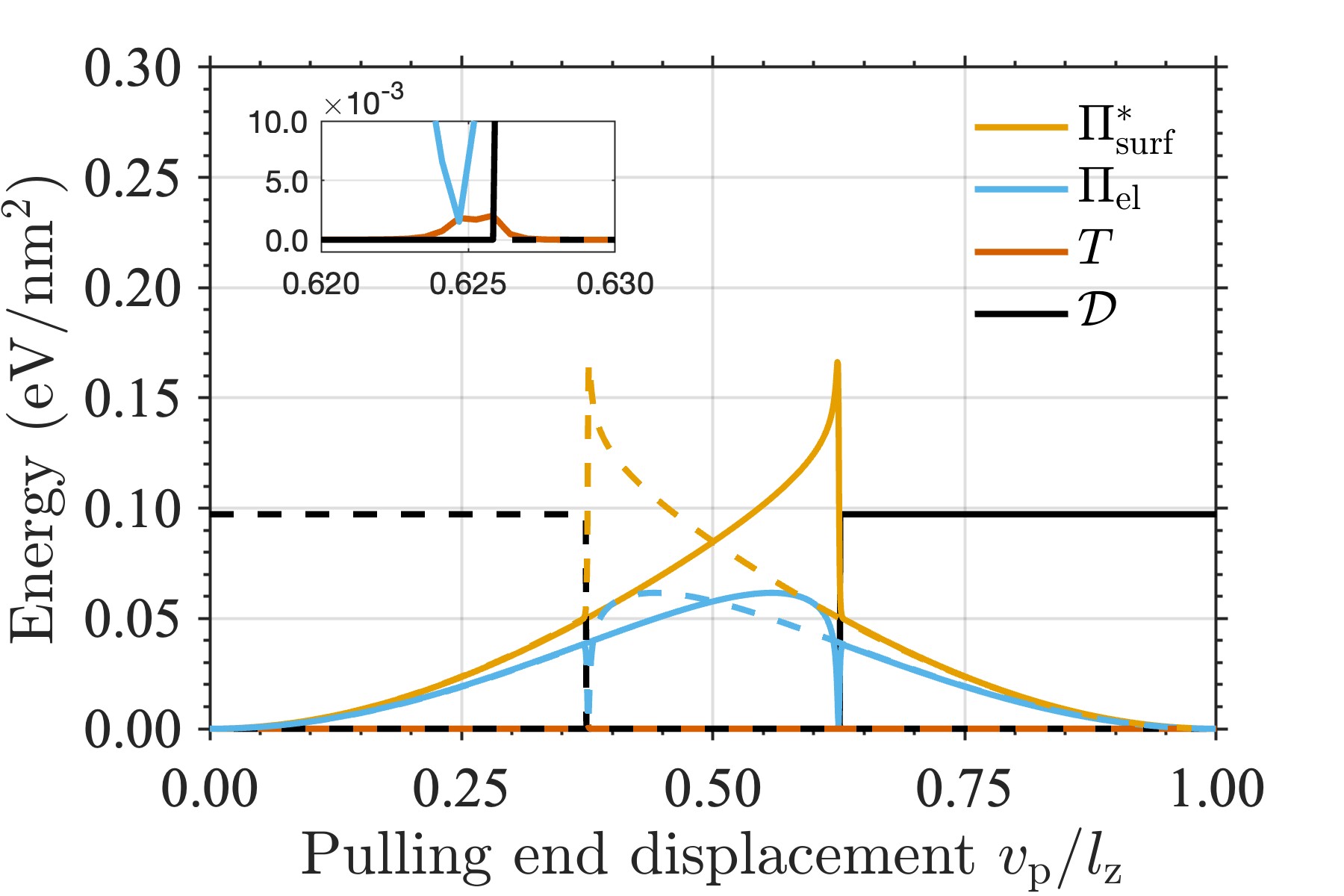}}
\put(-0.1,0){\includegraphics[height=54mm]{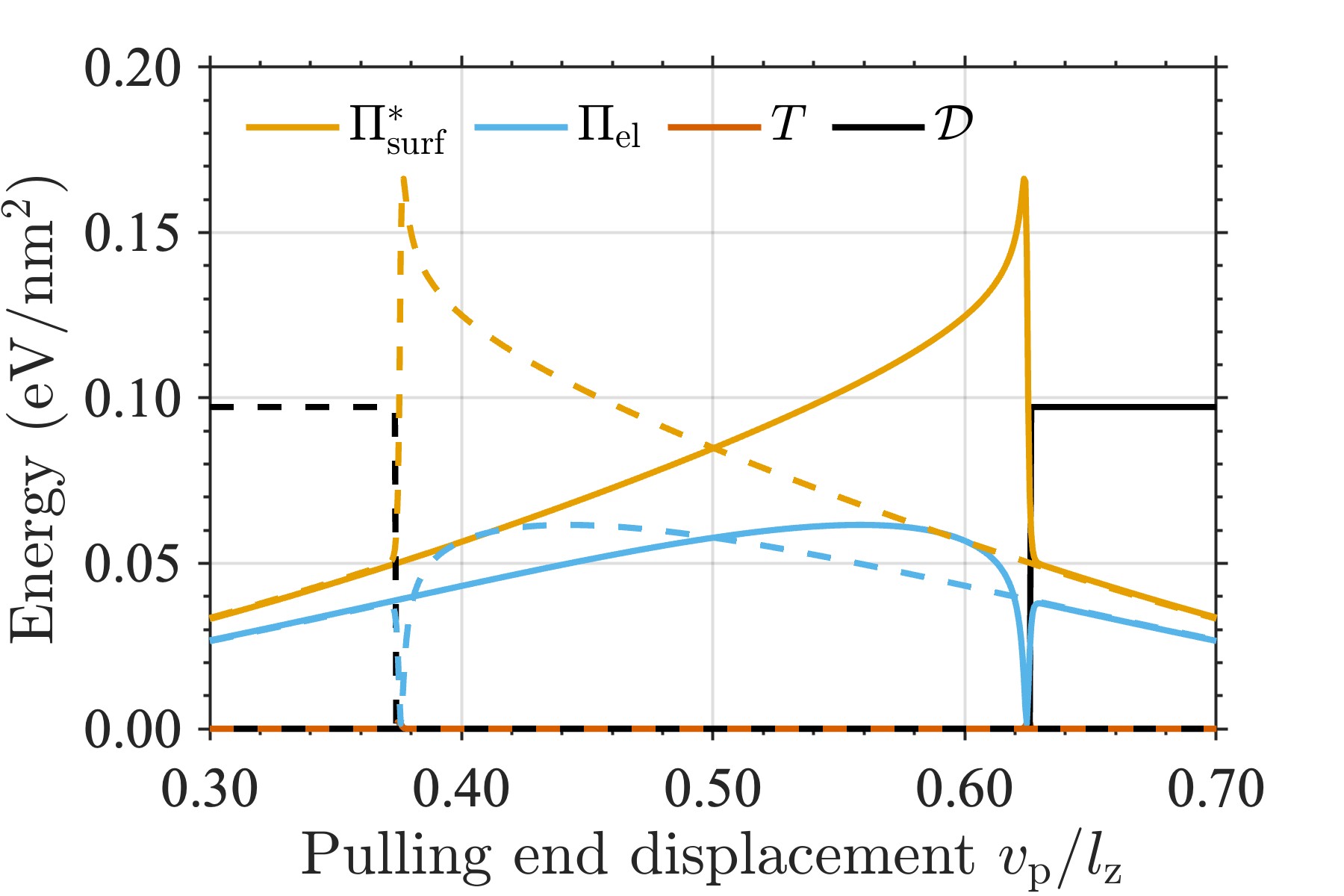}}
\put(-7,0){\fontsize{10pt}{20pt}\selectfont (a)  }
\put(0.3,0){\fontsize{10pt}{20pt}\selectfont (b) }
\end{picture}
\caption{Transition between different energies during one stick-slip wavelength of a 16 nm GNR sliding with BC 1 along the zigzag direction. The forward and backward sliding occur along the ${ \boldsymbol{e}}_\mathrm{z}$ (solid lines) and -${ \boldsymbol{e}}_\mathrm{z}$-directions (dashed lines), respectively. (a) Full energy evolution over the cycle; (b) Enlarged of (a). ($\Pi_{\mathrm{surf}}^*$ is the shift of surface potential energy of $\Pi_{\mathrm{surf}}$ to the zero level). }
\label{energy transitions}
\end{center}
\end{figure}

\begin{figure}[H]
\begin{center} \unitlength1cm
\begin{picture}(0,5.2)
\put(-7.8,0){\includegraphics[height=52mm]{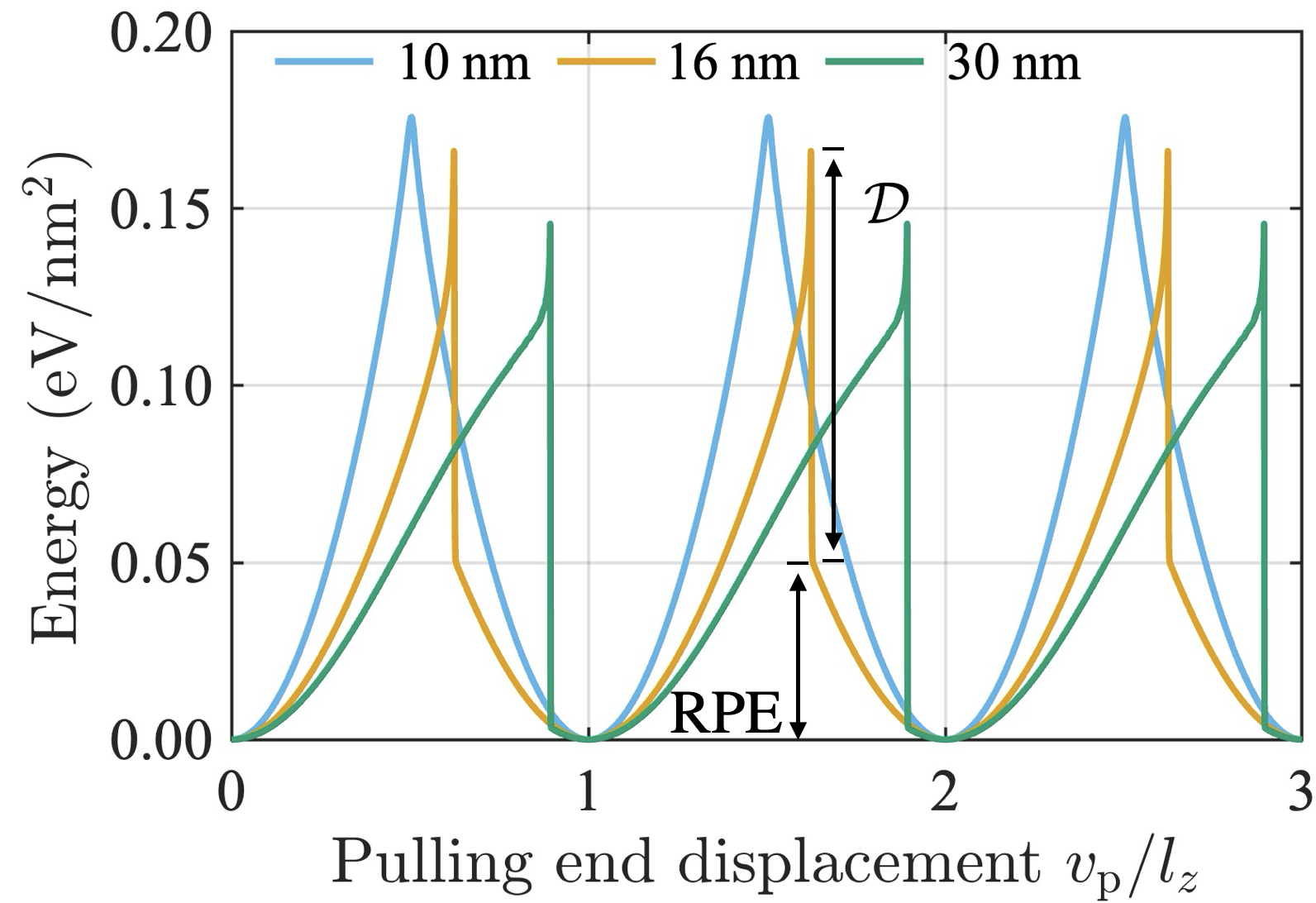}}
\put(-0.1,0){\includegraphics[height=54.2mm]{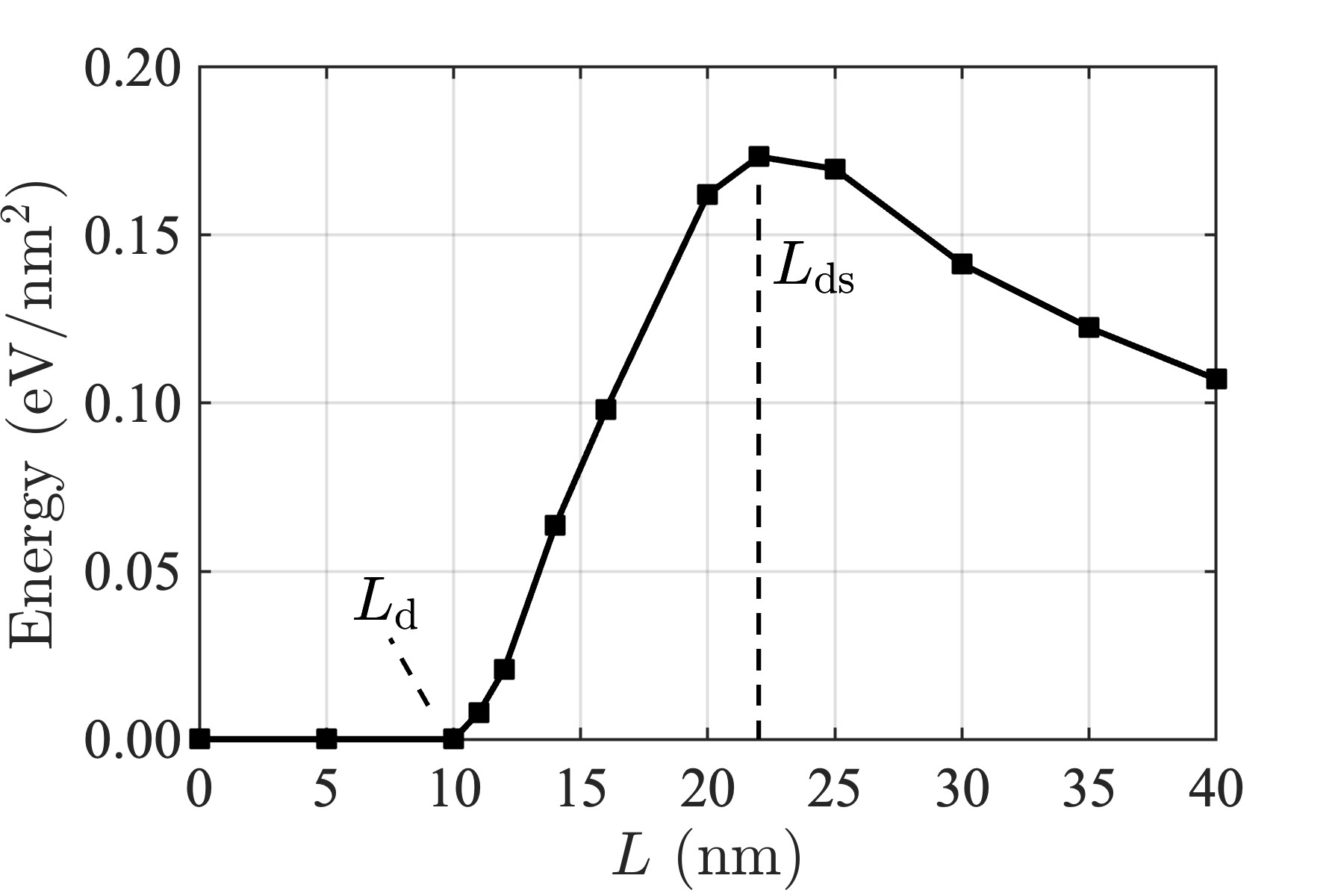}}
\put(-7,0){\fontsize{10pt}{20pt}\selectfont (a)  }
\put(0.3,0){\fontsize{10pt}{20pt}\selectfont (b) }
\end{picture}
\caption{(a) Variation of energy throughout the stick-slip motion for 10, 16, and 30 nm GNR lengths. (b) Variation of $\mathcal{D}$ with ribbon length.  }
\label{energy variation}
\end{center}
\end{figure}

  $\mathcal{D}$ is calculated from the area under the force-displacement curve in Fig.~\ref{force_slip_10nm}. Using Eq.~\eqref{interaction energy}-\eqref{Kinetic energy}, the energies are obtained, and the transition during a complete stick-slip period is shown in Fig.~\ref{energy transitions}. All of the energy dissipation occurs during the force drop, which leads to a sudden decrease in $\Pi_{\mathrm{surf}}$ (beyond 
$v_\mathrm{p}/l_\mathrm{z}$ =1/2) and a corresponding synchronous increase of $ T$, 
$\Pi_{\mathrm{el}}$, and $\mathcal{D}$. During force drop, $\Pi_{\mathrm{surf}}$ is not fully dissipated; a reversible portion remains, and it gradually decreases back to zero as the sliding reaches $v_\mathrm{p} = l_\mathrm{z}$. This remaining part of $\Pi_{\mathrm{surf}}$ is approximately 0.05 eV/$\mathrm{nm}^2$ (see Fig.~\ref{energy transitions}a and \ref{energy transitions}b). Toward the end of the force drop, $\Pi_{\mathrm{el}}$ increases abruptly. The sum of this increase in $\Pi_{\mathrm{el}}$ and the remaining portion of $\Pi_{\mathrm{surf}}$ constitutes the reversible potential energy, referred to as RPE. It is observed from Fig.~\ref{energy transitions}a and Fig.~\ref{energy transitions}b, that the energy transitions during forward and backward sliding are identical. Fig.~\ref{energy variation}a shows the variation of the system energy during the stick-slip motion for various ribbon lengths. It can be clearly noticed that as the GNR length increases, RPE decreases, while $\mathcal{D}$ increases.

It is clear from Fig.~\ref{energy variation}a and \ref{energy variation}b that $\mathcal{D}$ increases as the ribbon length increases. As shown in Fig.~\ref{energy variation}b, the variation of $\mathcal{D}$ with GNR length reveals a critical length, $L_\text{d}$, beyond which the sliding process becomes dissipative. Additionally, a saturation length $L_{\text{ds}}$ $\approx$ 22 nm, is identified, beyond which $\mathcal{D}$ begins to decrease. Here, length $L_\text{d}$  marks the GNR length beyond which snap-through occurs, while $L_{\text{ds}}$ corresponds to the critical length required for soliton formation.

\begin{figure}[H]
\begin{center} \unitlength1cm
\begin{picture}(0,6)
\put(-4,0){\includegraphics[height=54mm]{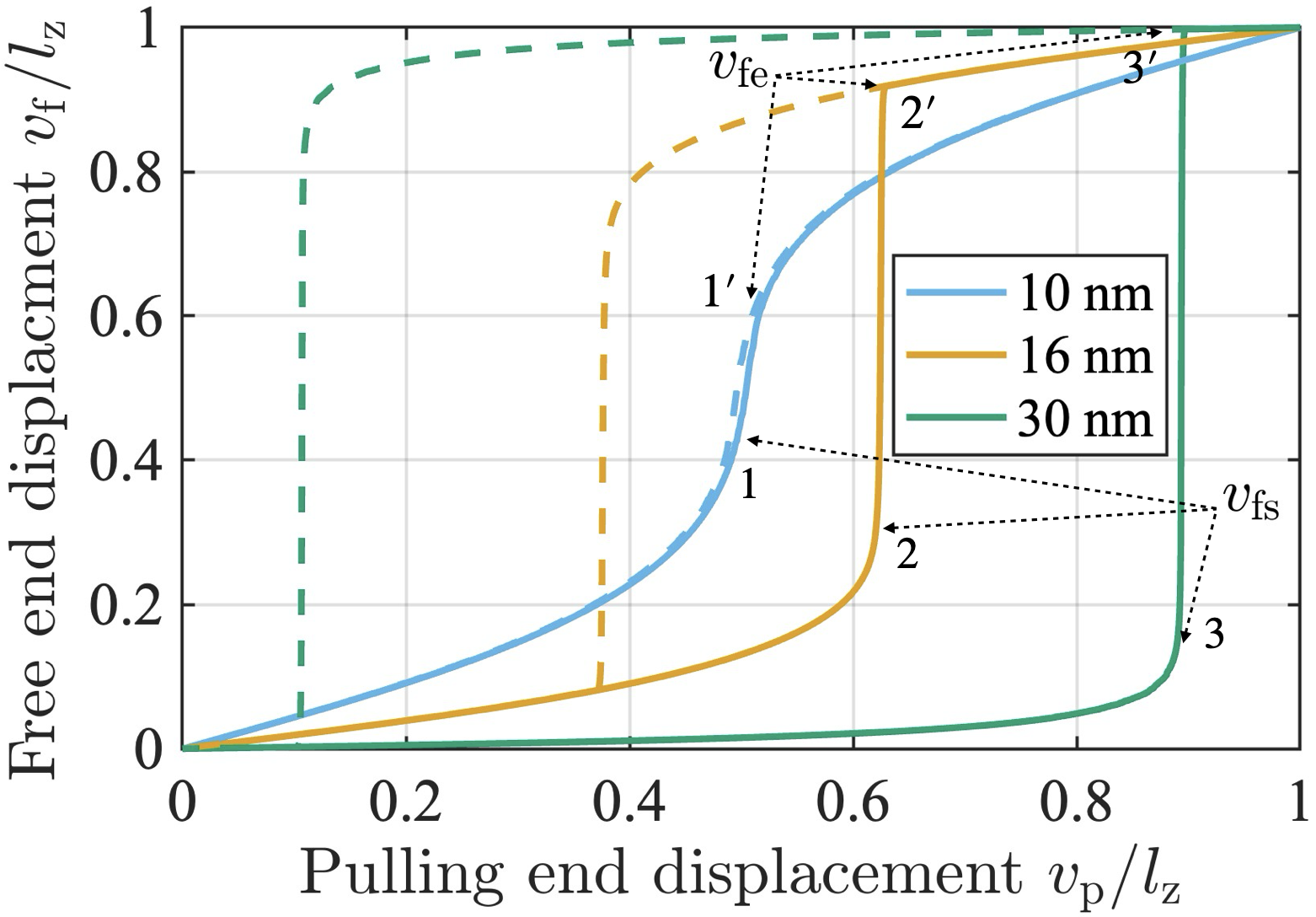}}
\end{picture}
\caption{ GNRs free end displacement $v_\mathrm{f}$ for different lengths $L$ in nm, showcasing the amount of slip during forward (solid lines) and backward (dashed lines) sliding along the ${ \boldsymbol{e}}_\mathrm{z}$-direction.}
\label{slip_trajectory}
\end{center}
\end{figure}

This energy transition is further examined with respect to the slip magnitude at the free end. The amount of slip is quantified by identifying the start point (\(v_\text{fs}\)) and end point (\(v_\text{fe}\)) of the sudden rise in the displacement of the free GNR end. Fig.~\ref{slip_trajectory} presents three pulling end trajectories during forward and backward loading for GNR lengths \(L = 10\), 16, and 30 nm. The slip length, marked by 1-$1^{\prime}$, 2-$2^{\prime}$, and 3-$3^{\prime}$, increases with the GNR length. Consequently, the amount of reversible potential energy (RPE) decreases and $\mathcal{D}$ increases as shown in Fig.~\ref{energy variation}a. Also, it is observed that during backward sliding, slip happens in a similar fashion as in forward sliding; this is conveyed by their similar trajectory shown in Fig.~\ref{slip_trajectory}.

\subsection{Strain transmission}\label{subsection 3.3}

To further investigate the interlayer sliding mechanism, extreme boundary conditions BC 1 and BC 3 are considered. This section focuses on strain transmission in zigzag GNRs under BC 1. Results for zigzag GNRs under BC 3 are presented in Appendix~\ref{Appendix G}, while those for armchair GNRs under both boundary conditions are provided in Appendix~\ref{Appendix H}.

\begin{figure}[H]
\begin{center} \unitlength1cm
\begin{picture}(0,5)
\put(-7.8,0.4){\includegraphics[height=45mm]{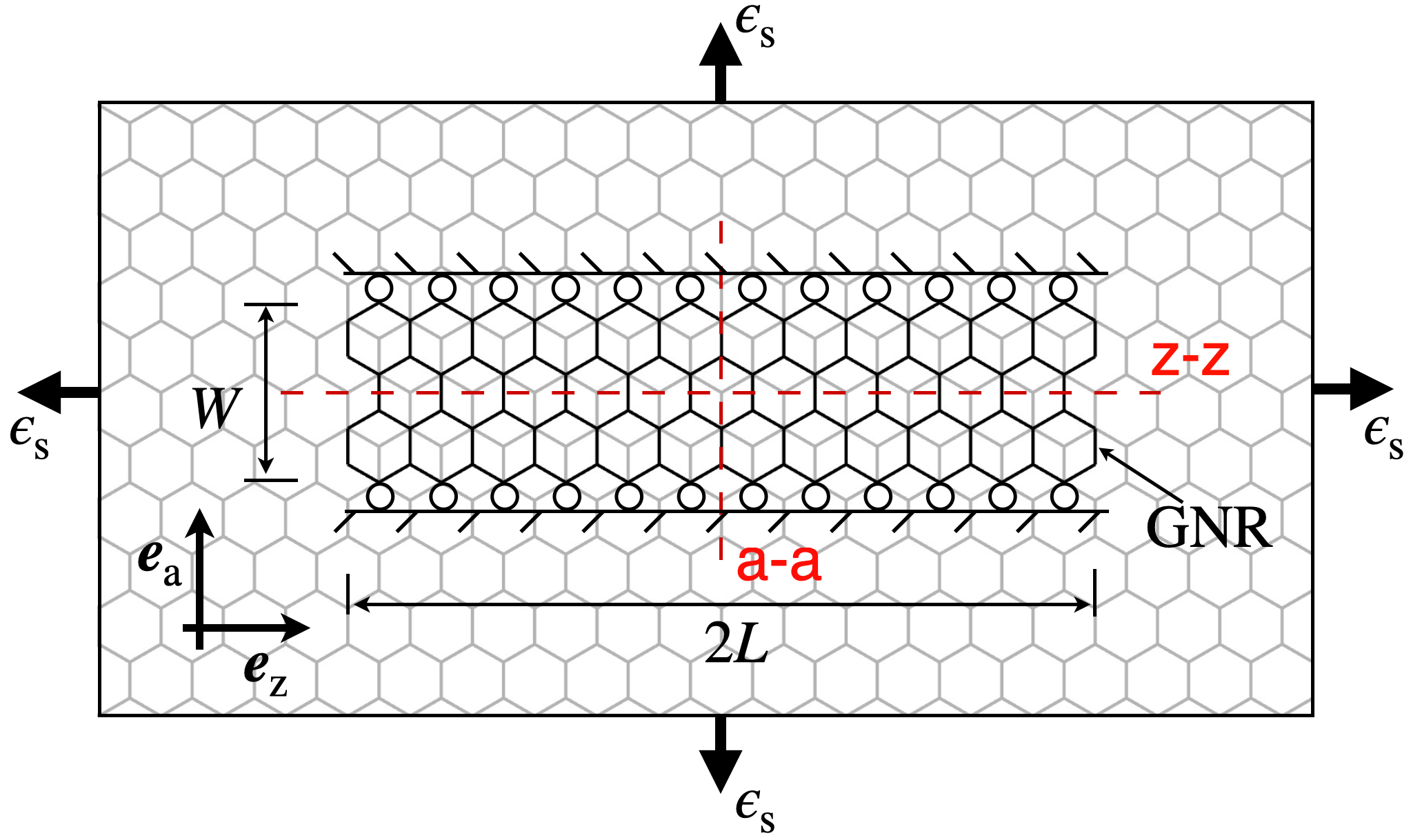}}
\put(0,0){\includegraphics[height=50mm]{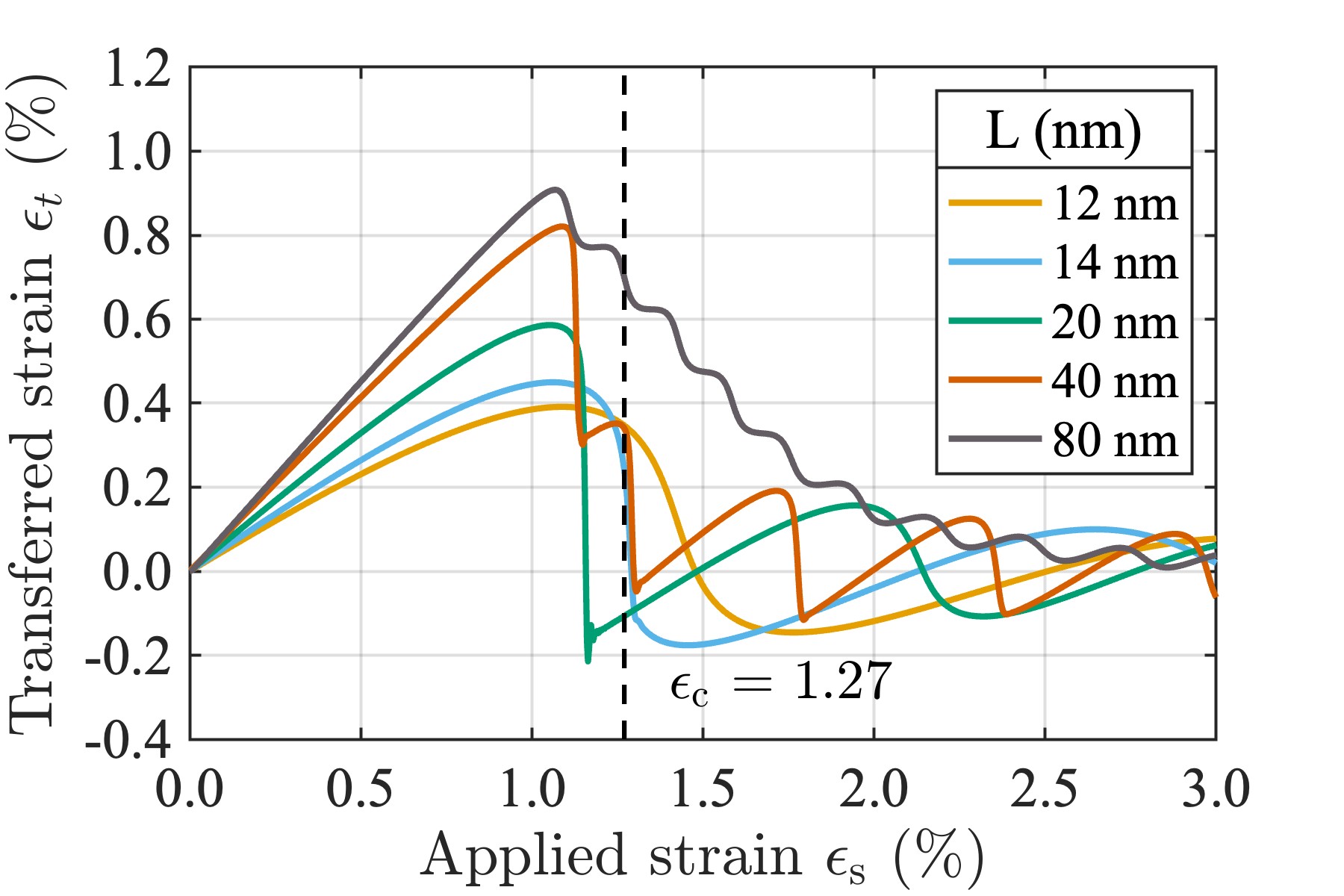}}
\put(-7.5,0){\fontsize{10pt}{20pt}\selectfont (a)  }
\put(0.3,0){\fontsize{10pt}{20pt}\selectfont (b) }
\end{picture}
\caption{(a) Schematic of strain transmission from substrate to GNR. (b) Variation of the average transferred strain $\epsilon_\text{t}$ with the applied strain $\epsilon_\text{s}$ for different zigzag GNR of length $L$.}
\label{Strain transfer}
\end{center}
\end{figure}

A zigzag GNR of length $2L$, initially kept in commensurate contact with the graphene substrate, under BC 1 is considered as illustrated in Fig.~\ref{Strain transfer}a. Only the substrate is uniformly and biaxially strained. FE simulations are carried out for various lengths of GNR to analyze the strain transfer characteristics as a function of the applied strain $\epsilon_{\mathrm{s}}$, and the observed behavior is presented in Fig.~\ref{Strain transfer}b. Since the entire system is symmetric about the a-a axis see Fig.~\ref{Strain transfer}a, the effective length for studying the interlayer mechanics is $L$.

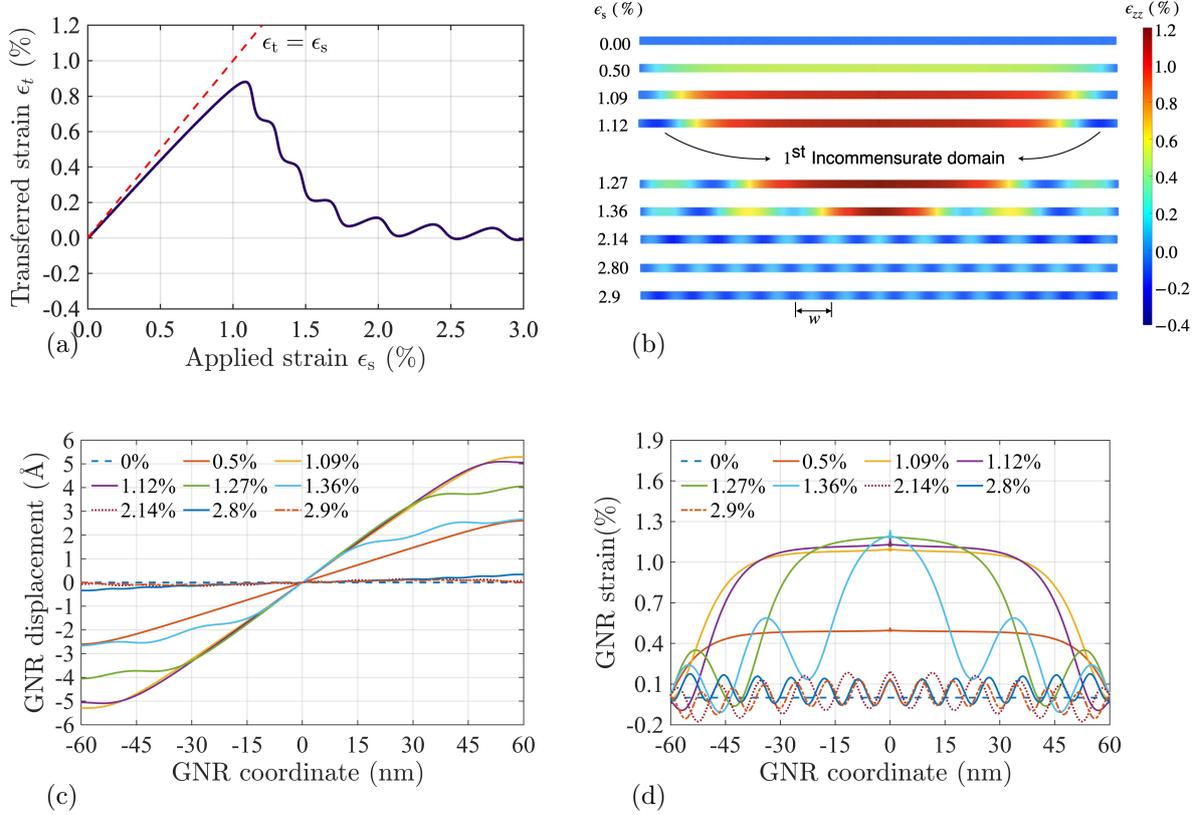
\begin{figure}[H]
\begin{center} \unitlength1cm
\begin{picture}(0,11.5)
\put(-7.7,6){\includegraphics[height=50mm]{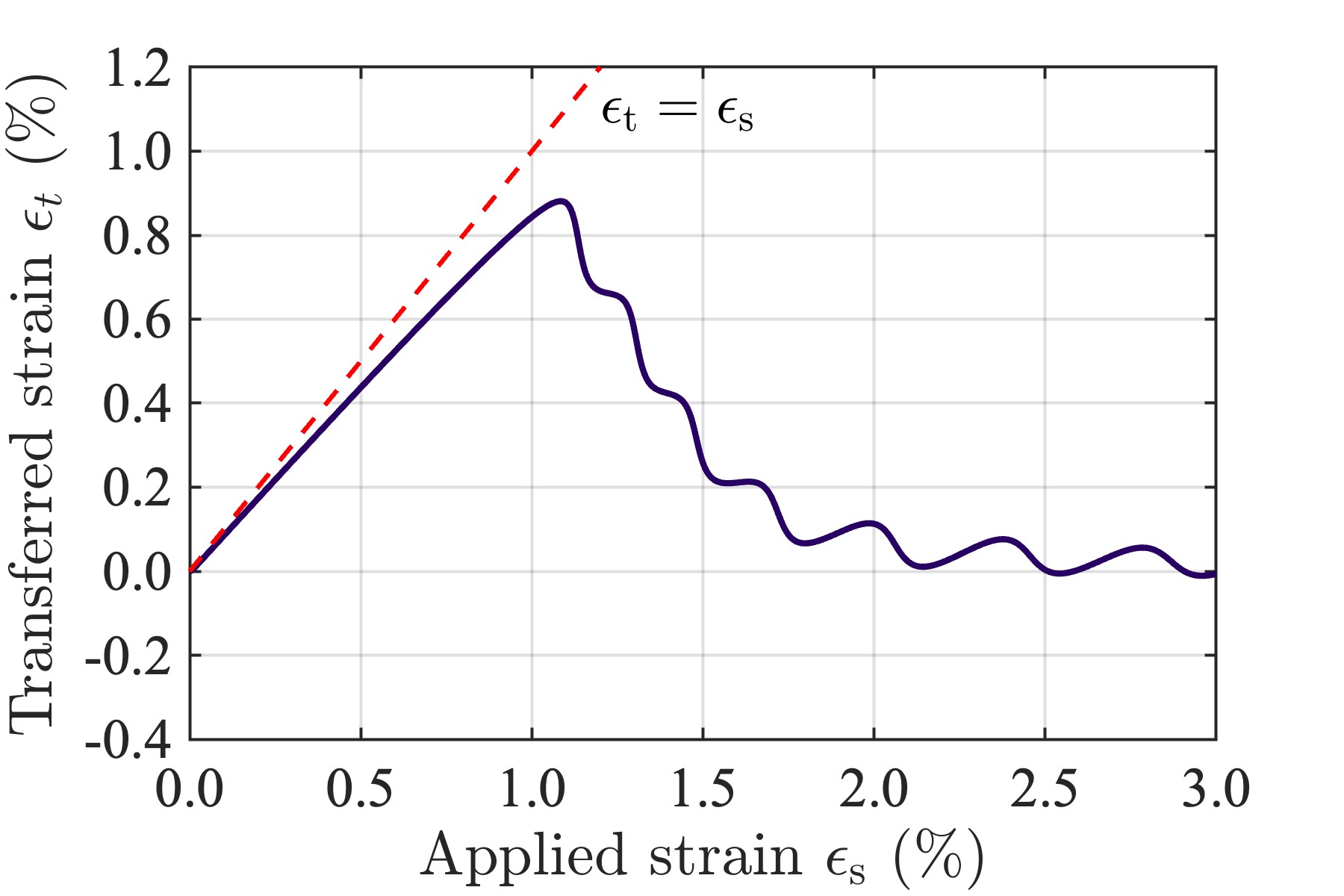}}
\put(0,6.5){\includegraphics[height=45mm]{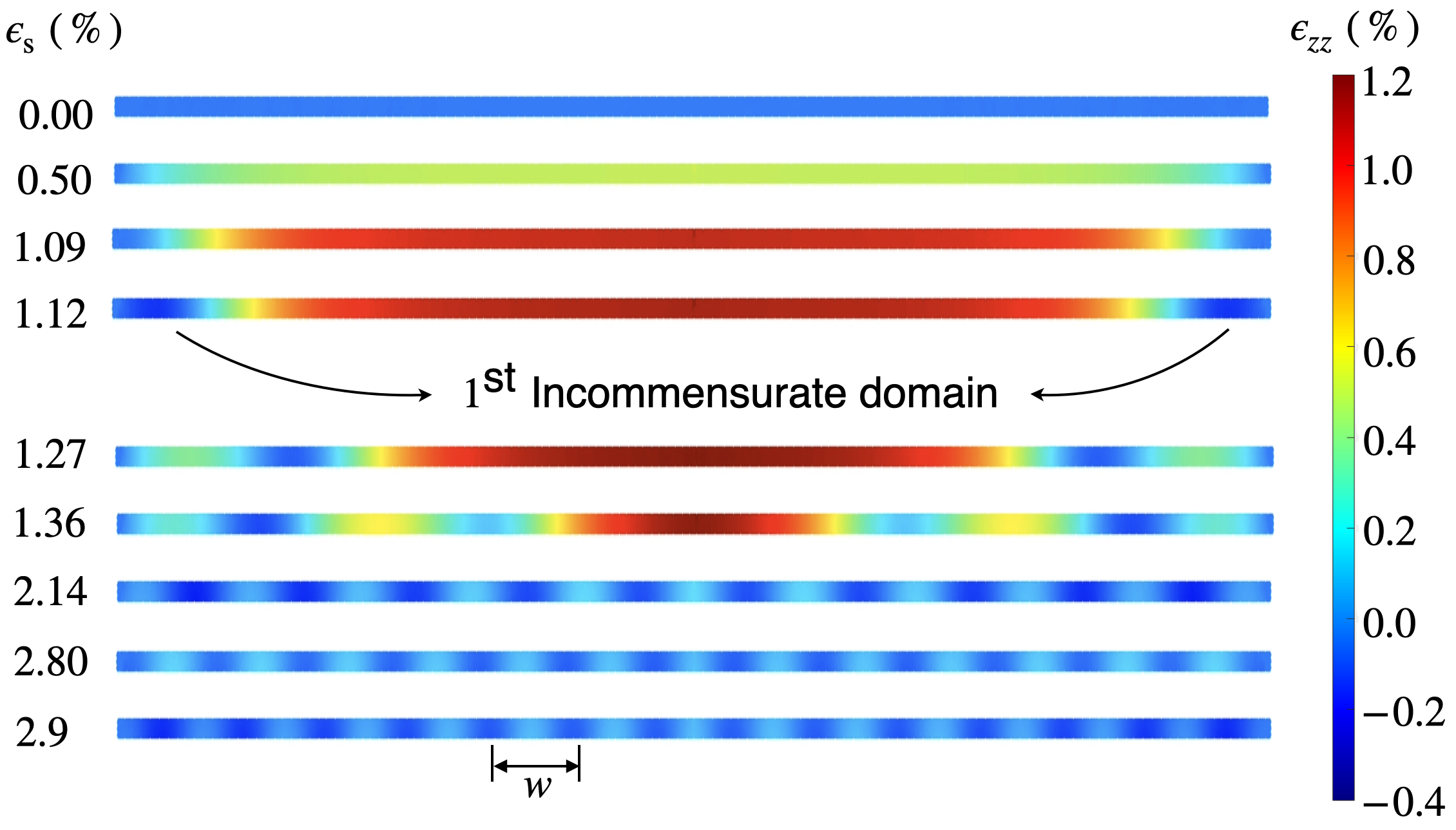}}
\put(-7.7,0.5){\includegraphics[height=50mm]{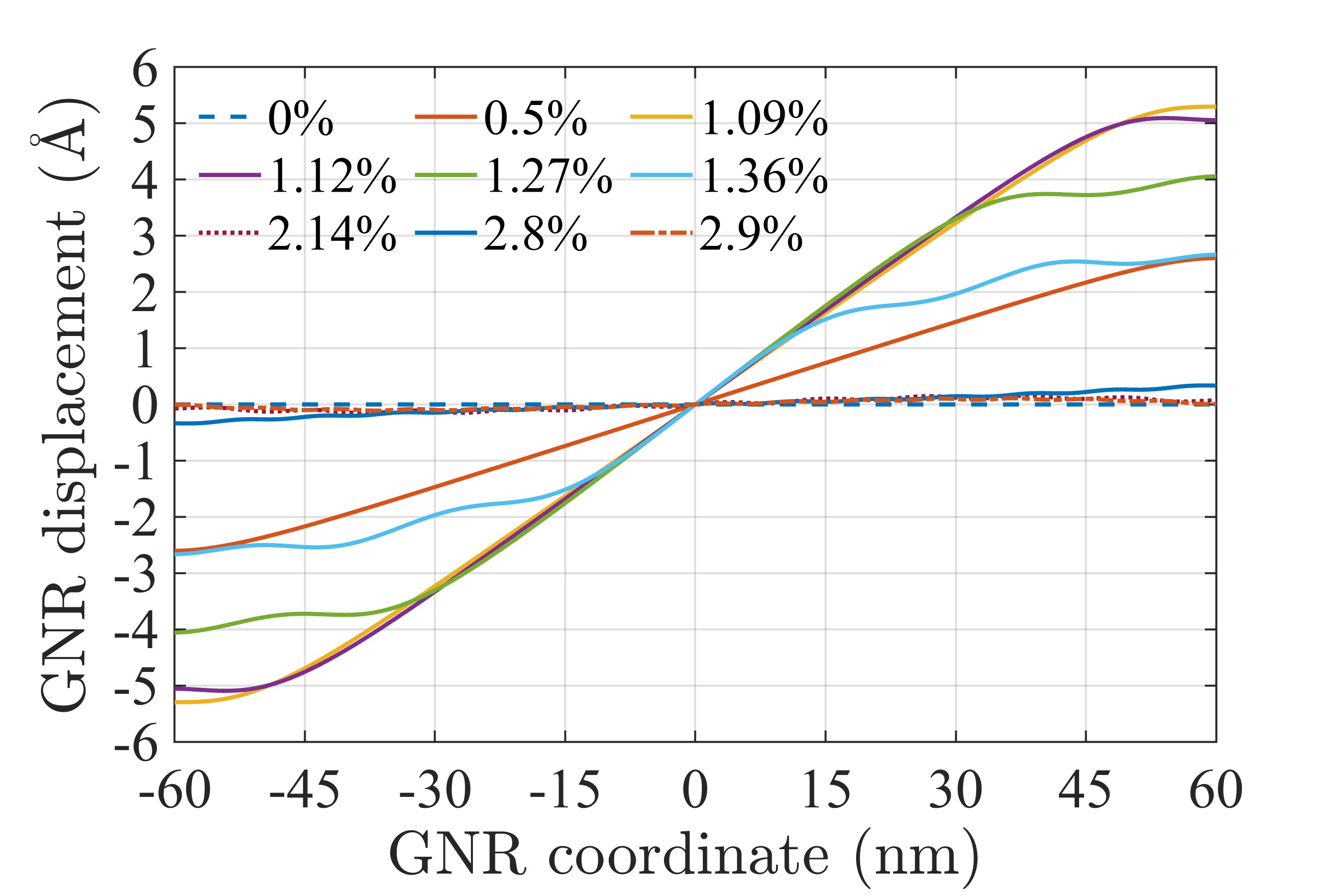}}
\put(0,0.5){\includegraphics[height=50mm]{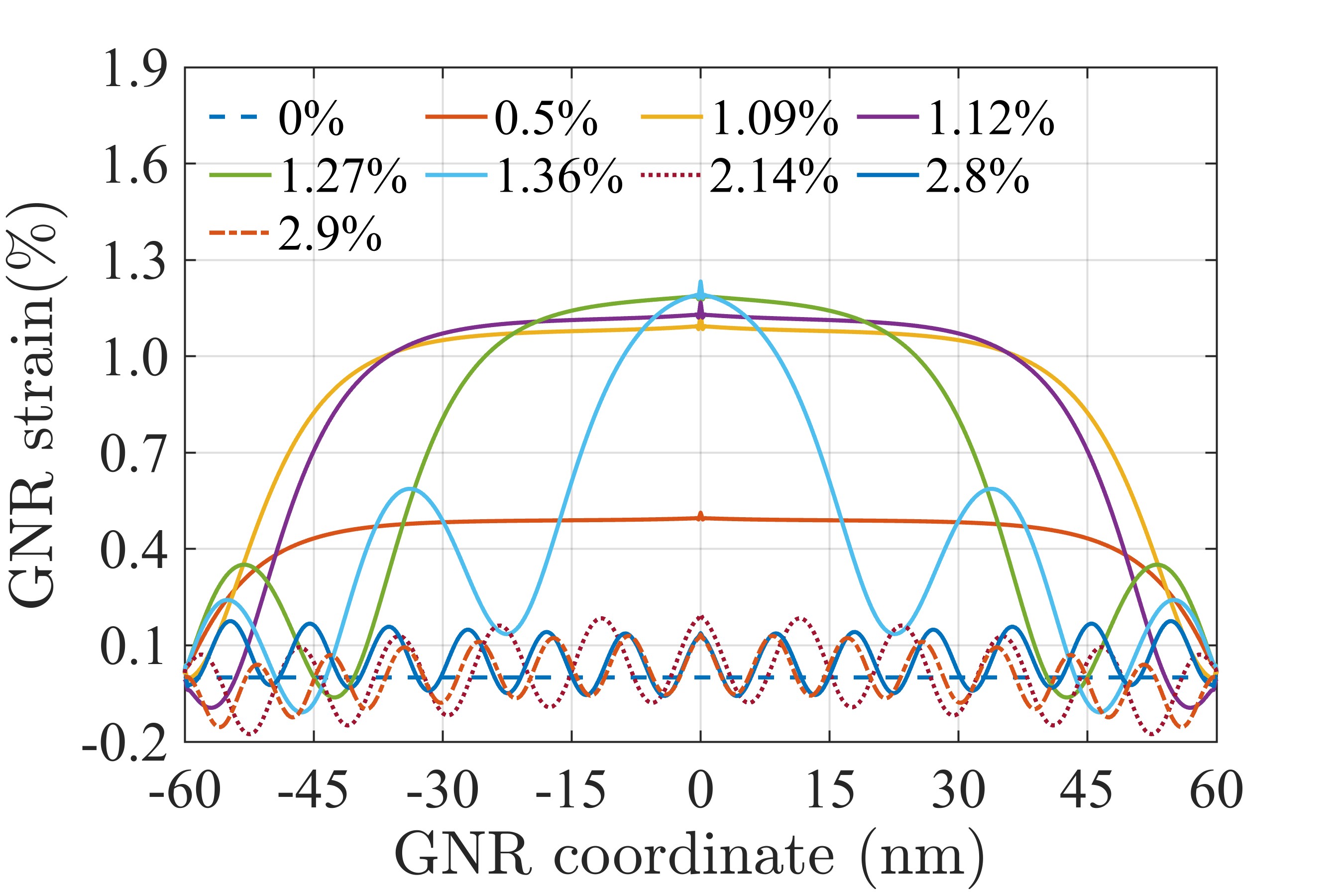}}
\put(-7.2,6.3){\fontsize{10pt}{20pt}\selectfont (a)  }
\put(0.5,6.3){\fontsize{10pt}{20pt}\selectfont (b) }
\put(-7.2,0.3){\fontsize{10pt}{20pt}\selectfont (c)  }
\put(0.5,0.3){\fontsize{10pt}{20pt}\selectfont (d) }
\end{picture}
\caption{Strain transmission with BC 1 in zigzag GNR of length 120 nm (=$2L$) aligned along the ${ \boldsymbol{e}}_\mathrm{z}$-direction of the substrate (a) Average strain transferred (dashed red line is for reference). (b) Simulation snapshots at key loading stages illustrating the distribution of $\epsilon_{zz}$ along the central axis z-z. The distribution of displacement field (c), and strain field (d)  in GNR for various substrate strains.}
\label{Strain distribution}
\end{center}
\end{figure}

Initially, for low values of the substrate strain, the two layers remain in a commensurate state, resulting in an almost linear increase in the average transferred strain $\epsilon_\text{t}$ from the substrate to the GNR. The actual transferred strain is slightly lower as seen in Fig.~\ref{Strain transfer}b and ~\ref{Strain distribution}a, due to the formation of incommensurate zones near the free edges. 
Once the applied strain exceeds a critical value $\epsilon_\text{c}$ (i.e, $\approx$ 1.1$\%$ for GNR of 40 nm), the elastic energy required to maintain commensurability outweighs the gain in interfacial vdW energy, thereby initiating the formation of localized incommensurate regions as shown in Fig~\ref{Strain distribution}b and \ref{Strain distribution}d (also see \href{https://doi.org/10.6084/m9.figshare.30061102.v1}{Supplementary Movie 2}). In these incommensurate domains, interlayer shear is significant and the strains are very small and slightly compressive as well, see Fig~\ref{Strain distribution}b and ~\ref{Strain distribution}d. Each observed dip in the strain transfer curve shown in Fig.~\ref{Strain transfer}b corresponds to the formation of such an incommensurate domain. For longer GNRs, as the applied strain increases, a saw tooth behavior in the transferred strain $\epsilon_\text{t}$ is observed. This occurs due to the sequential formation of localized incommensurate regions beginning at the free end of the nanoribbon, representing local relaxation events. The width $w$ of these regions, as shown in Fig.~\ref{Strain distribution}b, can be estimated as $\approx$ $l_\mathrm{z}(1+\epsilon_{\mathrm{s}})/\epsilon_{\mathrm{s}}$. Upon further application of strain, these incommensurate regions grow in number and extend towards the center of the GNR (z-z axis), eventually rendering the entire GNR effectively strain-free, Fig~\ref{Strain distribution}. 

Interestingly, for shorter GNRs $L$ $\leq$ $L_{\text{c}}$, the degree of initial commensuration is found to be minimal. The observed commensuration further diminishes with decreasing nanoribbon length, attributed to the disproportionately higher elastic energy cost relative to the vdW energy benefit. For such short GNRs, with continued strain application, slipping is observed to occur throughout the GNR. This slipping may even surpass the zero-strain state, temporarily entering a compressive strain regime before relaxing back to a zero-strain configuration, see Fig.~\ref{Strain transfer}b.

\begin{figure}[H]
\begin{center} \unitlength1cm
\begin{picture}(0,5.3)
\put(-7.8,0){\includegraphics[height=50mm]{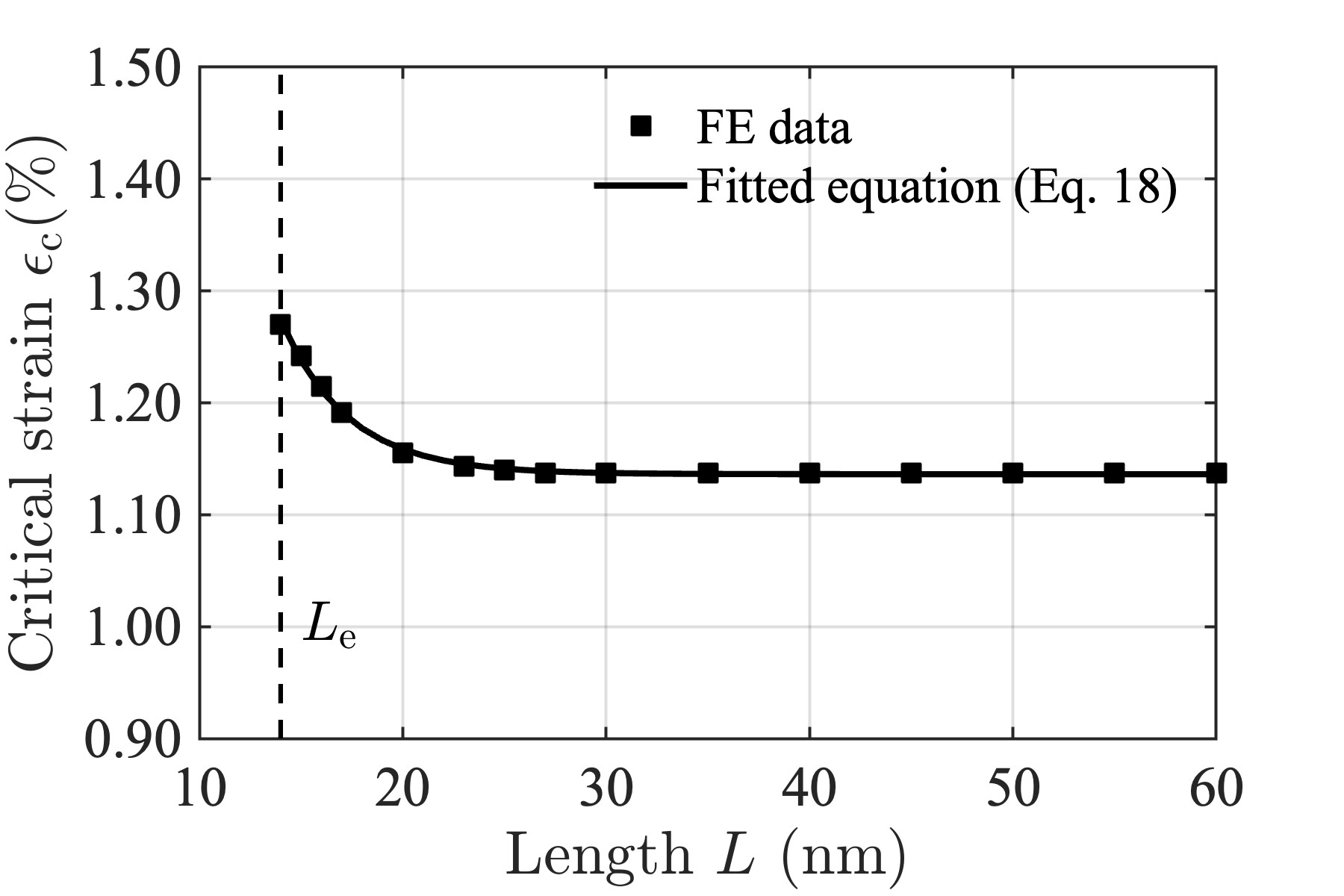}}
\put(0,0){\includegraphics[height=50mm]{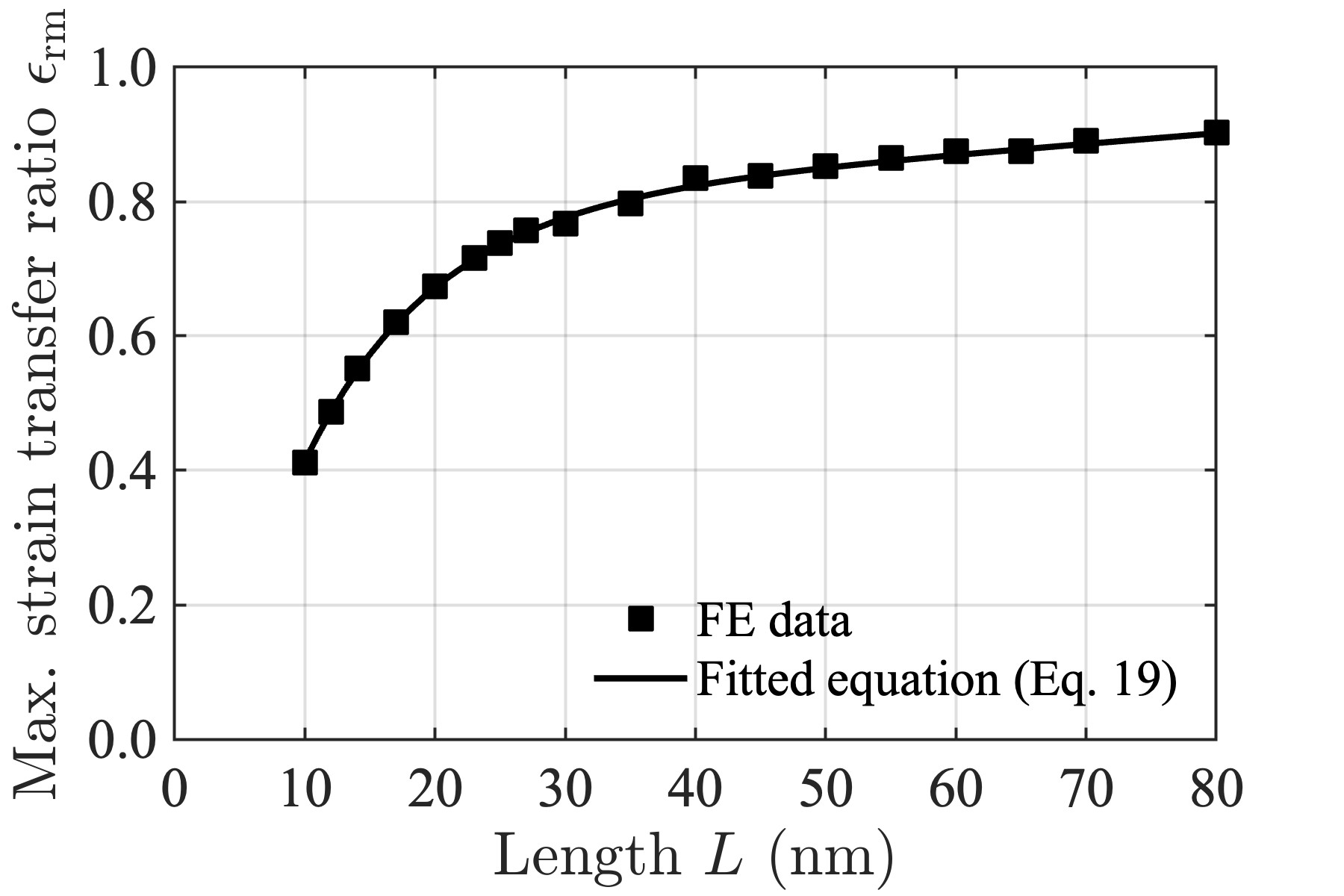}}
\put(-7.5,0){\fontsize{10pt}{20pt}\selectfont (a)  }
\put(0.3,0){\fontsize{10pt}{20pt}\selectfont (b) }
\end{picture}
\caption{(a) Critical strain for interlayer sliding with different GNR lengths (b) Influence of GNR length $L$ on the maximum strain transfer ratio. }
\label{critical strain transfer and max strain transfer}
\end{center}
\end{figure}

 It is found that this critical strain only arises when the GNR length exceeds approximately 14 nm ($\approx$ 3$L_{\text{c}}$). The corresponding strain value is 1.27$\%$ as highlighted in Fig.~\ref{Strain transfer}b. The corresponding length at which this strain occurs is referred to as the critical length for interlayer slipping, denoted by $L_{\text{e}}$. Furthermore, the critical strain $\epsilon_\text{c}$ decreases with increasing GNR length and approaches a saturation value of 1.15$\%$, as illustrated in Fig.~\ref{critical strain transfer and max strain transfer}a. Since the shear-lag theory leads to an exponential variation of the fields (see Appendix~\ref{Appendix F}) and $\epsilon_\text{c}$ exhibits saturation, its dependence on the GNR length $L$ (in nm) can be accurately represented by a hyperbolic cotangent function, given by
\begin{equation}
\ds     \epsilon_\text{c}  = a_1 \,\coth{(a_2\, L )} +a_3 ~.
\end{equation}

where $a_1$= 4.5199, $a_2$ = 0.15 $\text{nm}^{-1}$, and $a_3$ = -3.3837 are the curve fitting constants, obtained using MATLAB’s curve fitting toolbox, which determines them through nonlinear least-squares optimization. A similar curve fitting procedure is also employed for Eq.\eqref{strain_transmission_strained_zigzag_equation} and Eq.\eqref{critical_unstrained_zigzag_equation}.

To evaluate the efficiency of strain transfer across different lengths, the maximum strain transfer ratio $\epsilon_{\text{rm}}$ is studied, which is defined as the ratio of maximum $\epsilon_\text{t}$ to $\epsilon_{\text{s}}$. As indicated by the reference line in Fig.~\ref{Strain distribution}a, the applied strain is not fully transferred to the GNR due to losses at free edges. Consequently, perfect efficiency is achieved only in the limit of an infinite GNR length. The relationship between $\epsilon_{\text{rm}}$ and the GNR length $L$, as shown in Fig.~\ref{critical strain transfer and max strain transfer}b, can therefore be approximated by exponential functions, expressed as
\begin{equation}
\ds     \epsilon_{\text{rm}}  = b_1 \,\text{exp} (b_2\, L) + b_3 \,\text{exp} (b_4 \,L)~,
\label{strain_transmission_strained_zigzag_equation}
\end{equation}
with fitting constant $b_1 = 0.7873$, $b_2$ = 0.0017 $\text{nm}^{-1}$, $b_3 = -1.0515$, and $b_4$ = -0.1 $\text{nm}^{-1}$. These results clearly demonstrate the crucial role of the GNR length $L$ in governing both the extent of strain transfer and the overall structural stability of the vdW system. 

Similar to the pulling forces in the sliding case -- where the sliding force is reduced by approximately half when the boundary conditions are changed from BC 1 to BC 3, as shown in Fig.~\ref{All cases} and Fig.~\ref{DRM 10 nm all bcs} -- a similar trend is observed in the strain transmission behavior of the zigzag GNR see Fig.~\ref{Strain transfer}b and Fig.~\ref{Strain transmission zigzag BC 3}a. In contrast, this trend is less pronounced for armchair GNR (Fig.~\ref{armchair strain transmission}), owing to the double wavelength along the armchair direction, which causes the system to remain in a buckled configuration and lose alignment with the applied strain direction. Furthermore, analogous to the sliding case -- where the sliding force in the armchair direction under BC 1 is approximately 50$\%$ higher than in the zigzag direction (Fig.~\ref{DRM 10 nm all bcs}) -- the strain transmission results for armchair GNR shown in Fig.~\ref{armchair strain transmission}a also exhibit a similar directional dependence as zigzag GNR shown in Fig.~\ref{Strain transfer}b.

Our results are also in agreement with those reported in previous studies. \citet{Kumar2016}, using a one-dimensional modeling approach corresponding to the BC 1 configuration, reported 1.1$\%$ as the critical strain value. Similarly, \citet{Feng2021}, employing the Frenkel–Kontorova theory, predicted $\epsilon_\text{c}$ as 1.03$\%$ for BC 1 and 0.58$\%$ for BC 3. In comparison, our simulations yield $\epsilon_\text{c}$ = 1.15$\%$ for BC 1 and 0.59$\%$ for BC 3.

\section{Conclusion and outlook} \label{Section 4}
This work presents a finite element model based on Euler–Bernoulli beam formulation to study the interlayer interaction mechanics of a GNR sliding over a graphene substrate. The effects of size, sliding direction, and time dependency on the interfacial mechanics are examined. Using a continuation solver, unstable equilibria during GNR sliding are identified. Numerical results demonstrate good agreement with MD simulations, offering a computationally efficient alternative for studying large-scale nanostructures. The main findings are:
\begin{enumerate}
   
\item 
The continuum-based FE approach provides results consistent with MD simulations while significantly reducing computational cost, making it promising for large-scale modeling of two-dimensional materials.
\item 
The edge-pulling force saturates beyond a characteristic GNR length of $L_{\text{s}} \geq 17$ nm, indicating a length scale beyond which increasing ribbon size has negligible effect on the pulling force.
\item 
A distinct transition occurs at $L_{\text{d}} \approx 10$ nm, where sliding behavior changes from non-dissipative to dissipative. This highlights the role of system stability in the onset of energy dissipation.
\item 
A critical strain, $\epsilon_{\text{c}}$, emerges when the GNR length exceeds $L_{\text{e}}$ $\approx$ 14 nm, indicating a length-dependent activation of interlayer
deformation mechanisms.
\item 
This critical strain limits the maximum possible strain transferred to GNR, that is, 0.59$\%$ for BC 3 and 1.15$\%$ for BC 1.

\item 
While the present study focuses on graphene-graphene interactions, the developed FE model can be extended to other 2D material systems, enabling broad applicability in nanoscale interface mechanics.
\item 
Although computationally efficient, the present approach calls for further refinement of the contact mechanics representation to capture additional interfacial phenomena.
\item 
The findings provide essential design guidelines for graphene-based and other 2D nanomaterial systems, with implications for flexible electronics, tribology, strain sensing, and nanocomposite development.

\end{enumerate}

\section*{CRediT authorship contribution statement} 
\textbf{Gourav Yadav}: Writing – original
draft, Visualization, Validation, Methodology, Investigation, Formal
analysis, Data curation, Conceptualization. \textbf{Aningi Mokhalingam}: Writing – review and editing, Formal
analysis, Data curation, Investigation. \textbf{Shakti S. Gupta}: Writing – review and editing, Resources, Conceptualization, Formal
analysis, Supervision. \textbf{Roger A. Sauer}:
Writing – review and editing, Resources, Methodology, Conceptualization, Formal
analysis, Supervision.

\section*{Declaration of competing interest}
The authors declare no known competing financial or personal interests that could have influenced the work reported in this paper.

\section*{Data availability}
Data will be made available on request.

\section*{Acknowledgements} 
We acknowledge the support of the PARAM Sanganak supercomputer at IIT Kanpur, funded by the National Supercomputing Mission. Also, Gourav Yadav gratefully acknowledges the financial support from Ruhr University Bochum, Germany, received during his appointment as a guest researcher in the course of this research. 

\appendix
\setcounter{section}{0}
\renewcommand{\theequation}{\thesection\arabic{equation}}
\setcounter{equation}{0}

\titleformat{\section}[block]
  {\normalfont\Large\bfseries}
  {\appendixname~\thesection:}{0.5em}{}

\newpage
\section{Effect of time-step size and damping coefficient}\label{Appendix A}

\begin{figure}[H]
\begin{center} \unitlength1cm
\begin{picture}(0,5)
\put(-7.8,0){\includegraphics[height=50mm]{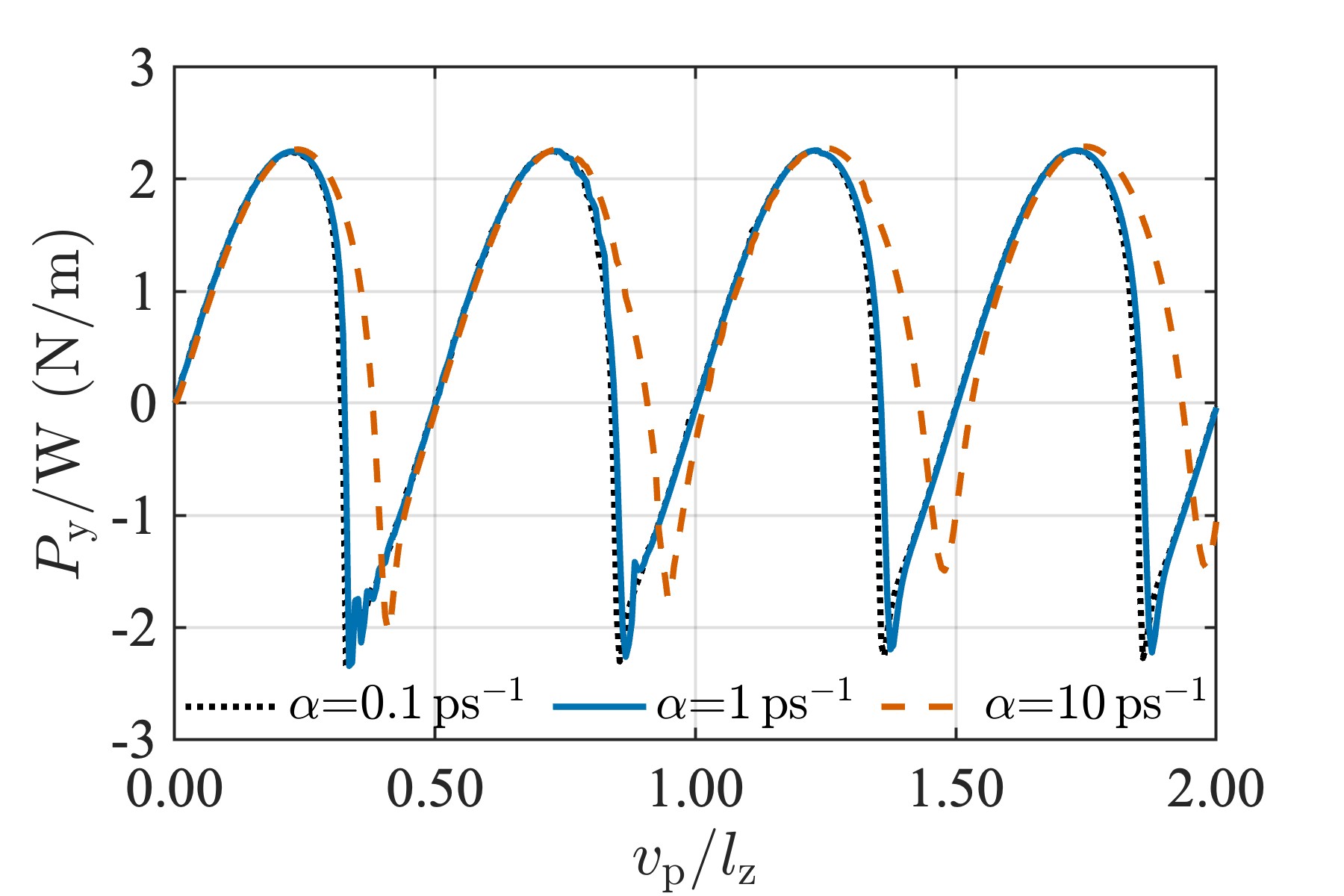}}
\put(0,0){\includegraphics[height=50mm]{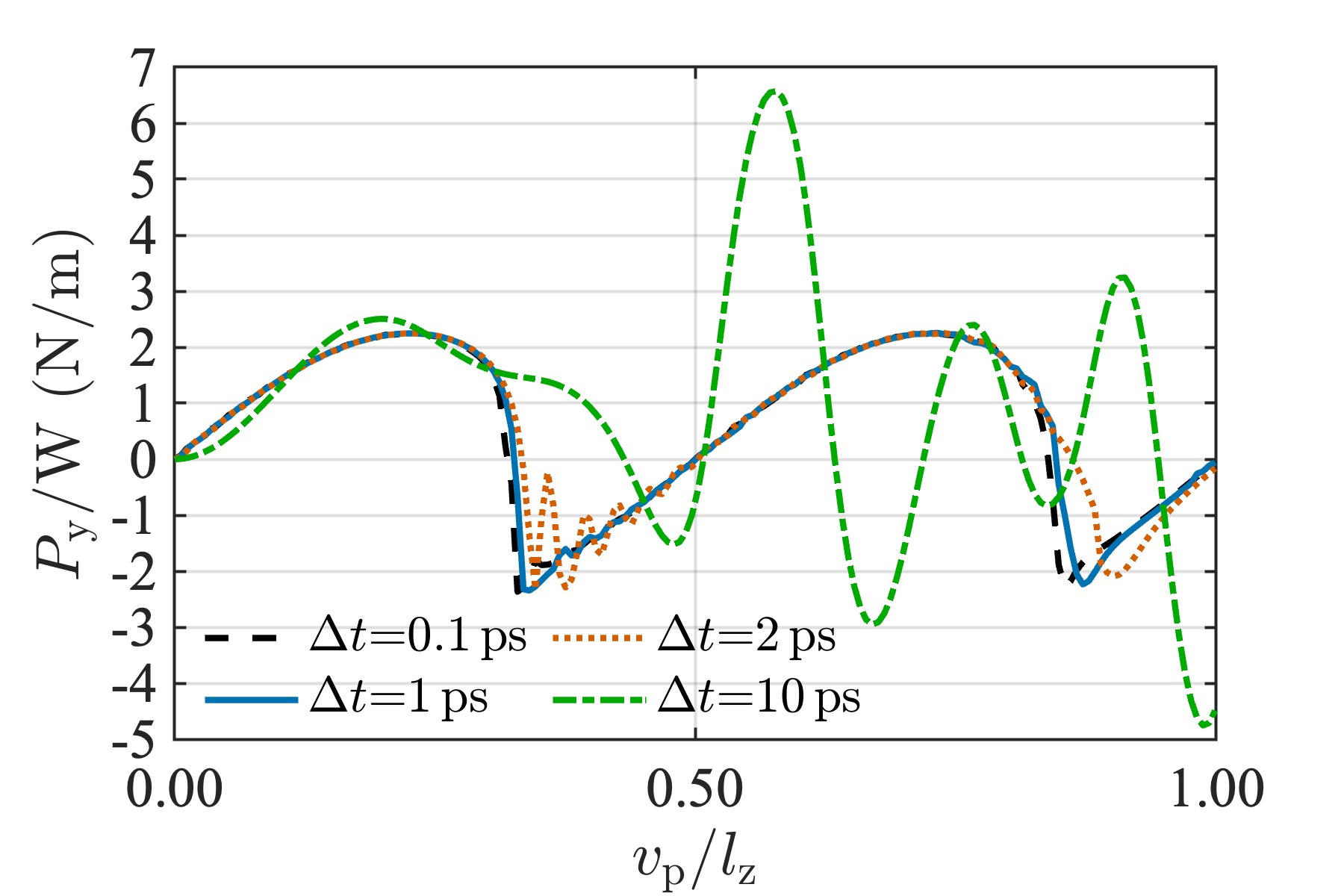}}  
\put(-7.5,0){\fontsize{10pt}{20pt}\selectfont (a)  }
\put(0.3,0){\fontsize{10pt}{20pt}\selectfont (b) }
\end{picture}
\caption{ Effect of dynamic FE simulation parameters on the pulling force variation of pulling a 16 nm GNR along the zigzag direction: (a) Rayleigh damping coefficient $\alpha$ for $\Delta t$ = 1 ps; (b) time step $\Delta t$ for $\alpha$ = 1 ps$^{-1}$. }
\label{Effect of time step}
\end{center}
\end{figure}

The time step size $\Delta t$ and the Rayleigh damping coefficient $\alpha$ significantly affect the frictional response. As shown in Fig.~\ref{Effect of time step}a, the effect of $\alpha$ is examined across three orders of magnitude. Similarly, the effect of time step is examined in Fig.~\ref{Effect of time step}b. We observe little difference between $\alpha = 0.1\text{ps}^{-1}$ and $\alpha$ = $1\text{ps}^{-1}$, and so the later value is used for the simulations reported in this work; and we observe that $\Delta t = 1$ps provides suitable accuracy. 

\section{GNR geometric parameters}\label{Appendix B}

Table \ref{table2} provides geometric details of the GNR studied here. The use of rounded (integer) values to represent the GNR lengths is adopted for clarity in presentation and discussion. A similar approach regarding the GNR dimensions is consistently followed throughout this work.

\begin{table}[H]
\centering
{\fontsize{9}{12}\selectfont
\renewcommand{\arraystretch}{1}
\begin{tabular}
{|p{1.1cm}|p{1cm}|p{1.5cm}|p{1cm}|p{1.5cm}|p{1.1cm}|p{1cm}|p{1.5cm}|p{1cm}|p{1.5cm}|}
\hline
\multicolumn{5}{|c|}{Armchair GNR} & \multicolumn{5}{c|}{Zigzag GNR} \\ \hline
Integer length $L$(nm)& Actual length (nm) & Unit cells along ${ \boldsymbol{e}}_\mathrm{a}$ & Width $W$(nm) & Unit cells along ${ \boldsymbol{e}}_\mathrm{z}$ &Integer length $L$(nm)& Actual length (nm) & Unit cells along ${ \boldsymbol{e}}_\mathrm{z}$ & Width $W$(nm) & Unit cells along ${ \boldsymbol{e}}_\mathrm{a}$ \\ \hline
5 &4.830 & 23 &0.725& 3&   5 & 5.092 & 21&0.70& 3 \\ 
10 &9.870 & 47&0.725&  3&  10 & 9.942 & 41&0.70& 3 \\ 
12 &11.970 & 57&0.725& 3&   12 & 12.124 & 50&0.70& 3 \\ 
15 &14.910 & 71 &0.725& 3&  14 & 14.064 & 58&0.70 & 3\\ 
18 &17.850  & 85 &0.725& 3&  15 & 15.034 & 62&0.70 & 3\\ 
20 &19.950 & 95 &0.725& 3& 16 &  16.004 & 66 &0.70& 3\\ 
25 &24.990 & 119 &0.725& 3&  20 & 20.126 & 83&0.70 & 3\\ 
30 &29.820 & 142 &0.725& 3& 30 &   30.068 & 124&0.70 & 3\\ 
40 &39.900 & 190 &0.725& 3&  40 & 40.010 & 165&0.70 & 3\\ 
50 &49.980 & 238 &0.725& 3&  50 & 50.194 & 207&0.70 & 3\\ 
60 &59.850 & 285 &0.725& 3&  70 & 70.078 & 289&0.70 & 3\\ 
70 &69.930 & 333 &0.725& 3&  80 & 80.020 & 330&0.70 & 3\\ 
\hline
\end{tabular}
}
 \caption{Geometric parameters of GNRs.}
    \label{table2}
\end{table}

\section{MD simulation for interlayer sliding}\label{Appendix C}
MD simulations are conducted to study the interlayer sliding behavior of GNR over a rigid graphene substrate for all three boundary conditions. The MD model consists of a GNR of width $\approx$ 0.7 nm and a wide range of lengths. 
For modeling the interlayer interaction, the Kolmogorov-Crespi (KC) potential \citep{Kolmogorov2005} and for intralayer atomic interactions, the AIREBO potential \citep{Stuart2000} are utilized. 
The relaxed configuration is obtained by energy minimization, using the FIRE algorithm, followed by thermal equilibration at 0.1 K, using the Nos\'e-Hoover thermostat \citep{Evans1985}. Loading along the ${ \boldsymbol{e}}_\mathrm{a}$
or ${ \boldsymbol{e}}_\mathrm{z}$ - direction for all MD simulations are carried out at a constant speed of 0.1 m/s with a time step of 0.2 fs at 0.1 K. Periodic boundary conditions are applied along the lateral directions in the plane to mitigate the effect of the longitudinal edges on the sliding behavior.  
All MD simulations are performed using the Large-scale Atomic/Molecular Massively Parallel Simulator (LAMMPS) \citep{Plimpton1995}.

\section{MD simulation of edge pulled sliding} \label{Appendix D}

\begin{figure}[H]
\begin{center} \unitlength1cm
\begin{picture}(0,5.5)
\put(-8,0){\includegraphics[height=5.5cm]{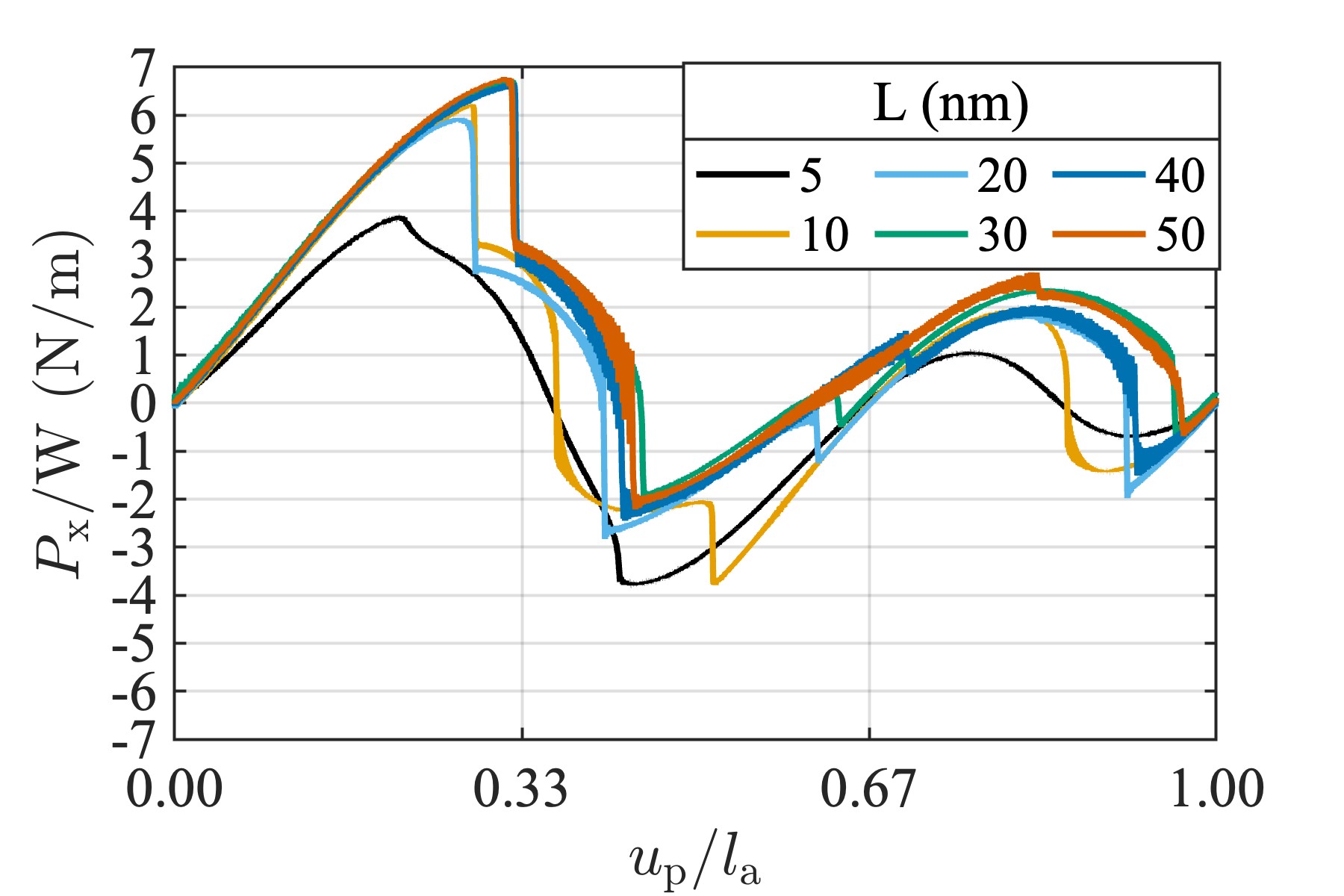}}
\put(0,0){\includegraphics[height=5.5cm]{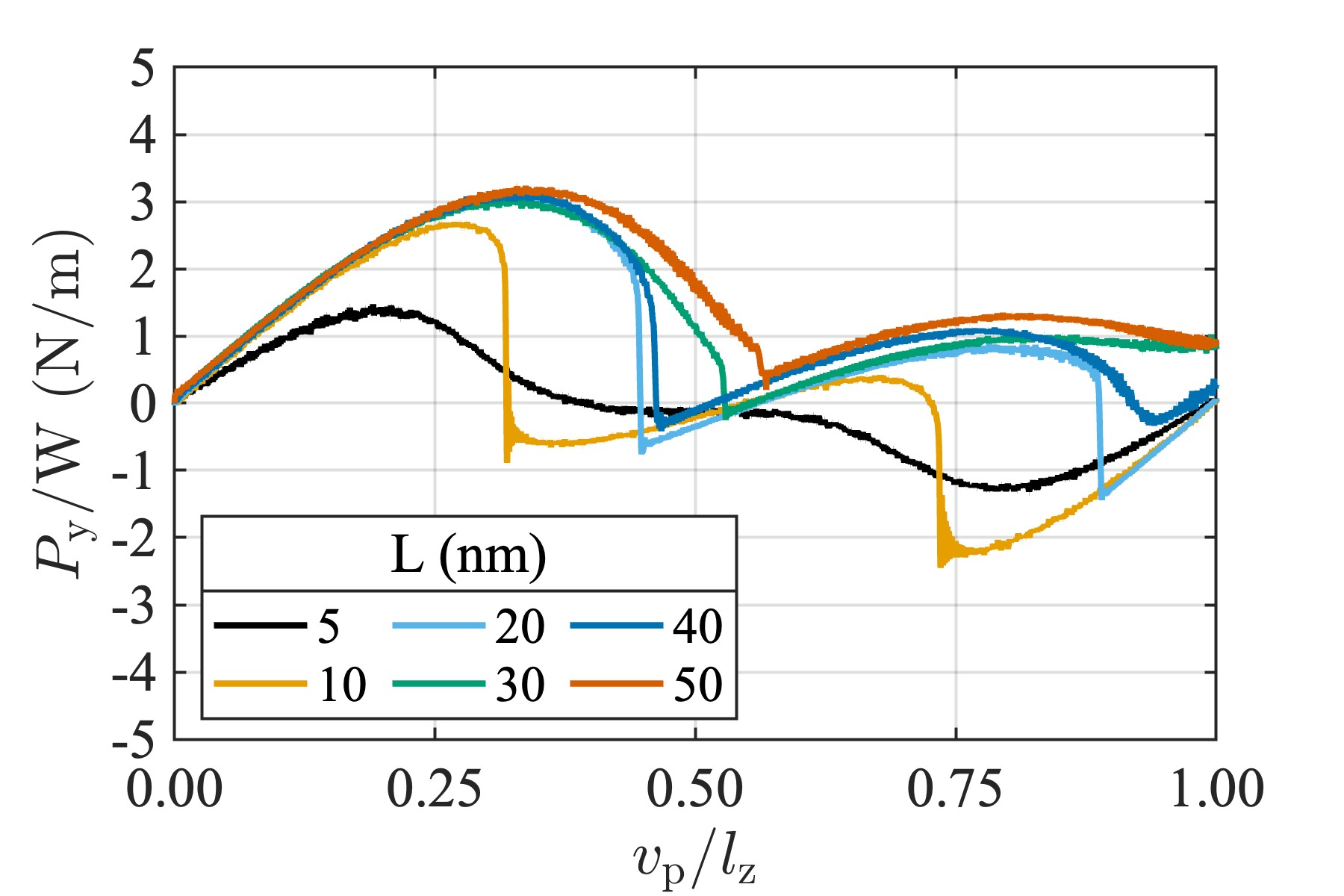}}
\put(-7.5,0){(a)}
\put(0,0){(b)}
\end{picture}
\caption{Effect of ribbon length $L$ on the sliding behavior for the BC 2, shown through the variation of pulling force with end displacement from MD simulations for (a) armchair and (b) zigzag direction. Here, $v_\mathrm{p}$ is a prescribed displacement in the zigzag direction applied at the pulling end.} 
\label{MD simualtion sliding results}
\end{center}
\end{figure}

 It is observed that for small ribbon lengths (\( L < 5 \)nm), the pulling forces 
$ {P}_{\textrm{x}} $
  and  $ {P}_{\textrm{y}} $, corresponding to GNR sliding along the armchair and zigzag directions of the graphene substrate, respectively, exhibit a smooth transition from positive values (indicative of a resistive response of the graphene nanoribbon, GNR) to negative values (indicative of a driving response). For short ribbons $ {P}_{\textrm{x}} $
  and  $ {P}_{\textrm{y}} $ vary similary with sliding displacement as the force acting on the carbon atom, see Fig.~\ref{gap_vector_schematic}b and Fig.~\ref{MD simualtion sliding results}. This behavior highlights the conservative nature of the interlayer shear forces for shorter ribbons. However, for longer ribbons (\( L > 5 \) nm), sudden transitions and abrupt drops in shear force from positive to negative are observed, indicating snap-through instability.

A longer ribbon (\( 5 \,\text{nm}< L < 20 \) nm) corresponds to increased interlayer shear resistance to sliding, leading to localized deformation near the pulling edge. At the same time, the rest of the GNR remains largely undeformed. As a result, the system attempts to equilibrate itself using the available degrees of freedom, particularly in the case of unconstrained lateral sliding. This results in in-plane bending of the ribbon, following Path 1 for the armchair direction (see Fig.~\ref{Sliding paths}a) and Path 3 for the zigzag direction (see Fig.~\ref{Sliding paths}c). For even longer ribbons (\( L > 20 \) nm), the sudden transition from positive to negative shear force is delayed, and force peaks become saturated, indicating a critical length beyond which force saturation occurs. In these long GNRs, in-plane deformation becomes so significant that atoms near the free edge remain in `AB' stacking configuration, effectively balancing the force exerted by the rest of the GNR. As the pulling head reaches a displacement equal to the lattice period (\( l_\textrm{a} \) or \( l_\textrm{z} \)), a strain soliton propagates toward the free end, resulting in a global displacement of the GNR by one lattice wavelength, see \href{https://doi.org/10.6084/m9.figshare.30061102.v1}{Supplementary Movie 1}. Also, captured in \href{https://doi.org/10.6084/m9.figshare.30061102.v1}{Supplementary Movie 1} is the occurrence of a slip (sudden large displacement of the free end) as the soliton glides down the free end. Here, $ {P}_{\textrm{x}} $ and  $ {P}_{\textrm{y}} $ values are obtained as the sum of the reaction forces due to the rest of the GNR on the region of prescribed velocity.

\section{Perturbation effect on static armchair sliding}\label{Appendix E}
\begin{figure}[H]
\begin{center} \unitlength1cm
\begin{picture}(0,5.5)
\put(-8,0){\includegraphics[height=5.5cm]{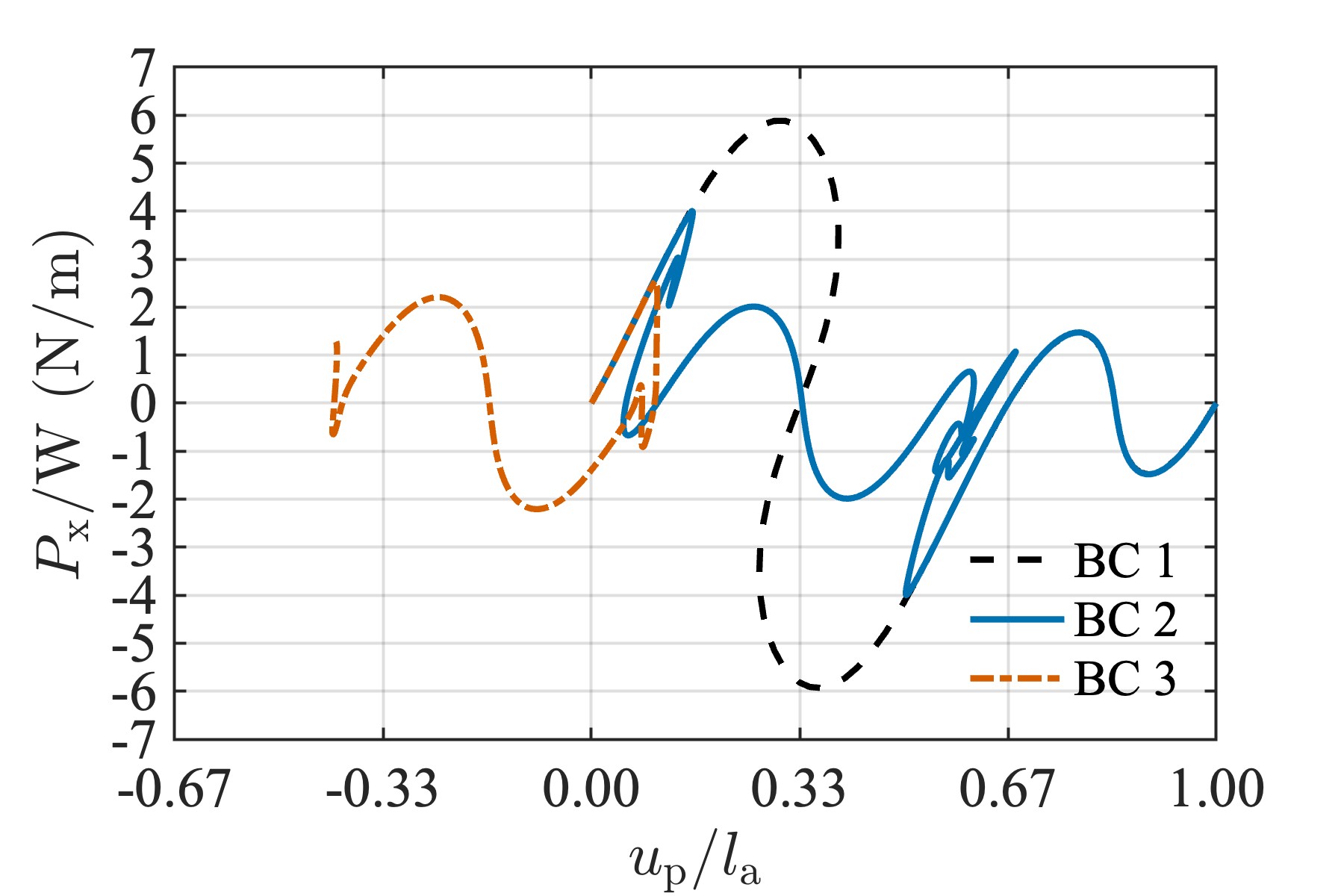}}
\put(0,0){\includegraphics[height=5.5cm]{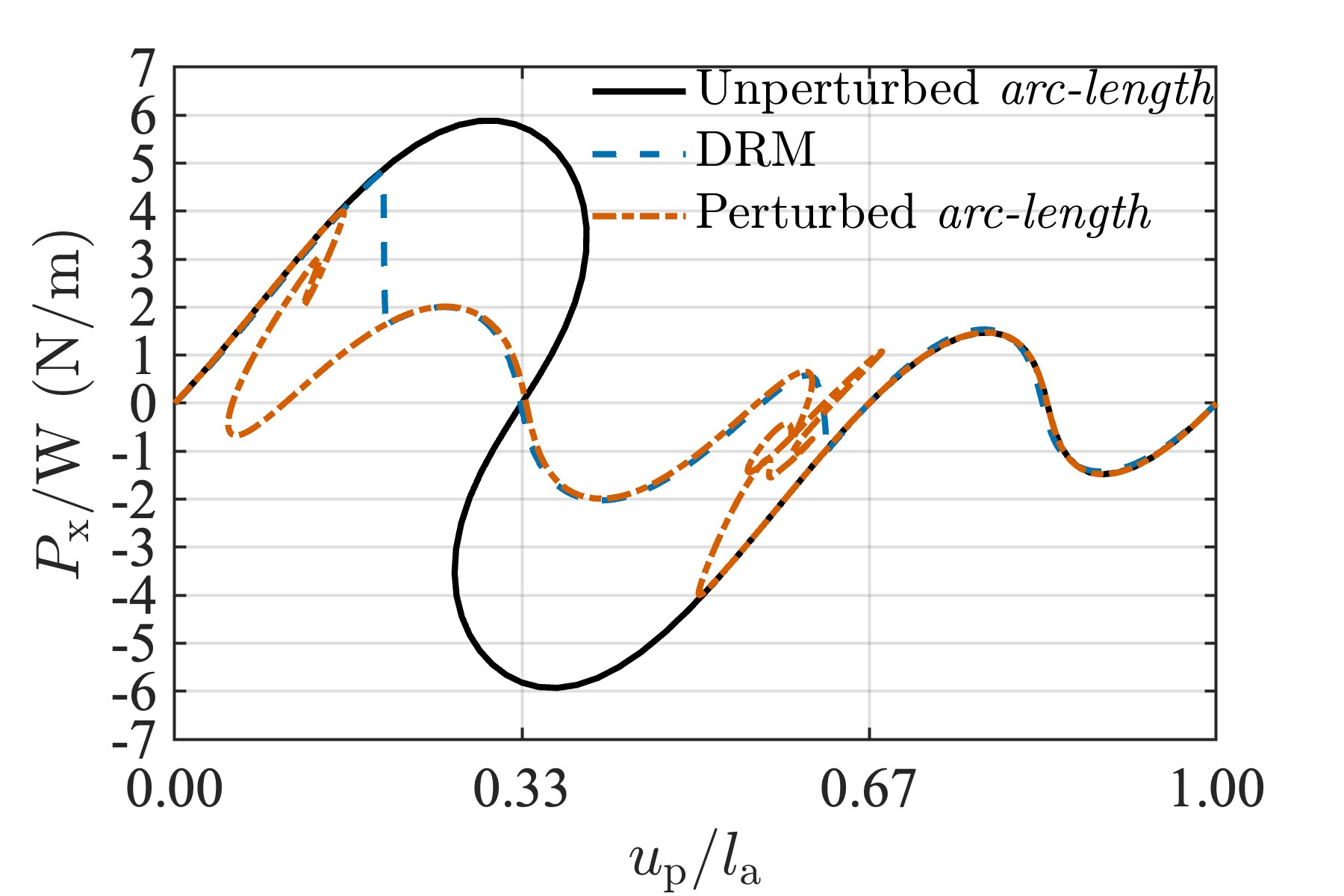}}
\put(-7.5,0){(a)}
\put(0,0){(b)}
\end{picture}
\caption{Sliding results of a 10 nm armchair GNR under lateral perturbation using the arc-length solver. (a) Effect of boundary conditions. (b) Comparison with DRM under BC 2.} 
\label{arclength_bcs_effect_perturbed}
\end{center}
\end{figure}

Here, the effect of perturbations on static sliding is examined. In the unperturbed static, perfectly symmetric sliding of an armchair GNR along the ${\boldsymbol{e}}_\mathrm{a}$ - direction, the symmetric energy barrier ensures that path 2 is always followed, as shown in Fig.~\ref{All cases}. However, when the system is laterally perturbed -- thus breaking the symmetry -- a boundary condition dependent response emerges, as illustrated in Figs.~\ref{arclength_bcs_effect_perturbed}a and \ref{arclength_bcs_effect_perturbed}b.

 Lateral perturbation results in the in-plane buckling and the system follows a path similar to Path 1, with the response matching that observed under dynamic conditions (see Figs.~\ref{DRM 10 nm all bcs}a and \ref{arclength_bcs_effect_perturbed}b ).
For the laterally perturbed static case with BC 3, the GNR initially buckles exactly at the expected point (as in the DRM case -- half of $P_{\text{y}}^{\text{max}}$ under BC 1) but then moves in the opposite direction, owing to the existence of multiple equilibrium configurations. It should also be noted that, for the present perturbation magnitude, a slight difference is observed between the perturbed arc-length results and those from DRM (Fig.~\ref{arclength_bcs_effect_perturbed}b), which is likely due to the specific perturbation amplitude applied.

\section{Characteristic length scale}\label{Appendix F}
Under the small deformation assumption, an analytical solution of the resulting displacement is provided for sliding along ${\boldsymbol{e}}_\mathrm{z}$ - direction (with $g_\mathrm{a}$ considered as zero). The shear traction component $t_\text{z}$ along ${\boldsymbol{e}}_\mathrm{z}$ - direction is given by
\begin{equation} 
\ds   t_\text{z}( g_\text{z}) := - \frac{\partial \Psi}{\partial g_z} =  t_{\textrm{zm}} \sin\Bigg(\frac{2\pi g_\textrm{z}}{l_\textrm{z}}\Bigg)~,
\label{zigzag traction}
\end{equation}  
where $t_{\textrm{zm}}$ = $16\pi\Psi_2/(l_\textrm{z} A_0)$ is the shear strength. Following the shear lag model, the mechanical equilibrium of the 1D ribbon can be expressed as 
\begin{equation}    
\ds  \text{E}_{\text{2D}}\frac{\text{d}^2 g_\mrz}{\text{d}x^2}= t_\text{z}(g_\text{z})~,
\label{shear lag equillibrium} 
\end{equation} 
where $\text{E}_{\text{2D}}$ = 2247 $\text{eV}\, \text{nm}^{-2}$, is the material in-plane stiffness. The solution to this equilibrium equation for small values of $g_{\text{z}}$ ($\ll a_{\text{cc}}$) can be obtained by linearizing the shear traction at the AB configurations 
\begin{equation}   
\ds  t_\text{z} \approx \frac{32\pi^2\Psi_2}{l_\textrm{z}^2A_0} \mrg_{\text{z}} = \text{K}_{\text{z}} g_{\text{z}} ~,
\label{traction linearisation} 
\end{equation} 
where $\text{K}_{\text{z}}$ represents the interfacial stiffness. The solution is then given by
\begin{equation}   
\ds  g_{\text{z}} (x) = B_1 \cosh \left(\frac{x}{L_\text{c}}\right) +  B_2 \sinh \left(\frac{x}{L_{\text{c}}}\right) ~,   
\label{general solution along zigzag direction} 
\end{equation}
where $B_1$ and $B_2$ are determined from the boundary conditions, and $L_{\text{c}} = \sqrt{\frac{\text{E}_{\text{2D}}}{\text{K}_{\text{z}}}}$ is the characteristic length scale or stress decay length for shear interactions. For the current parameters, value of $L_{\text{c}}$ is 4.8 nm.

\section{Strain transmission in zigzag GNR under BC 3}\label{Appendix G}
\begin{figure}[H]
\begin{center} \unitlength1cm
\begin{picture}(0,5.5)
\put(-7.8,0){\includegraphics[height=50mm]{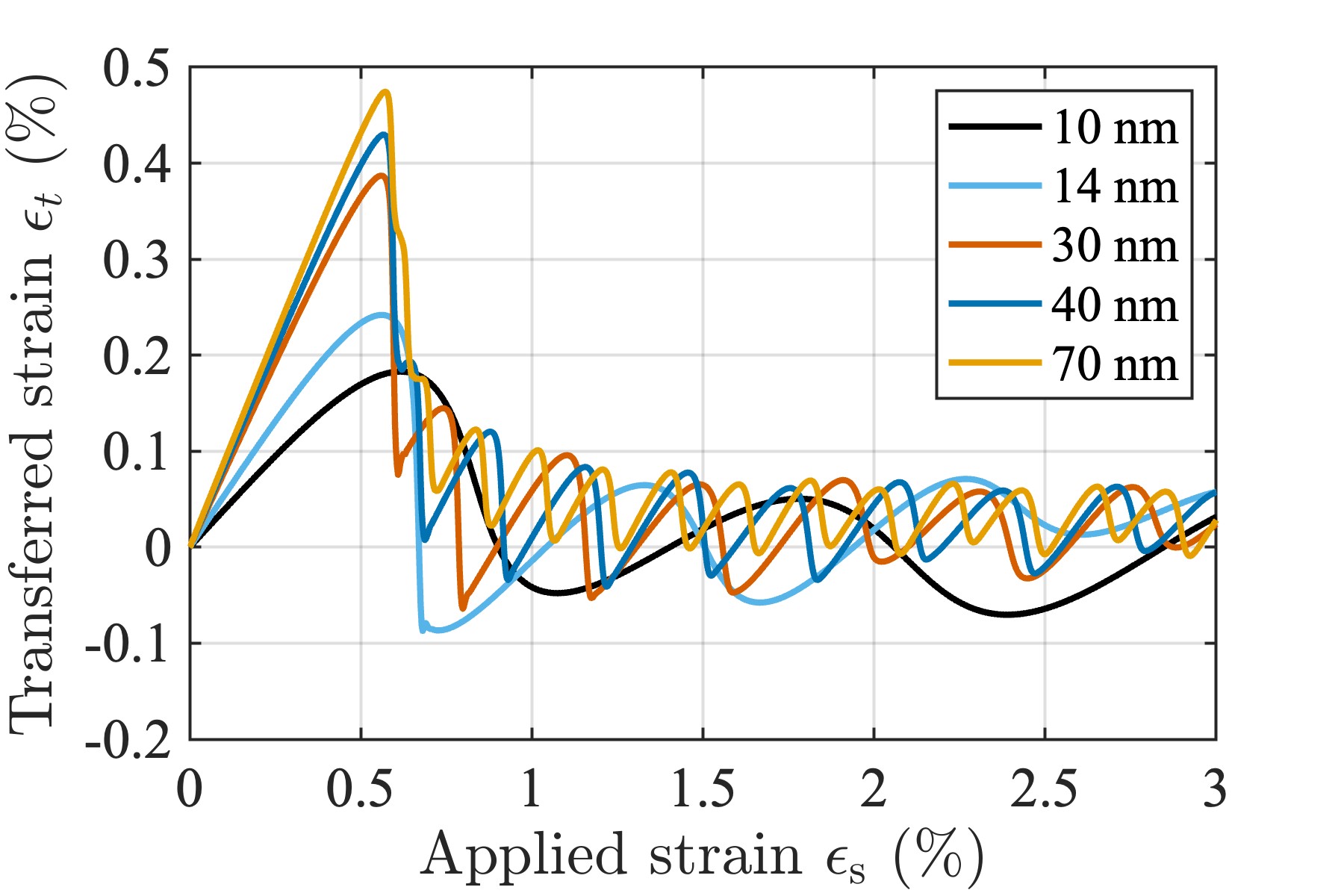}}
\put(0,0){\includegraphics[height=50mm]{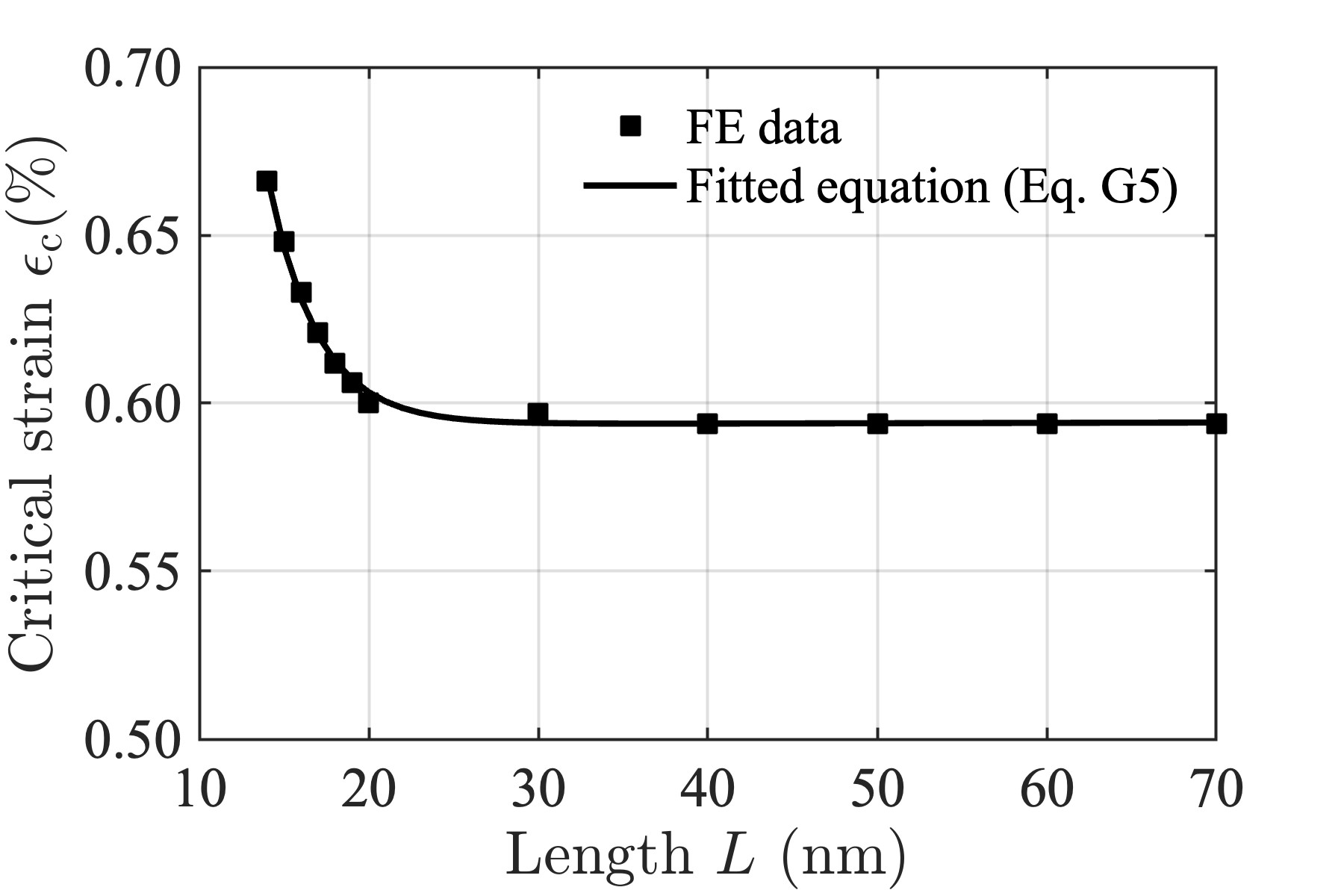}}
\put(-7.5,0){\fontsize{10pt}{20pt}\selectfont (a)  }
\put(0.3,0){\fontsize{10pt}{20pt}\selectfont (b) }
\end{picture}
\caption{Strain transfer with BC 3 in zigzag GNRs. (a) Variation of the average transferred strain $\epsilon_\text{t}$ with the applied strain $\epsilon_\text{s}$ for GNR of  different lengths $L$ (in nm).(b) Critical strain for interlayer sliding with different GNR lengths.}
\label{Strain transmission zigzag BC 3}
\end{center}
\end{figure}

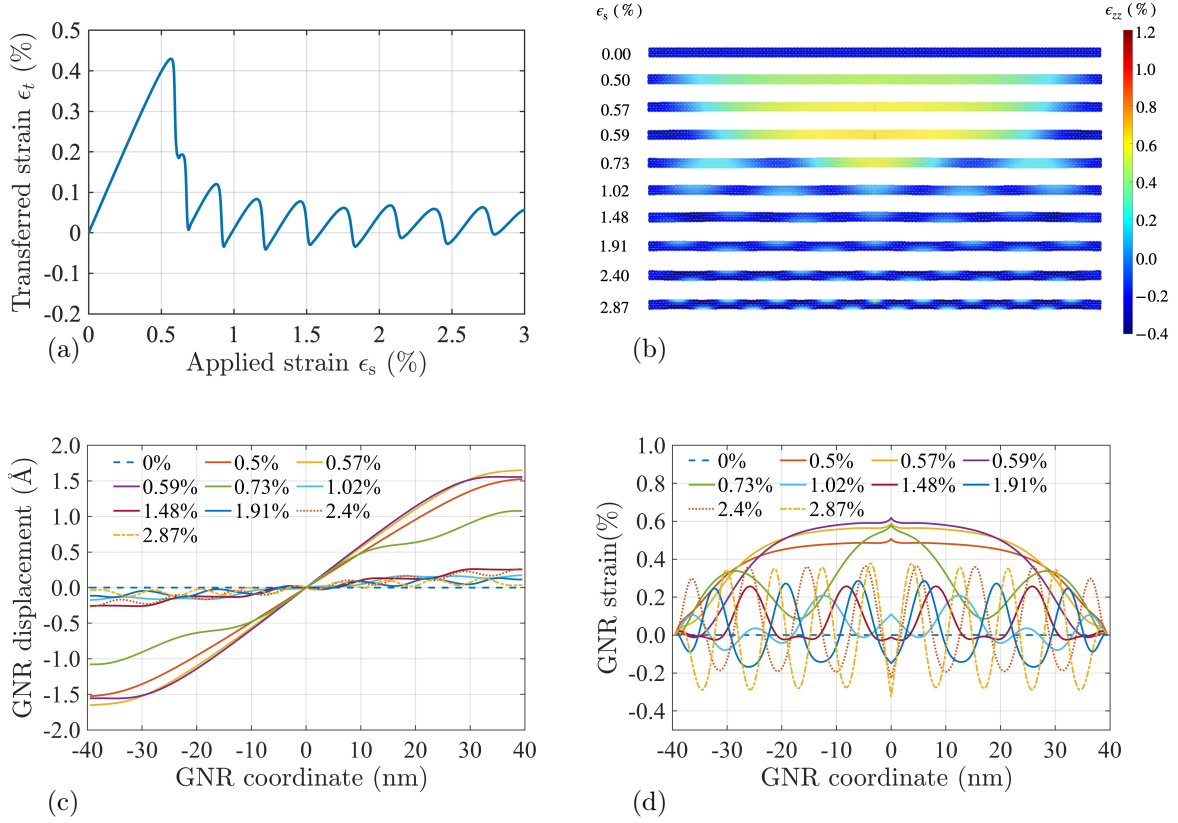
\begin{figure}[H]
\begin{center} \unitlength1cm
\begin{picture}(0,11.3)
\put(-7.7,6){\includegraphics[height=50mm]{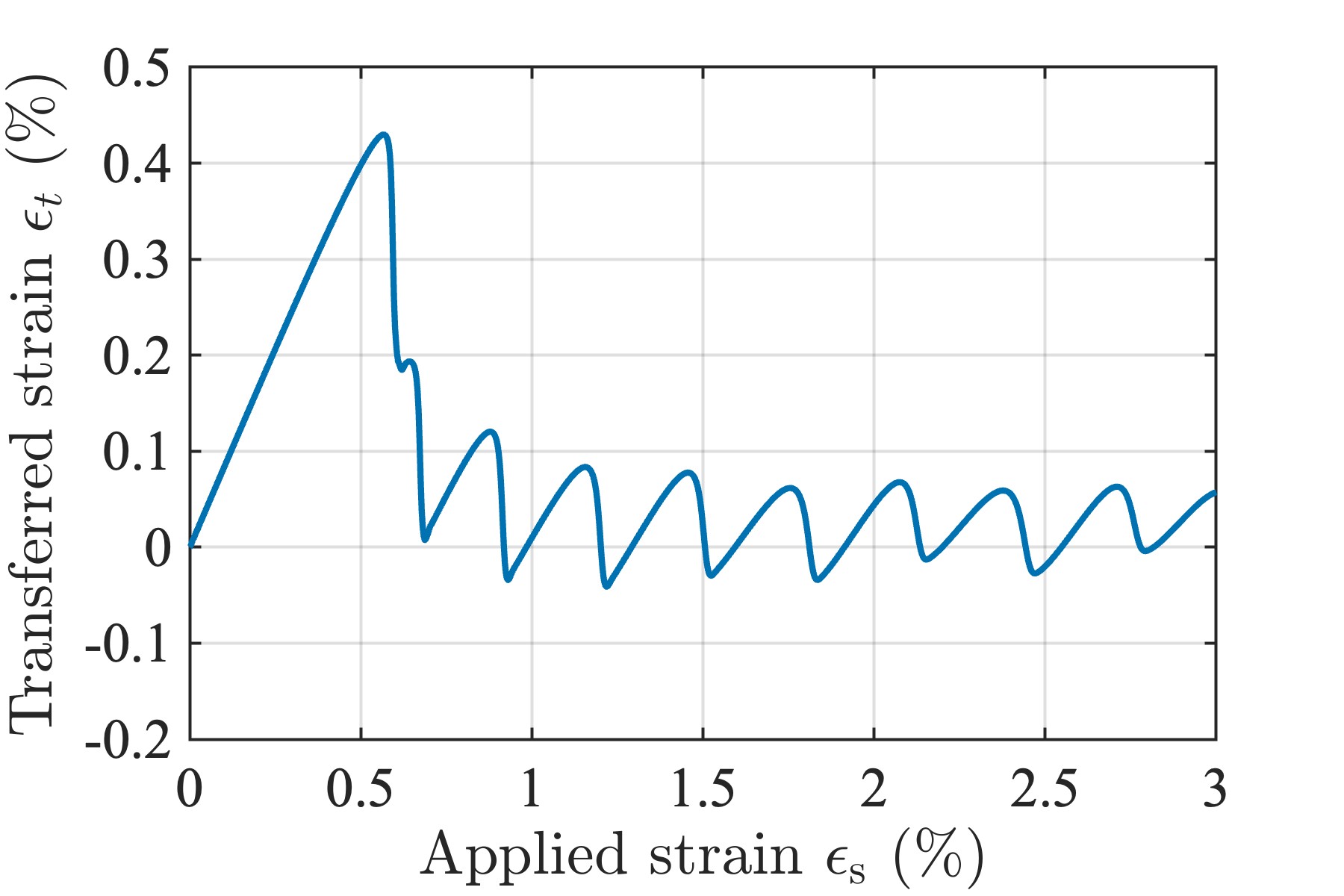}}
\put(0,6.5){\includegraphics[height=45mm]{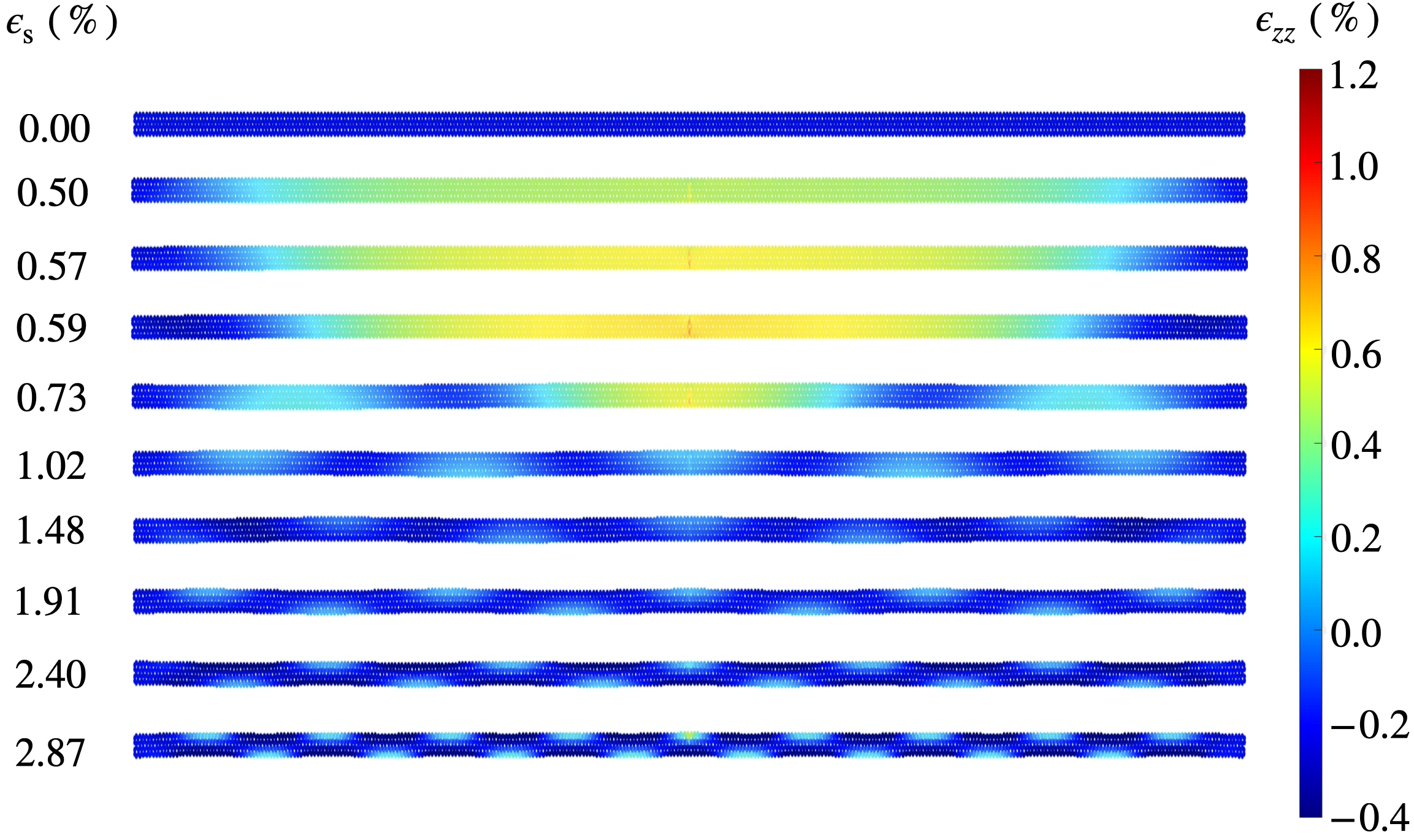}}
\put(-7.7,0.5){\includegraphics[height=50mm]{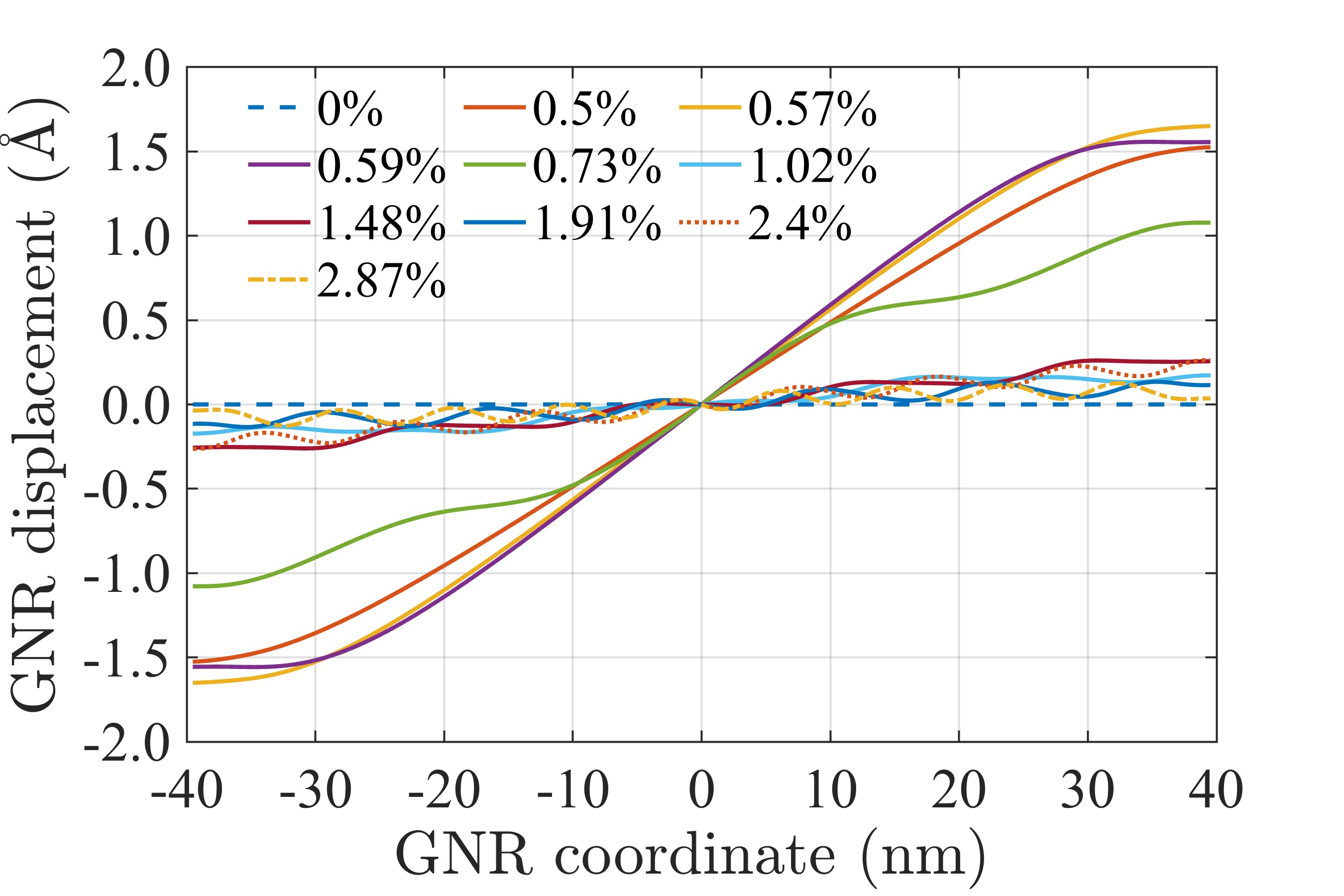}}
\put(0,0.5){\includegraphics[height=50mm]{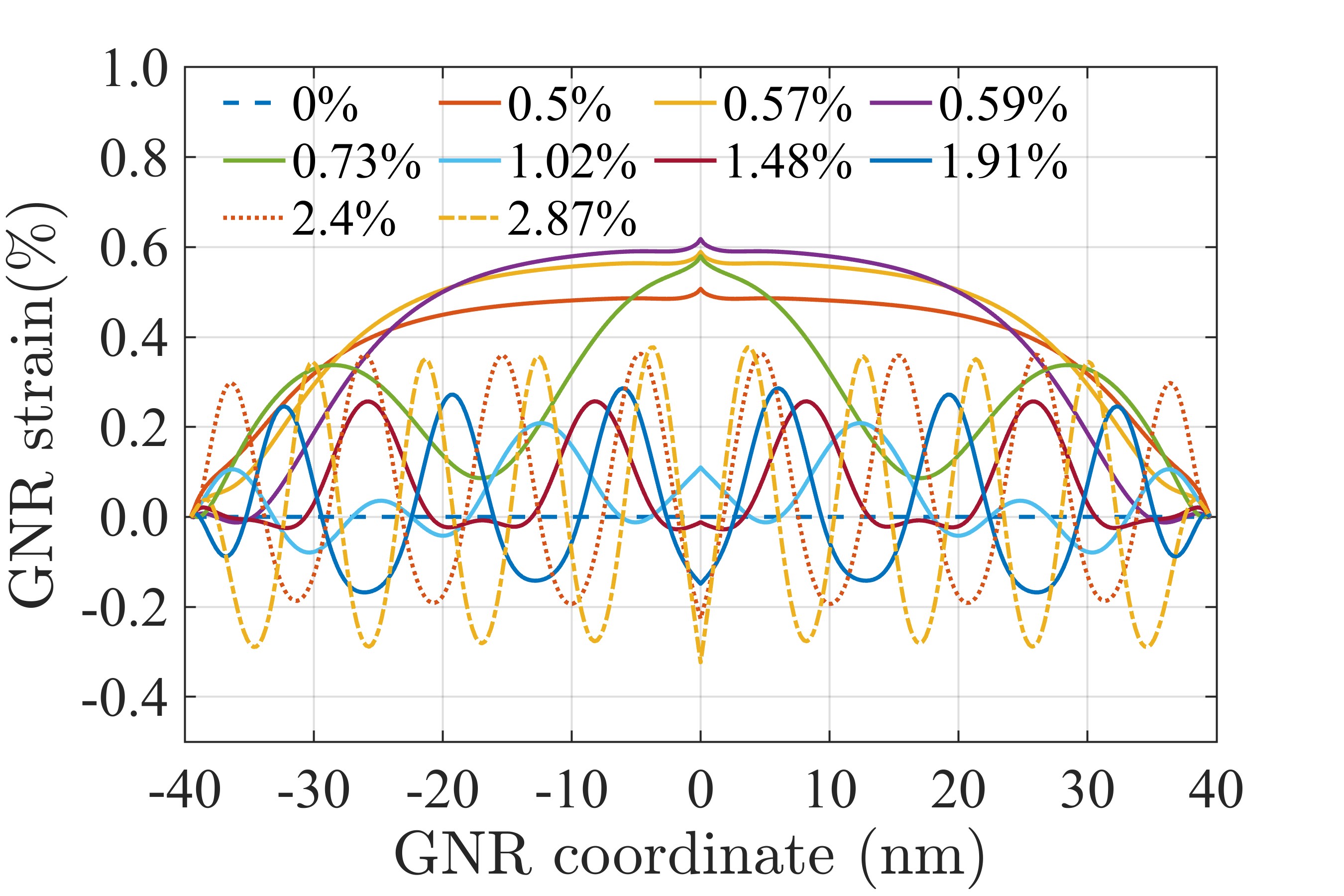}}
\put(-7.2,6.3){\fontsize{10pt}{20pt}\selectfont (a)  }
\put(0.5,6.3){\fontsize{10pt}{20pt}\selectfont (b) }
\put(-7.2,0.3){\fontsize{10pt}{20pt}\selectfont (c)  }
\put(0.5,0.3){\fontsize{10pt}{20pt}\selectfont (d) }
\end{picture}
\caption{Strain transmission with BC 3 in zigzag GNR of length 80 nm (=$2L$) aligned along the ${ \boldsymbol{e}}_\mathrm{z}$-direction of substrate (a) Average strain transferred. (b) Simulation snapshots at key loading stages illustrating the distribution of $\epsilon_{zz}$ along the central axis z-z. The distribution of displacement field (c), and strain field (d)  in GNR for various substrate strains.}
\label{Strain transmission zigzag BC 3 complete analaysis}
\end{center}
\end{figure}

For the completely unconstrained condition (BC 3), the strain transmission is also studied (see \href{https://doi.org/10.6084/m9.figshare.30061102.v1}{Supplementary Movie 3}). The observed behavior, as presented in Figs.~\ref{Strain transmission zigzag BC 3} and \ref{Strain transmission zigzag BC 3 complete analaysis}, is found to be qualitatively similar to that of the zigzag GNR with BC 1, whose results are shown in Figs.~\ref{Strain transfer}--\ref{critical strain transfer and max strain transfer}. Quantitatively, it is observed that $\epsilon_\text{t}$ values are $\approx$ 50$\%$ of those from BC 1.

The obtained critical strain is approximated by exponential relation, given as
\begin{equation}
  \ds     \epsilon_\text{c}  = c_1 \,\text{exp} (c_2\, L) + c_3\,\text{exp} (c_4\, L)~,
  \label{critical_unstrained_zigzag_equation}
\end{equation}
where the fitting constants take the values $c_1 = 8.8162$, $c_2 = -0.3415$ $\text{nm}^{-1}$, $c_3 = 0.5935$, and $c_4 = 0.000001$ $\text{nm}^{-1}$.

\section{Strain transmission in armchair GNR}\label{Appendix H}
\begin{figure}[H]
\begin{center} \unitlength1cm
\begin{picture}(0,4.5)
\put(-7.8,0){\includegraphics[height=50mm]{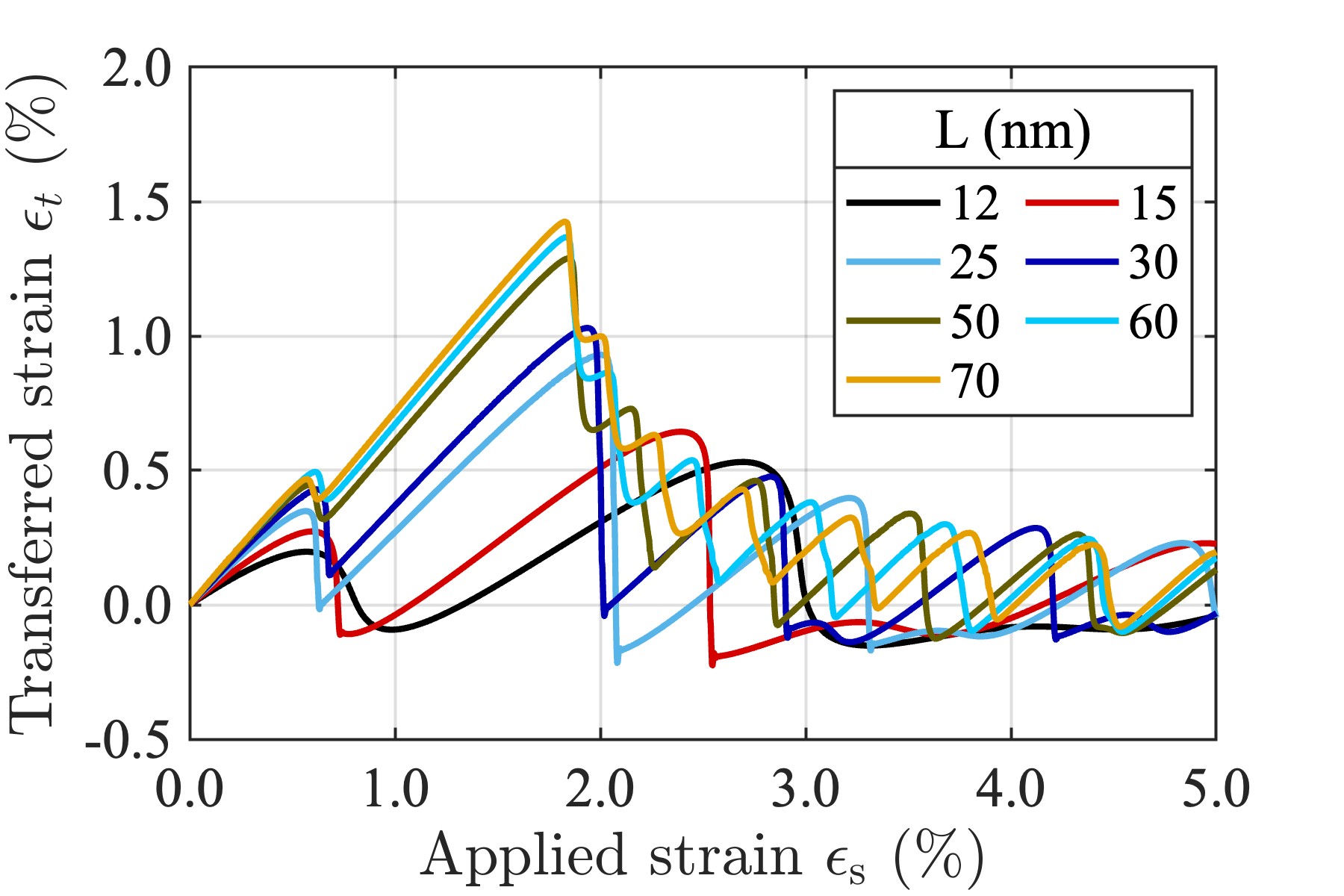}}
\put(0,0){\includegraphics[height=50mm]{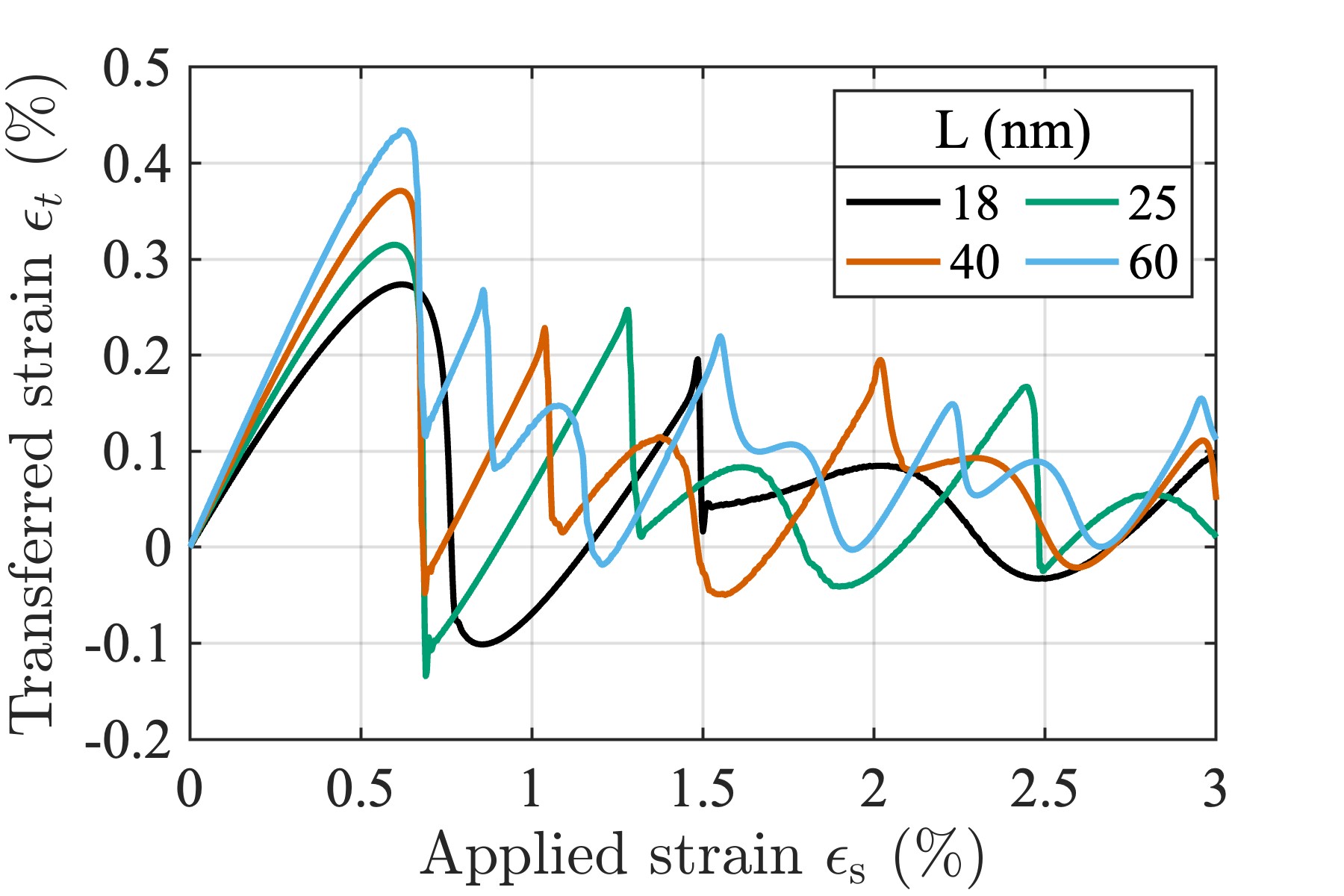}}
\put(-7.5,0){\fontsize{10pt}{20pt}\selectfont (a)  }
\put(0.3,0){\fontsize{10pt}{20pt}\selectfont (b) }
\end{picture}
\caption{Variation of the average transferred strain $\epsilon_\text{t}$ in armchair GNRs with the applied strain $\epsilon_\text{s}$ for different lengths $L$, under BC 1 (a), and under BC 3  (b).}
\label{armchair strain transmission}
\end{center}
\end{figure}

Similar to the strain transmission in zigzag GNRs described in Section~\ref{subsection 3.3} and Appendix~\ref{Appendix G}, the strain transmission in armchair GNRs is also investigated, as shown in Fig.~\ref{armchair strain transmission}. The study is conducted under both BC 1 and BC 3. The observed behavior is qualitatively similar to that of zigzag GNRs, where a linear increase in $\epsilon_\text{t}$ is recorded up to a critical applied strain $\epsilon_\text{c}$, followed by a sudden drop in $\epsilon_\text{t}$ with continued loading of the graphene substrate $\epsilon_\text{s}$. Notably, in the armchair configuration, a double drop in $\epsilon_\text{t}$ is observed, corresponding to the double periodicity along the armchair direction, as seen in Fig.~\ref{armchair strain transmission}a. Furthermore, $\epsilon_\text{c}$ exhibits a clear dependence on ribbon length. However, these effects are less pronounced under BC 3, where the absence of lateral constraints causes the system to lose its alignment with the applied strain direction, resulting in a slightly irregular response (see Fig.~\ref{armchair strain transmission}b).

\bibliographystyle{unsrtnat}
\bibliography{bib_completion_abbreivated}

\end{document}